\documentclass{nsr}

\usepackage{amsmath,graphicx,array}
\usepackage{dcolumn,soul}%

\usepackage{amsthm}
\usepackage{multirow}
\usepackage{adjustbox}
\usepackage{gensymb}
\usepackage[figuresright]{rotating}
\usepackage{listings}%
\usepackage{hyperref}
\hypersetup{
    colorlinks=true,
    citecolor=blue,
    linkcolor=blue,
    urlcolor=blue
}
\usepackage{aas_macros}

\makeatletter
\def\uns{\ifmmode\,\else$\,$\fi}%

\makeatother
 
\jvol{XX}
\jnum{X}
\jyear{Year}
\doi{10.1093/nsr/XXXX}
\received{XX XX Year}
\revised{XX XX Year}
\accepted{XX XX Year}

\usepackage{amsfonts}%
\usepackage{mathrsfs}%
\usepackage[title]{appendix}%
\usepackage{booktabs}%
\usepackage{float}
\usepackage{pdflscape}
\usepackage{pgf,pgffor}
\usepackage{textcomp}
\usepackage{threeparttable} 
\usepackage{xcolor}
\usepackage{lineno}
\usepackage{caption}
\usepackage[caption = false]{subfig}
\usepackage{academicons}
\usepackage{pifont}
\usepackage{longtable}
\usepackage{cleveref}
\definecolor{orcidlogocol}{HTML}{A6CE39}
\makeatletter

\newcommand{\AFF}[1]{{\foreach\d[count=\ni]in{#1}{\ifnum\ni=1\ref{\d}\else,\ref{\d}\fi}}}

\begin{document}

\dhead{RESEARCH ARTICLE}

\subhead{PHYSICS}

\title{The complex periodic variability of PKS 0301-243: hints of a supermassive black hole binary candidate}

\author{Jintong Wang$^{1,2}$}
\author{Yunguo Jiang$^{3,4,*}$}
\author{Shifeng Huang$^{5,6}$}

\affil{$^1$SDU-ANU Joint Science College, Shandong University, Weihai, 264209, China}
\affil{$^2$Department of Physics and Astronomy, University of Bologna, Via Gobetti 93/2, Bologna, 40129, Italy}
\affil{$^3$Shandong Key Laboratory of Space Environment and Exploration Technology, Institute of Space Sciences, Shandong University, Weihai, 264209, China}
\affil{$^4$School of Space Science and Technology, Shandong University, Weihai, 264209, China}
\affil{$^5$Department of Astronomy, University of Science and Technology of China, Hefei 230026, China}
\affil{$^6$School of Astronomy and Space Sciences, University of Science and Technology of China, Hefei 230026, China}

\authornote{\textbf{Corresponding authors.} Email: jiangyg@sdu.edu.cn}

\abstract[ABSTRACT]{In hierarchical galaxy formation, supermassive black hole binaries are a ubiquitous consequence of frequent galaxy mergers\cite{Haehnelt_2002}, yet subparsec binaries remain difficult to be confirmed. Periodic flux modulations may arise from the orbital motion of black holes in close binaries \cite{D'Orazio_2015,Graham_2015,Mingarelli_2025}, while relativistic jet precession can provide an additional source of periodic variability \cite{Britzen_2023}. Such periodic signatures therefore motivate us to identify close binary candidates. Here, we report an optical periodicity of $269^{+13}_{-14}$ days in both Zwicky Transient Facility $g$ and $r$ bands of the blazar PKS 0301-243, with corrected global significances of $2.75\sigma$ and $2.18\sigma$, respectively. This signal persists over 7 cycles across a baseline of $\sim2000$ days. We interpret this period as a possible modulation associated with the orbital motion of a supermassive black hole binary. A longer optical periodicity of $889^{+47}_{-46}$ days is identified in Katzman Automatic Imaging Telescope, with corrected global significance of $1.97\sigma$. The longer periodicity spans approximately 5 cycles over a baseline of $\sim 4500$ days, potentially associated with jet precession. Both periodicities are independently identified using multiple methods.
A supermassive black hole binary incorporating a relativistic jet and a circumbinary disc could reproduce the observed light curve ensemble. 
For a total mass of $10^{8.95}M_\odot$ and a mass ratio $q=0.1$, the derived milli-parsec separation and a short gravitational wave driven coalescence time, $t_{\rm c} = 1122^{+144}_{-144}$ years, imply that PKS 0301-243 may be a rare source of nano-hertz gravitational waves and a potential multimessenger laboratory for the terminal stages of supermassive black hole coalescence\cite{Charisi_2022,Mingarelli_2025}.} \keywords{supermassive black hole binary, blazar, time-domain astronomy, gravitational wave, multimessenger astronomy}

\maketitle

\section{INTRODUCTION} Active galactic nuclei (AGNs) harbouring a supermassive black hole (SMBH) are among the most luminous objects in the universe. Blazars are a special subtype of AGN, characterized by the high variability caused by relativistic jets that directed towards observers\cite{Urry_1995}. 
The quasi-periodic variability across different bands of blazars are also intriguing observational phenomena \cite{Ackermann_2015,Sarkar_2021,ONeill_2022}, and may originate from the orbital motion in a SMBH binary (SMBHB) system\cite{Sobacchi_2017}. A prominent, though debated, candidate is OJ 287, which exhibits complex optical outbursts with a quasi-periodic structure. Although it was traditionally modeled as a binary system with a 12-year orbital cycle\cite{Laine_2020}, recent observational campaigns have sparked discussions regarding its true physical nature and the precise timing of its flares\cite{Komossa_2023,Valtonen_2024}. For blazars, the stochastic jet emission could overwhelm the regular orbital signal of SMBHB. The credibility of a periodic signal therefore generally increases with the number of observed cycles, the duration of the monitoring baseline, and its consistency across different methods\cite{Vaughan_2016,Haiman_2023}.

In the hierarchical galaxy formation scenario, SMBHBs are expected to form following galaxy mergers \cite{Haehnelt_2002}. Dual active galactic nuclei with kiloparsec separations are commonly observed in X-ray, optical, and infrared bands\cite{Komossa_2003,Comerford_2011,DeRosa_2023,Chen_2026}. At sufficiently small separations, numerical simulations predict that gas from large radii gradually spirals inward though viscous dissipation, while the accretion disc is truncated at a few binary separations by the gravitational torques exerted by the SMBHB, forming a cavity. Subsequently, gas from the inner edge of the circumbinary disc is transported through accretion streams and feeds mini-discs around the individual SMBHs \cite{D'Orazio_2013,Farris_2014,Lai_2023}. Within this accretion geometry, periodic variability can arise from the modulation of the accretion rate by the binary orbital motion \cite{D'Orazio_2013,Farris_2014,Miranda_2017}. Orbital motion can also produce sinusoidal  light curves  through relativistic Doppler boosting, particularly when one of the two mini-discs  dominates the observed flux \cite{D'Orazio_2015,Charisi_2022}. Periodic variability therefore provides the indirect evidence for  close SMBHB  candidate, but its significance estimation requires careful consideration of both stochastic variability and alternative physical processes.

Here, we report the discovery of an additional $\sim270$-day periodic signal from the Zwicky Transient Facility (ZTF) observations, traced over 7 cycles, in the BL Lac object PKS 0301-243 ($z=0.266$ \cite{Pita_2014}). Previously, this source was identified as one of only 6 out of 155 blazars ($\sim4\%$), showing a $\sim894$-day periodic signal above $3\sigma$ significance  from the  the Katzman Automatic Imaging Telescope (KAIT) light curve \cite{Chen_2024}. The significance estimation was based on a simple power-law (SPL) red noise model through both the generalized Lomb–Scargle periodogram (GLSP) and the weighted wavelet Z-transform (WWZ). 

Among the available datasets, we focus primarily on the KAIT light curve and the ZTF $g$ and $r$ bands light curves, which provide the best temporal coverage and sampling for us robust characterization of the periodic variability. The  KAIT observations span approximately 4500 days, while the ZTF observations cover approximately 2000 days. Additional optical, near Infrared (IR), and $\gamma$-ray light curves are presented in the Supplementary Material.


\section{RESULTS}

\begin{figure*}[h!]
\centering
\includegraphics[width=0.32\textwidth]{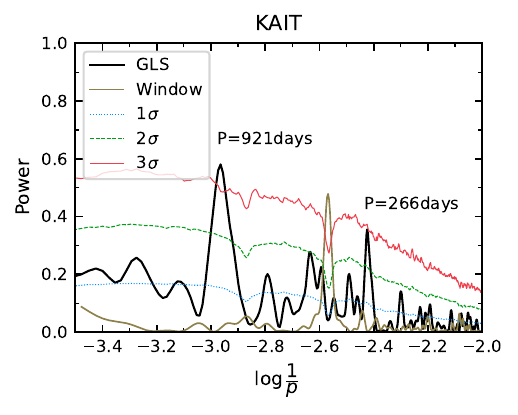}
\includegraphics[width=0.32\textwidth]{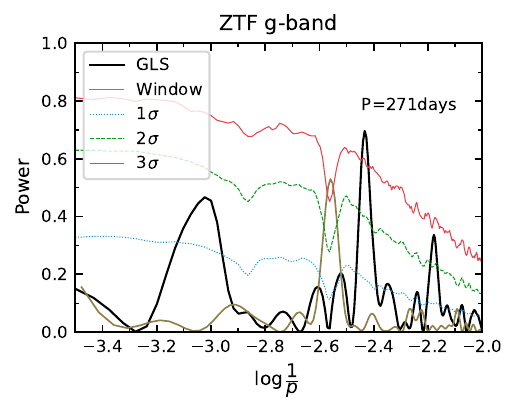}
\includegraphics[width=0.32\textwidth]{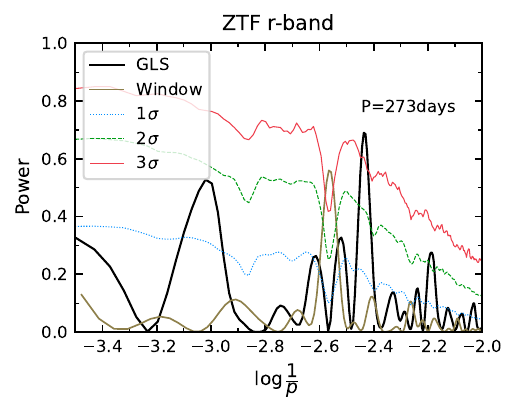}
\includegraphics[width=0.32\textwidth]{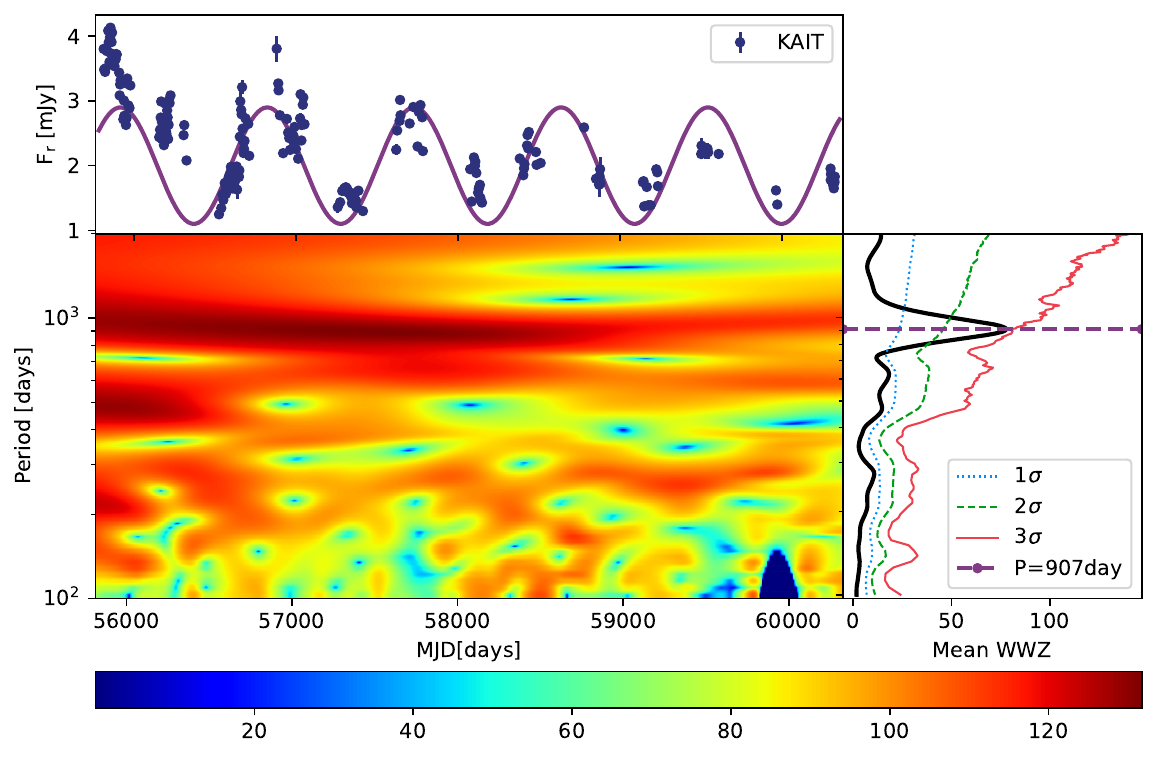}
\includegraphics[width=0.32\textwidth]{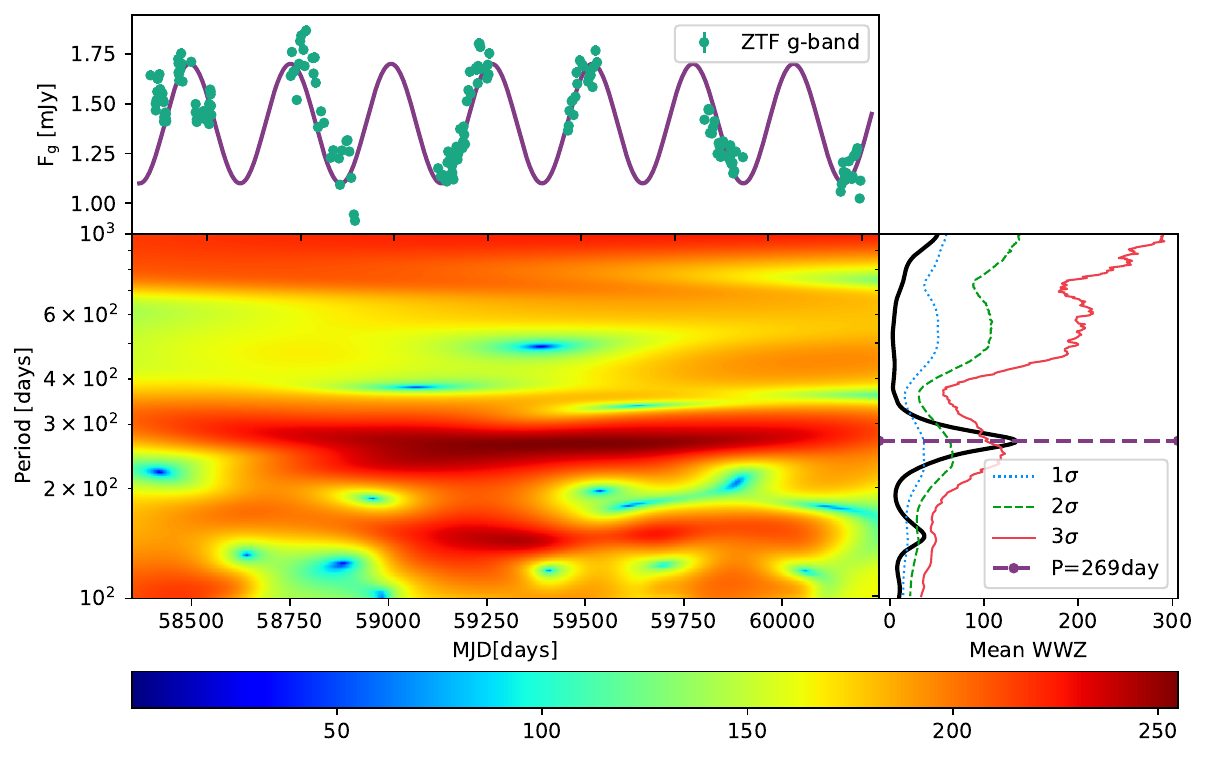}
\includegraphics[width=0.32\textwidth]{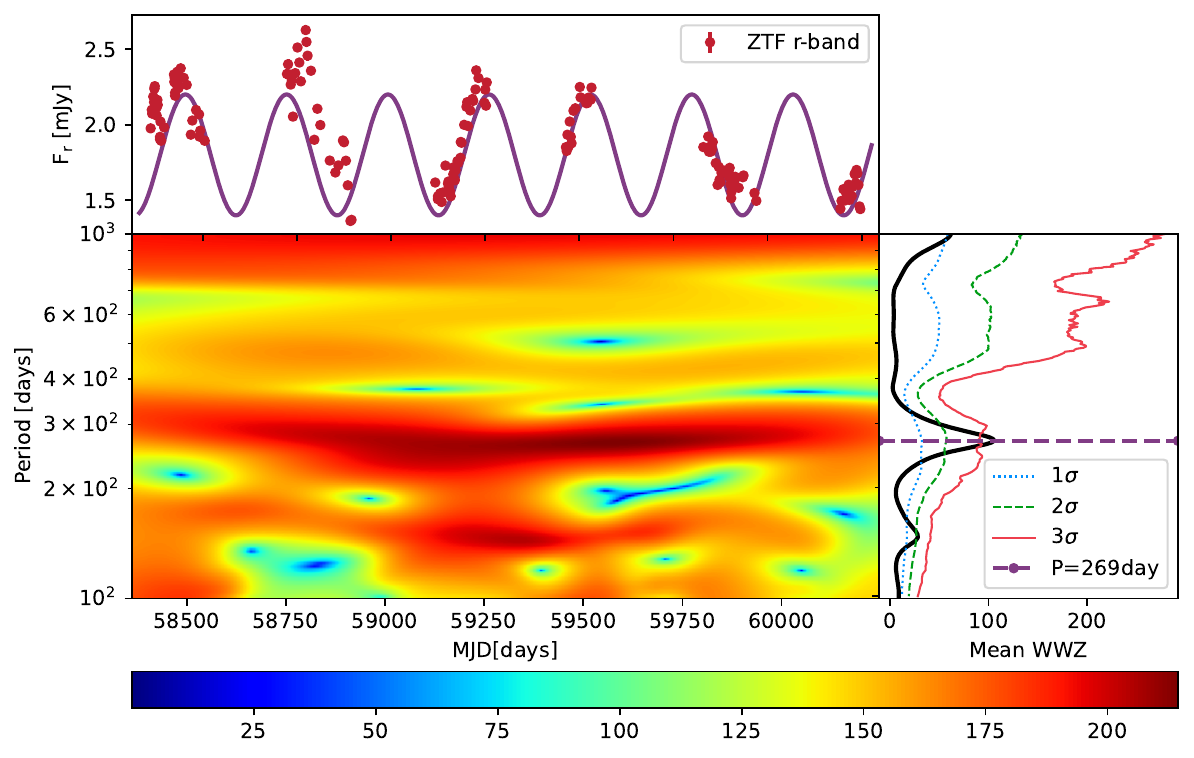}
\includegraphics[width=0.32\textwidth]{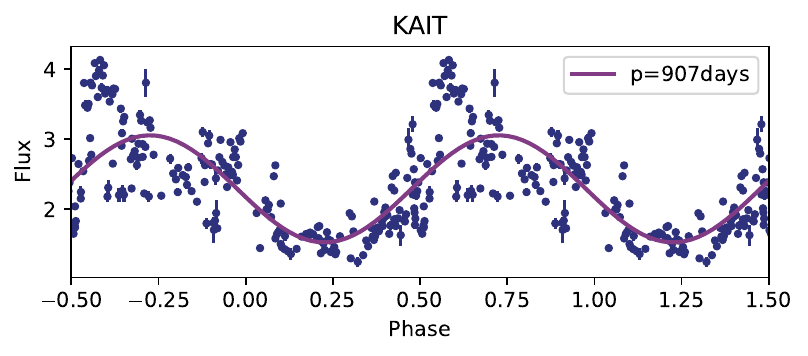}
\includegraphics[width=0.32\textwidth]{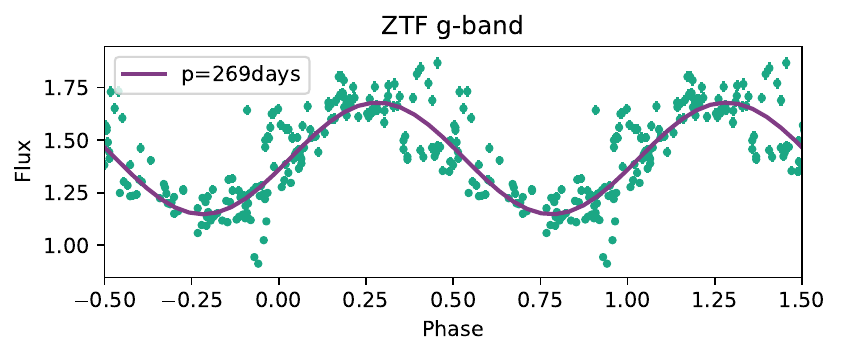}
\includegraphics[width=0.32\textwidth]{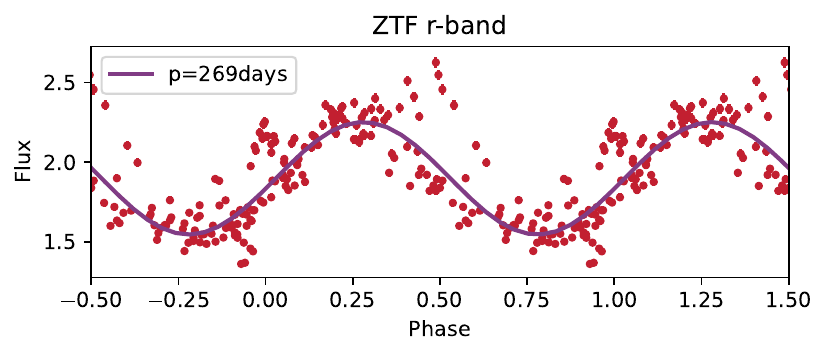}
\caption{Detection of optical periodicities in the combined KAIT and ZTF $r$ band, ZTF $g$ band, and ZTF $r$ band light curves of PKS 0301-243. The GLSP are shown in the upper row of each set of panels. The middle row of each set shows the WWZ periodograms. The subpanels show (from top left, anticlockwise): the light curve and the sinusoidal curve fitting results, the WWZ time-frequency map, and the mean WWZ power spectrum. The bottom panels show the phase-folded light curves corresponding to different identified periodic signals.}
\label{fig:QPO-KAIT+ZTF}
\end{figure*}

We collect the multiwavelength light curve data of PKS 0301-243 (see Section \ref{sec:Data}), and plot the optical, and infrared (IR), and $\gamma$-ray light curves in Figure \ref{fig:LC}. 
To identify periodic signals, we employ three methods including the GLSP\cite{Lomb_1976,Scargle_1982,Zechmeister_2009} (GLSP), the WWZ\cite{Foster_1996}, and the Gaussian process (GP) regression\cite{Gurpide_2025}. All of them can handle the inhomogeneous time series\cite{Adhikari_2024,Gurpide_2025}. The GLSP efficiently identifies periodic signals, but it is sensitive to sampling aliases and red noise \cite{VanderPlas_2018}.
The WWZ can trace the temporal evolution of periodic signals. GP regression models periodic and stochastic variability simultaneously, depending on the assumed covariance kernel.
The local and global significance are obtained through the Monte Carlo (MC) method \cite{ONeill_2022} based on $10^4$ light curves, which are simulated by the damped random walk (DRW) red noise process (see Section \ref{sec:period}). 

For KAIT and ZTF light curves, we compute both the GLSP and WWZ.
The $\sim270$-day periodic signal detected in the ZTF $g$ band and $r$ band light curves, presented in Figure \ref{fig:QPO-KAIT+ZTF}, show that the local significance exceeds the $3\sigma$ level in the GLSP and WWZ analyses. For ZTF $g$ band, the global significance is 98.77\% in GLSP and 99.70\% in WWZ, whereas for ZTF $r$ band, the corresponding values are 98.05\% and 98.53\%, respectively. Besides, the KAIT light curve also show a candidate periodicity at approximately $\sim890$-day, consistent with the 894-day periodicity preciously reported\cite{Chen_2024}. Its local significance exceeds $3\sigma$ level in GLSP and approaches to 3$\sigma$ level in WWZ, while its global significance is  97.53\% and 96.68\% in GLSP and WWZ, respectively. 

The candidate periodic signals and their corresponding local and global significance are summarized in Table \ref{table:QPO}. 
As we have employed two different period-search techniques, we apply the Bonferroni correction \cite{Abdi_2007_Bonferonni} to account for multiple testing and obtain a corrected global p-value for each light curve. This conservative correction is applied when multiple independent tests are performed on a dataset, thereby increasing the chance of at least one false positive. If $p$ denotes the smallest global p-value among $n$ tests, the corrected global p-value is given by $n \times p$. For $n=2$, we obtain corrected global p-values of $6 \times 10^{-3}$ ($2.75\sigma$), and $2.98\times 10^{-2}$ ($2.18 \sigma$) for the ZTF $g$ and $r$ bands, respectively. For KAIT, the corrected global p-value is $4.94\times10^{-2}$ ($1.97\sigma$).
For other supplementary light curves, the WWZ results are illustrated in Figure \ref{fig:QPO-ASAS-SN}, \ref{fig:QPO-ATLAS}, \ref{fig:QPO-IR}, and \ref{fig:QPO-Fermi}. These periodic signals are summarised in Table \ref{table:additional-QPO}.

Chen \& Jiang\cite{Chen_2024} assumed that the red noise described by SPL model and reported a local significance above $3\sigma$ in both GLSP and WWZ. Our MC analysis adopts a DRW process as the stochastic null model and yields the lower local and global significances. The difference therefore does not arise from the absence of the $\sim890$-day periodic modulation in the light curve, but from the different assumption adopted for the underlying stochastic variability \cite{Robnik_2024}.
The KAIT observations span near 4500 days, corresponding to approximately 5 nominal cycles of the $\sim 890$-day periodicity, and the first 3 cycles are clearly sampled in the light curve. By comparison,  the approximately 2000-day ZTF baseline covers about 7 cycles of the $\sim270$-day periodicity (Figure \ref{fig:QPO-KAIT+ZTF}).

We also adopt the GP process to model the observed light curves, and consider the Bayes factors to select the best fitted one. The GP modelling identified the two candidate periodic signals in the KAIT and ZTF light curves (see Section \ref{sec:GP}). For the KAIT data, 
the Bayesian evidence show that 2$\times$DRW model is the best fitted model. Adding a  Lorentzian or periodic component to the best  model introduced an additional peak near 930 days in the power spectral density (PSD) plot, consistent with the empirical period of $889^{+47}_{-46}$ days (see Section \ref{sec:inject}). The corresponding Bayes factors are 2.2 or 2.4, which provides modest support for the Lorentzian or periodic component.
In both ZTF bands, the period posterior of the additional the Lorentzian or periodic component concentrated near 270 day across the DRW stochastic backgrounds. 
The GP results therefore  reproduce the two periods found by the GLSP and WWZ analyses. However, the data do not support both components added to the null model simultaneously, since the null+ 2 $\times$ Lorentzian models have no greater Bayesian evidence than the corresponding single period models. The modest support for single periodic component in the KAIT and ZTF light curves is consistent with the global significance inferred from the GLSP and WWZ analyses.

We model the periodic modulations using both a pure sinusoidal model (Eq. (\ref{eq:sine})) and an exponentially decayed sinusoidal model (Eq. (\ref{eq:sine+exp})). The former provides a phenomenological description of approximately coherent periodic variability expected from orbital modulation, the latter reproduces the long-term decreasing trend visible in both $\gamma$-ray and optical light curves (see Section \ref{sec:MCMC}). Such a long-term decline may reflect a gradual decrease in the intrinsic jet emission or in accretion rate. This decreasing trend can also be found in supplementary light curves in different bands. The best-fitting parameters are constrained by maximizing the likelihood function\cite{ONeill_2022}, and the uncertainties are estimated using the \texttt{emcee} sampler\cite{Foreman-Mackey_2013}. The fitting parameter estimates are summarised in Table \ref{tab:MCMC-LC}. The phase-folded light curves (Figure \ref{fig:QPO-KAIT+ZTF}) and the corresponding Markov Chain Monte Carlo (MCMC) model fits (Figure \ref{fig:MCMC-KAIT}, \ref{fig:MCMC-Zg}, and \ref{fig:MCMC-Zr}) reproduce the dominant long-term modulations and provide a phenomenological characterization of the inferred periodicities.

The empirical period uncertainties are estimated by the injection-recovery test (see Section \ref{sec:inject}), and the recovered period distributions and their edge accumulation fractions are reported in Table \ref{tab:MCMC-recover}. 
For the $\sim890$-day modulation detected in KAIT light curve, we obtain an empirical period estimate of $889^{+47}_{-46}$ days. For the $\sim270$-day modulation detected in the ZTF light curves, the adopted exponentially decayed sinusoidal fit to the ZTF $g$ band gives  $269^{+13}_{-14}$ days\footnote{We choose the exponentially decayed sinusoidal model of ZTF $g$ band to represent ZTF light curve results since it show more accepted fits and low edge accumulation fraction $f_{\rm edge}$ (see Section \ref{sec:inject})}. These empirical period estimates and uncertainties are adopted in the subsequent physical calculations.

\section{DISCUSION}
\subsection{A jetted SMBHB interpretation of the periodic signals}

The investigation on the origin of the two optical periodic modulations helps us to understand the substructure of the target.
The  $\sim 890$ days optical periodicity are most probably jet-related. Such interpretation is supported by two similar morphology of the optical and $\gamma$-ray light curves. A $\gamma$-ray quasi-periodic oscillation (QPO) of $2.1\pm0.3$ years ($767\pm110$ days) with  a significance above $5\sigma$ was reported for PKS 0301-243 \cite{Zhang_2017}, broadly consistent within the uncertainties with the optical period found here. Our independent analysis of $\gamma$-ray also recovers power at 740 days (see Figure \ref{fig:QPO-Fermi}). Moreover, both the optical and $\gamma$-ray light curves show a long-term decreasing trend, and their WWZ maps indicate that the corresponding periodic modulation weakens or vanishes after MJD 58,000 (Figure \ref{fig:QPO-KAIT+ZTF} and \ref{fig:QPO-Fermi}). Since the $\gamma$-ray emission of blazars is dominated by the relativistic jet, the similar periodic modulation and temporal evolution trend provides supporting evidence that the longer optical modulation is also jet-related. It is widely accepted that both jet precession\cite{Sobacchi_2017,Britzen_2023} and helical trajectory of blob within the jet\cite{Zhou_2018,Sarkar_2021} could produce periodicity. However, QPO caused by blob motion is typically of months, and the emission of blobs last only a few hundred days\cite{Zhou_2018,Sahakyan_2022,Prince_2023_qpo}. Thus, the jet precession provides a possible interpretation for the optical $\sim 890$-day periodic signal \cite{Sobacchi_2017,ONeill_2022,de_la_Parra_2025}.

\begin{figure}[t]
\centering
    \centering
    \includegraphics[width=0.45\textwidth]{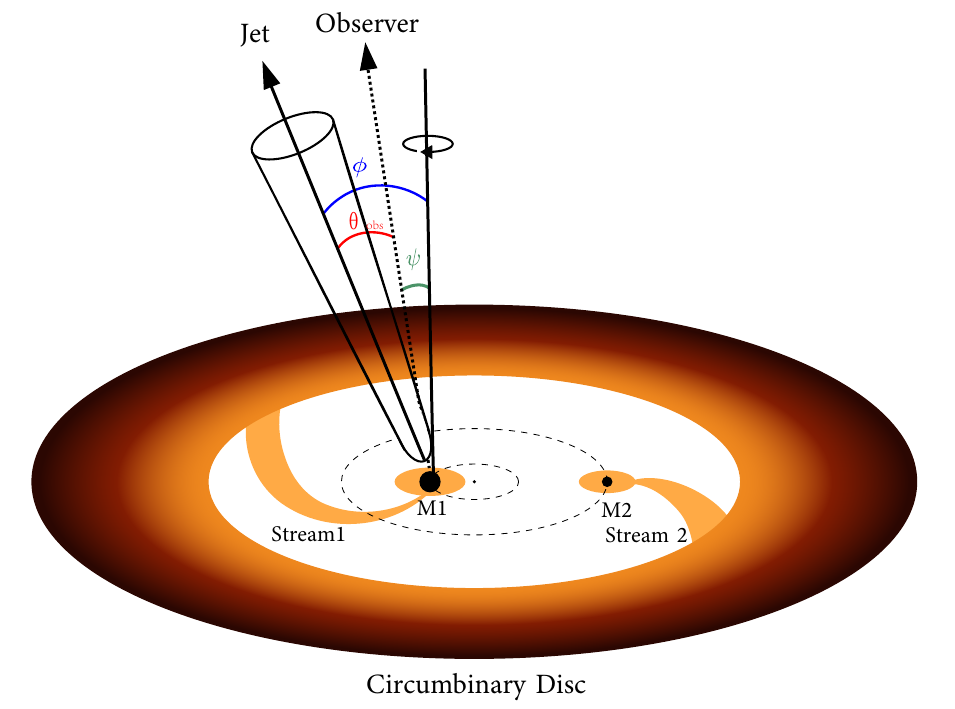}
    \caption{Cartoon of the configuration of an SMBHB system with a relativistic jet associated with the primary SMBH. 
    The $\phi$ is the angle between the orbital angular momentum and the jet axis, $\psi$ denotes the angle between the orbital angular momentum of the primary black hole and the line of sight. The schematic is not to scale.}
    \label{fig:model}
\end{figure}

We next consider the physical origin of the $\sim270$-day optical periodicity. Its persistence over 7 cycles and approximately sinusoidal modulation are naturally compatible with an interpretation associated with the orbital motion of an SMBHB system. If the observed period traces the binary orbit, Kepler's law implies a milli-parsec separation of PKS 0301-243 (see below), placing the putative binary in a compact regime. Hence, this source is possibly  in the inspiral stage of orbital evolution.
Within the circumbinary accretion geometry for such compact SMBHB, two kinds of modulation can produce such orbital timescale \cite{Kelley_2019}. First, the binary torque can periodically modulate the gas streams penetrating the central cavity, leading to variations in the accretion rates of the mini-discs, and further their luminosities \cite{D'Orazio_2013,Farris_2014}. Such periodically modulated accretion can, under suitable conditions, generate approximately sinusoidal variability \cite{Miranda_2017} (see Section \ref{sec:SMBHB}).
Second, the orbital motion can modulate the observed flux through relativistic Doppler boosting \cite{D'Orazio_2015,Charisi_2022}. If the optical emission is dominated by the second mini-disc in an assumed small mass ratio system, the  projected velocity offset due to orbital motion could lead to the periodic variation of the Doppler factor and therefore of the observed flux. The detailed flux modulation by Doppler factor will be presented in Section \ref{sec:DB}.

Adopting the circumbinary disc SMBHB model, the separation between the two SMBHs can be estimated based on Kepler’s law of motion. The intrinsic orbit periodicity of the binary system $P_{\rm int}$ is given by\cite{Fan_2010}
\begin{equation}
    P_{\rm int} \sim 1.72 \left(\dfrac{M_{\rm tot}}{10^8 M_\odot}\right)^{-1/2} \left( \dfrac{r}{10^{16} {\rm cm}} \right)^{3/2},
\end{equation}
where $P_{\rm int} = P_{\rm orb}/(1 + z)$ is the intrinsic orbital period corrected for redshift in units of years, $M_{\rm tot}$ denotes the total mass of binary, and $r$ is the binary separation. For PKS 0301-243, the SMBH mass was estimated by the host galaxy bulge luminosity using the empirical $M_{\rm BH}$-bulge luminosity relation\cite{Paliya_2021}. Assuming that this estimated mass is the total mass of the putative binary, $M_{\rm tot}=10^{8.95} M_\odot$, and adopting $P_{\rm orb} = 269^{+13}_{-14}$ days and the redshift $z$ = 0.266\cite{Pita_2014}, the separation is derived as $r \sim 3.26^{+0.10}_{-0.11}\times 10^{-3}$ pc. Note that the binary separation $r$ is independent on $q$. This indicates that the SMBHB system in PKS 0301-243 has evolved into the milli-parsec scale. A schematic toy model illustrating the accretion geometry for such an SMBHB is shown in Figure \ref{fig:model}.

Therefore, we adopt $P_{\rm jet} = 889^{+47}_{-46}$ days as the jet precession period, and $P_{\rm orb} = 269^{+13}_{-14}$ days as the binary orbital period. Although the $\sim270$-day periodic modulation is detected in the ZTF light curve (and also in supplementary data), its long-term persistence cannot yet be established in the KAIT data. The GLSP of KAIT data shows only a weak signal (global significance of 84.91\%). However, no statistically significant counterpart was detected with the WWZ and GP methods. Therefore, the KAIT light curve  neither independently confirms nor rules out the presence of the $\sim270$-day signal given the sparse sampling and small amplitude. We consequently conclude that the SMBHB scenario is a plausible interpretation. 

\begin{figure*}[ht!]
\centering
\includegraphics[width=0.9\textwidth]{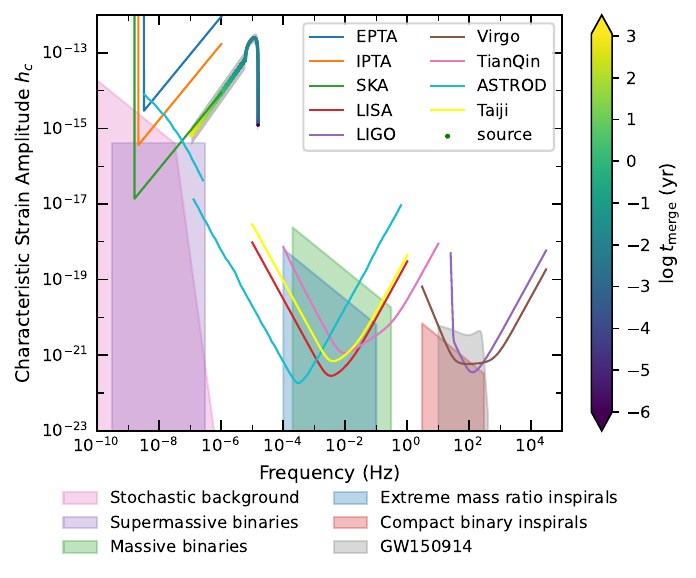}
\caption{The sensitivity curves of current and future GW detectors and the ranges of different GW sources. The coloured track shows the characteristic strain amplitude $h_c$ evolution of PKS 0301-243 assuming a fiducial mass ratio $q=0.1$, where the colour indicates the remaining time to merger. The gray shaded region represents the uncertainty range obtained by varying the mass ratio between $q=0.05$ and $q=0.2$. The characteristic strain is calculated as $h_c=h \sqrt{f T_{\rm obs}}$ with $T_{\rm obs}=15$ yr.}
\label{fig:GW_detect}
\end{figure*}

\subsection{Gravitational Wave estimation}
If the above analysis holds, this source provides us a rare opportunity to study the final stages of SMBHB evolution, and provides clues for final parsec problem. Such close SMBHB system is expected to emit nano-Hertz (nHz) gravitational wave (GW)\cite{Becsy_2022,Agazie_2023,Petrov_2025}. For PKS 0301-243, the rest-frame GW frequency for a circular binary orbit is $f_{\rm GW} = 2 / P_{\rm int} = 1.09^{+0.06}_{-0.05} \times10^{-7}$ Hz.

The GW strain amplitude from a circular orbit, averaged over binary inclination, depends on the intrinsic orbit frequency $f_{\rm int} =1/P_{\rm int}$, the luminosity distance $D_{\rm L}$, and the chirp mass $\mathcal{M}$ determined by the masses of two SMBHs $M_1$ and $M_2$\cite{Maggiore_2007} (Section \ref{sec:GW}). Adopting the $\Lambda$CDM model with cosmological constants $H_0=70 \rm{km~s^{-1}~Mpc^{-1}}$ and $\Omega_m = 0.3$\cite{Wright_2006}, the redshift $z=0.266$ yields a luminosity distance of $D_{\rm L} = 1352.47$ Mpc. Since the mass ratio $q$ is currently unconstrained by observations, we evaluate the continuous GW strain for three representative values to encapsulate the possible range of binary configurations. For periodic accretion rate variation driven by the secondary SMBH in the circumbinary disc, the intermediate mass ratio is expected in the range $0.01 \lesssim  q \lesssim  0.3$\cite{Kelley_2019}. In addition, the Doppler boosting interpretation also favours a relatively low mass ratio, $0.03\le q \le 0.2$ \cite{Charisi_2022}. Thus,  we set $q=0.1$ as the fiducial value, and consider both $q=0.05$ and $q=0.2$ cases. Correspondingly, we obtain $h_{\rm GW} = 8.61^{+0.26}_{-0.30} \times 10^{-17}$ for $q=0.1$, $h_{\rm GW} = 4.73^{+0.14}_{-0.17} \times 10^{-17}$ for $q=0.05$, and $h_{\rm GW} = 1.45^{+0.04}_{-0.05} \times 10^{-16}$ for $q=0.2$. 

The coalescence time for a circular SMBHB can be derived from the semimajor axis of orbit $a$, $M_1$, and $M_2$ \cite{Peters_1964}. We approximate the semimajor axis by the binary separation $a \sim r$. Under this assumption, the predicted coalescence time are $t_{\rm c} = 1122^{+144}_{-144}$ years for $q=0.1$, $t_{\rm c} = 2045^{+ 262}_{-263}$ years for $q=0.05$, and $t_{\rm c} = 668^{+86}_{-86}$ years for $q=0.2$, suggesting a remarkably short coalescence timescale for PKS 0301-243. We note that $t_c$ is sensitive to the assumed mass ratio $q$, spanning a factor of $\sim 3$ across the range considered.

To study temporal evolution of the GW signal, we simulate it using the Python package \textit{riroriro} code\cite{Buskirk_2019,van_Zeist_2021}.
The package generates an analytic compact binary waveform model \cite{Buskirk_2019}, with the inspiral evolution described using post-Newtonian approximations. The simulated evolution track is obtained by adopting each value of $q$ and the $D_{\rm L} = 1352.47$ Mpc. We set $t=0$ at the moment of SMBHB coalescence, so the present epoch corresponds to $t\approx -1122$ years for the fiducial value $q=0.1$.  To reflect the possible range of $q$, we show  a gray shaded region bounded by the $q=0.05$ and $q=0.2$ cases in Figure \ref{fig:GW_detect}. Because the GW frequency evolves negligibly over any foreseeable detector mission lifetime ($\Delta f_{\rm GW}/f_{\rm GW} \ll 1$ over decades), this source can be treated as a monochromatic continuous wave source.

The detectability of the GW from this source is meaningful for the future nHz GW programs \cite{Mingarelli_2025}.
To compare the GW signal with the published sensitivity curves of GW instruments\cite{Moore2015,Ni_2013,Liu_2023}, we convert the simulated GW strain $h_{GW}$ to GW characteristic strain $h_c$ via $h_c = h_{\rm GW} \sqrt{f_{GW}T_{\rm obs}}$. We adopt  $T_{\rm obs} = 15$ years to match the typical duration of current PTA datasets. As shown in Figure \ref{fig:GW_detect}, the predicted characteristic strain of PKS 0301-243 indicates that this source lies within the sensitivity of future GW detector ASTROD\cite{Ni_2013} and is on the edge of detectability for SKA\cite{Feng_2020}. It is possible that this source could be detected by SKA with longer observation baseline \cite{Sesana_2009}. We explored the plausible mass ratio parameter space ($0.05\le q \le 0.2$) and found that the detectability conclusions remain unchanged.

\section{Conclusion}
In this work, we conducted a comprehensive investigation of the optical periodicity of the SMBHB candidate PKS 0301-243. The $\sim270$-day periodic modulation is detected in the ZTF light curves and persists over 7 cycles, while a longer $\sim890$-day periodicity spanning approximately 5 cycles is recovered in the KAIT light curve, consistent with the previously reported periodicity \cite{Chen_2024}. The two periodicities are recovered using complementary time-series analyse methods, including GLSP, WWZ,  MCMC modelling, and injection-recovery tests. GP modelling and Bayesian model selection also provide additional statistical support for the two optical periodicities. Our analysis further shows that the inferred significance of the $\sim890$-day periodic signal is sensitive to the adopted stochastic process model, highlighting the importance of noise modelling and complementary Bayesian model comparison in searches for periodicity in AGNs. 

The two optical periodicities are plausible associated with distinct variability components within a compact SMBHB scenario incorporating a relativistic jet. The $P_{\rm orb} =269^{+13}_{-14}$ days periodicity may arise from orbital motion in a milli-parsec SMBHB, whereas the $P_{\rm jet}  =889^{+47}_{-46}$ days periodic signal is plausible related to jet precession. The corresponding binary parameters and GW properties are consistent with a subparsec system potentially emitting in the nHz frequency regime. The detailed origin of the orbital modulation, including periodic accretion and relativistic Doppler boosting, remains uncertain, while alternative mechanisms such as Lense-Thirring precession cannot be completely excluded. Overall, the temporal, statistical, and physical evidence favours the SMBHB interpretation. Nevertheless, the available time series data do not yet establish the long-term coherence of the $\sim 270$-day periodicity, and the SMBHB scenario should therefore be regarded as a plausible interpretation. PKS 0301-243 therefore represents a particularly promising candidate for continued time-domain monitoring and future multimessenger studies of compact SMBHB. Follow-up optical monitoring of ZTF and the longer and more densely sampled light curves from the Vera C. Rubin Observatory Legacy Survey of Space and Time (LSST) will provide a critical test of whether the $\sim270$-day modulation remains coherent over more cycles.
 
\section{METHODS}
Detailed methods are available in the \href{ https://doi.org/10.1093/nsr/XXXX}{online 
supplementary material}.

\section*{DATA AVAILABILITY} 
Observational data in this work are available in observatories archives.
The KAIT optical photometry data could obtained from KAIT Fermi AGN Light-Curve Reservoir (\url{http://herculesii.astro.berkeley.edu/kait/agn}).
The ZTF optical photometry data could obtained from ZTF Public Data Releases (\url{https://www.ztf.caltech.edu/ztf-public-releases.html}).
The ASAS-SN optical photometry data could obtained from Sky Patrol V2.0 (\url{https://www.astronomy.ohio-state.edu/asassn/index.shtml}).
The ATLAS optical photometry data could obtained from ATLAS Forced Photometry (\url{https://fallingstar-data.com/forcedphot}).
The NEOWISE IR photometry data could obtained from NASA/IPAC Infrared Science Archive (\url{https://irsa.ipac.caltech.edu/frontpage}).
The $\gamma$-ray data could obtained from Fermi-LAT Lightcurve Repository (\url{https://fermi.gsfc.nasa.gov/ssc/data/access/lat/LightCurveRepository}).

\section*{CODE AVAILABILITY}
The software packages to generate the results of this paper are available from \\
\texttt{GLSP} (\url{https://github.com/mzechmeister/GLS/tree/master/idl})\\
\texttt{WWZ} (\url{https://github.com/skiehl/wwz}) \\
\texttt{Phase-fold} (\url{https://github.com/jakevdp/PracticalLombScargle})\\
\texttt{mind the gaps} (\url{https://github.com/andresgur/mind_the_gaps}) \\
\texttt{dynesty}
(\url{https://github.com/joshspeagle/dynesty}) \\
\texttt{emcee} (\url{https://github.com/dfm/emcee}) \\
\texttt{riroriro} (\url{https://github.com/wvanzeist/riroriro}) \\

\section*{SUPPLEMENTARY DATA}
Supplementary data are available below and at \href{https://doi.org/10.1093/nsr/XXXX}{NSR} online.

\section*{ACKNOWLEDGEMENTS}
We thank Cristian Vignali, Paola Severgnini, Fabio Rigamonti, Dong Lai and Zhen Pan for helpful discussions. We also thank Yuanming Wang for checking the radio data.

\section*{FUNDING}
This work has been funded by the National Natural Science Foundation of China (12673023), and by the Shandong Provincial Natural Science Foundation (ZR2026MS0040). Shifeng Huang acknowledges the support of the Fundamental Research Funds for Central Universities (WK2030000097).

\section*{AUTHOR CONTRIBUTIONS} 
Jintong Wang performed the data reductions and analyses, carried out MCMC fits, model interpretation, GW estimation and simulation, and prepared the initial manuscript. Yunguo Jiang conceived and supervised the project, proposed the research outline and revised the manuscript. Shifeng Huang studied the LT precession model, discussed SMBHB models with other authors, and revised the manuscript. All authors contributed to the text.

\paragraph{\it Conflict of interest statement} None declared.

\clearpage
\onecolumn
\setlength{\oddsidemargin}{0.5cm}
\setcounter{figure}{0}
\setcounter{table}{0}
\setcounter{equation}{0}
\setcounter{section}{0}
\setcounter{secnumdepth}{2}
\renewcommand*{\thefigure}{S\arabic{figure}}
\renewcommand*{\thetable}{S\arabic{table}}
\renewcommand*{\theequation}{S\arabic{equation}}
\renewcommand*{\thesection}{S\arabic{section}}
\renewcommand*{\thesubsection}{S\arabic{section}.\arabic{subsection}}

{\textbf{\LARGE\centering Supplementary Materials for\\ The complex periodic variability of PKS 0301-243: hints of a supermassive black hole binary candidate} \par}

\section{Source selection and data collection} \label{sec:Data}
Previous work searched for optical quasi-periodic oscillations (QPOs) in a sample of 155 blazars using optical light curves from the Katzman Automatic Imaging Telescope (KAIT), the Small and Moderate Aperture Research Telescope System (SMARTS), and the Steward Observatory\cite{Chen_2024}. Among these sources, 18 targets showed possible QPOs with timescale between 200 and 1400 days, while only 6 sources, including PKS 0301-243, reached a significance of $3\sigma$ in both the generalized Lomb–Scargle periodogram (GLSP) and the weighted wavelet Z-transform (WWZ) under a simple power-law (SPL) red noise model \cite{Chen_2024}. PKS 0301-243 therefore belongs to the small fraction ($6/155\simeq4\%$) of sources in the original sample showing robust periodic variation according to both methods. In particular, a significant periodicity of 894 days was reported in the KAIT light curve \cite{Chen_2024}, motivating our selection of PKS 0301-243 for further investigation.

\begin{figure}
\centering
\includegraphics[width=\textwidth]{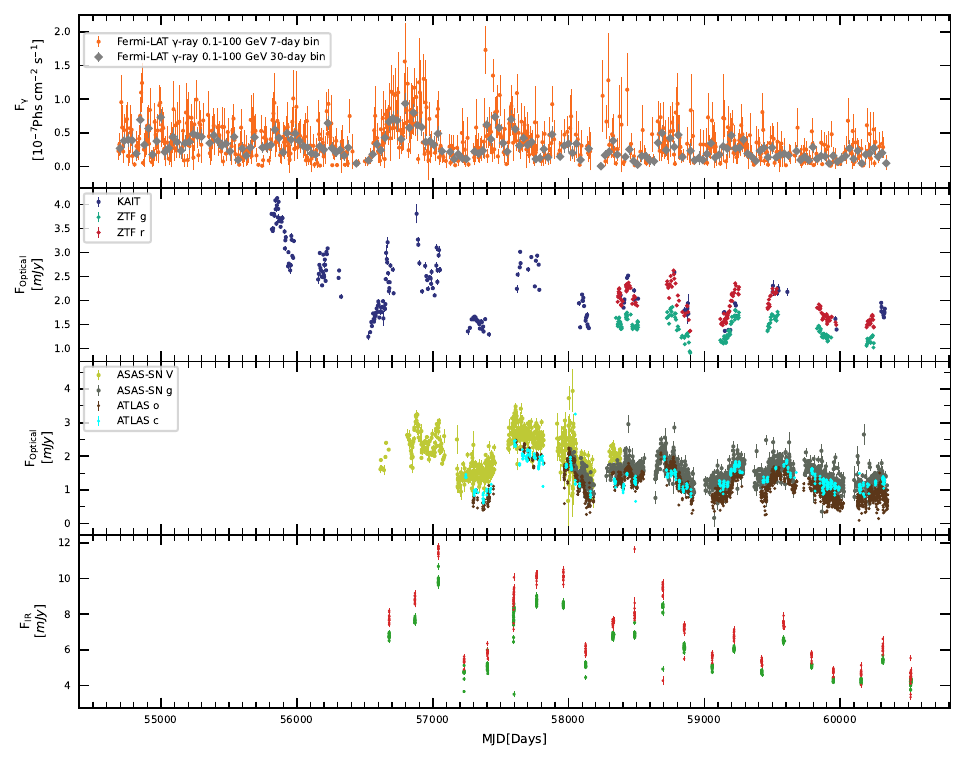}
\caption{Multiwavelength light curves of PKS 0301-243 covering $\gamma$-ray, optical and IR bands. From top to bottom panels, the light curves of 7-day binned and 30-day binned $\gamma$-ray of 0.1-100 GeV, KAIT, optical $V$ band from ASAS-SN, $g$ band from ASAS-SN and ZTF, $r$ band from ZTF, o and c from ATLAS, and $W1$ and $W2$ bands from NEOWISE are plotted, respectively.}
\label{fig:LC}
\end{figure}

In this work, we collected multiwavelength data of the target from public data archives. These include optical photometry data from KAIT, Zwicky Transient Facility (ZTF), All-Sky Automated Survey for Supernovae (ASAS-SN), and Asteroid Terrestrial-impact Last Alert System (ATLAS), infrared photometry data from Near-Earth Object Wide-field Infrared Survey Explorer (NEOWISE), and $\gamma$-ray data from the Fermi Large Area Telescope (LAT).  These collected light curves are represented in Figure \ref{fig:LC}. We focus primarily on the KAIT light curve and the ZTF $g$ and $r$ bands light curves, which provide the best temporal coverage and sampling for the robust characterization of the periodic variability. In the following content, we elaborate on the processing of various data.

\subsection{Optical photometry observations.}
Optical photometry data from the KAIT\footnote{\url{http://herculesii.astro.berkeley.edu/kait/agn}}\cite{Li_2003} has a duration from MJD 55,813 to 60,328 (2011 September 9 to 2024 January 18). The data are approximately in the $R$ band.  

The ZTF, the deepest time-domain survey of the northern sky, operates with the Palomar 48-inch Schmidt Telescope, which has a 47 $\deg^2$ field of view\cite{Bellm_2019,Masci_2019}. This large field of view enables ZTF to scan the entire northern sky every night. We collect the $g$ band and $r$ band photometric data from ZTF\footnote{\url{https://www.ztf.caltech.edu/ztf-public-releases.html}} covering MJD 58,348 to 60,247 (2018 August 18th to 2023 October 30th). To obtain high-quality observational data, we retain data with catflags score = 0\cite{Wang_2025}. 

ASAS-SN is a ground-based all-sky survey that observes the entire visible sky daily\cite{Shappee_2014,Kochanek_2017}. We download the $V$ band and $g$ band data from ASAS-SN Sky Patrol V2.0\footnote{\url{https://www.astronomy.ohio-state.edu/asassn/index.shtml}}\cite{Hart_2023}, covering MJD 56,615 to 58,386 and MJD 58,013 to 60,343, respectively. 

The ATLAS\cite{Tonry_2018} observes the northern sky (dec. $>$ -45$\degree$) every two nights using two optical filters: c (4157.33-6556.44 $\text{\AA}$) and o (5582.07-8249.18 $\text{\AA}$). We obtain light curve of this source from the ATLAS Forced Photometry Service\footnote{\url{https://fallingstar-data.com/forcedphot}}, covering MJD 57,243 to 60,349. To improve data quality, we applied filters to the errors and the limiting magnitude as follows \cite{Hern_2023ATLAS}. We obtained the cumulative distributions of the flux error and limiting magnitude and determined the percentiles at which the slope of the distributions changed. This resulted in a 90 per cent percentile for the flux error (DUJY $< 40.00$) and 5 per cent percentile for the limiting magnitude (MAG5SIG $< 17.51$). Finally, we retain 697 points in ATLAS c, and 2580 points in ATLAS o. A 1.5 mJy offset is added to flux values in both bands to account for calibration differences.

\subsection{Infrared photometry observations}
PKS 0301-243 was observed by the NEOWISE\cite{Mainzer_2014} in two infrared (IR) bands including $W1$ and $W2$, centred at 3.4 and 4.6 $\mu \rm{m}$, respectively. IR data are obtained from the NASA/IPAC Infrared Science Archive\footnote{\url{https://irsa.ipac.caltech.edu/frontpage}}. A total of 349 and 346 data points are obtained in the $W1$ and $W2$ bands, respectively, spanning from 56,679 MJD to 60,520. As multiple observations are made within a single day, We rebin the data into 1-day bins, resulting in 56 data points in each band.

\subsection{$\gamma$-ray observations.}
The Fermi Large Area Telescope (Fermi LAT) is a space-based telescope that continuously monitors the sky for high energy photons in the 20 MeV to 300 GeV range from astrophysical sources\cite{Atwood_2009}. The $\gamma$-ray light curve of PKS 0301-243, designated as J0303.4-2407 in the 4FGL catalogue\cite{Abdollahi_2020}, were obtained from the Fermi LAT Lightcurve Repository\footnote{\url{https://fermi.gsfc.nasa.gov/ssc/data/access/lat/LightCurveRepository}}\cite{Abdollahi_2023ApJS}. The light curves was obtained using the analysis mode in which the photon spectral index of the source is free. We consider two time series with time bins of 7 days and 30 days, spanning approximately 15.5 years (from August 8, 2008, to February 1, 2024), in the 0.1-100 GeV energy range. These data are filtered by $\rm TS \ge 10$, which corresponds approximately to a $3\sigma$ rejection criterion for one degree of freedom\cite{Abdollahi_2023ApJS}.

\section{Search for periodicity}\label{sec:period}
\subsection{GLSP}
In order to detect potential periodicities in heterogeneously sampled multiwavelength light curves, we employ the GLSP\cite{Lomb_1976,Scargle_1982,Zechmeister_2009}. The GLSP accounts for observational uncertainties and it is suitable for unevenly spaced time series\cite{Penil_2020}. It analyses the data by fitting sine waves, including an offset and weights, and normalizes the resulting power of the Fourier transform. 
We perform GLSP analyses using the IDL programme\cite{Zechmeister_2009}. The GLSP frequency range is determined by the temporal span of each light curve. The maximum frequency is 0.01 $\rm d^{-1}$, and the minimum frequency is $(N-1)/(2\Delta T)$ $\rm d^{-1}$, where $\Delta T$ is the total observational duration and $N$ is the number of data points. The average frequency spacing is given by $\Delta f = 1/(\zeta \Delta T)$ $\rm d^{-1}$. For this work, we adopt a sampling factor of $\zeta = 10$ to ensure sufficient frequency resolution across the observing window \cite{ONeill_2022}. The results of the GLSP are illustrated in Figure \ref{fig:QPO-KAIT+ZTF}.

\subsection{WWZ}
\begin{figure}[htbp!]
\centering
\includegraphics[width=0.32\textwidth]{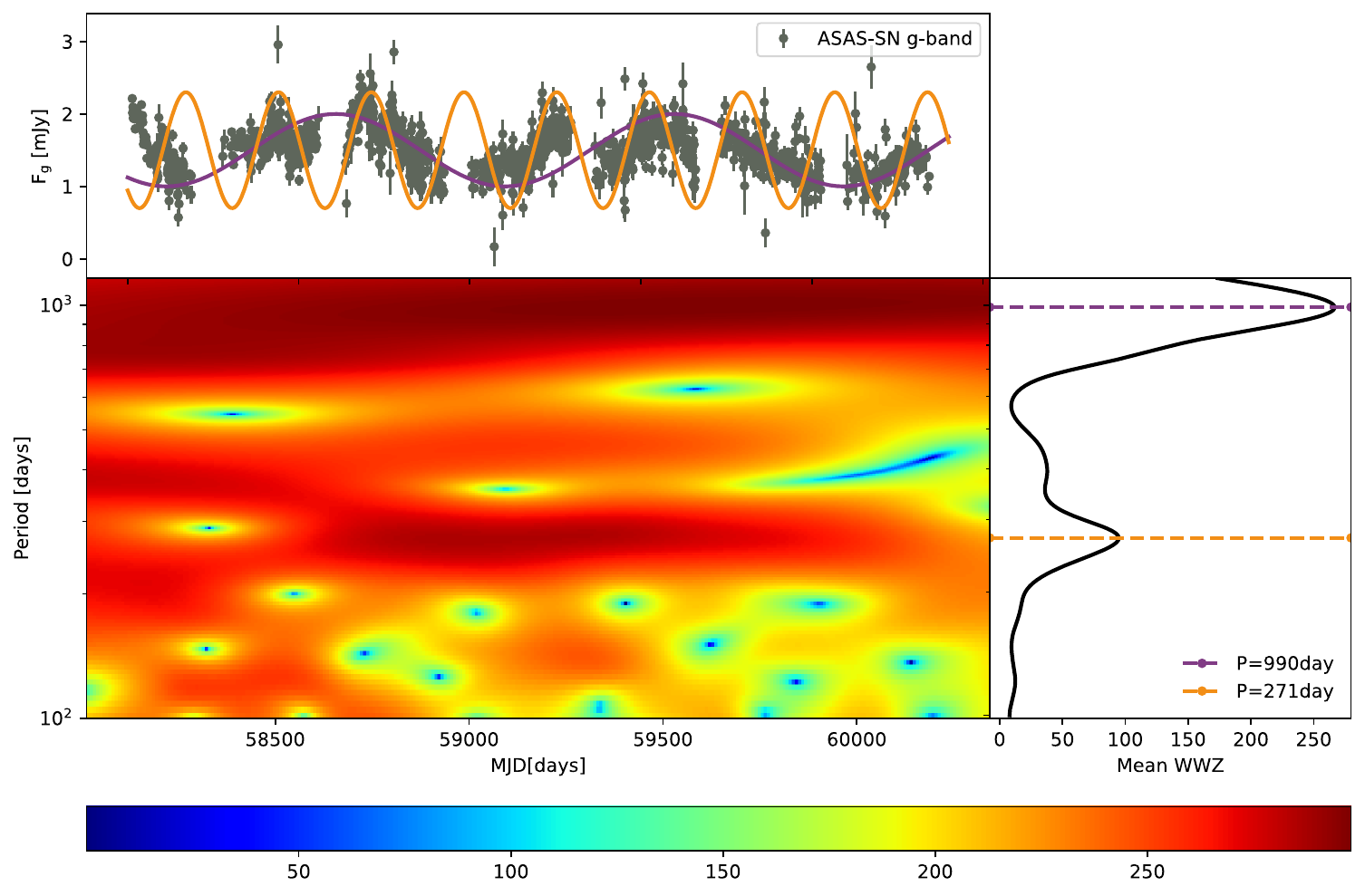}
\includegraphics[width=0.32\textwidth]{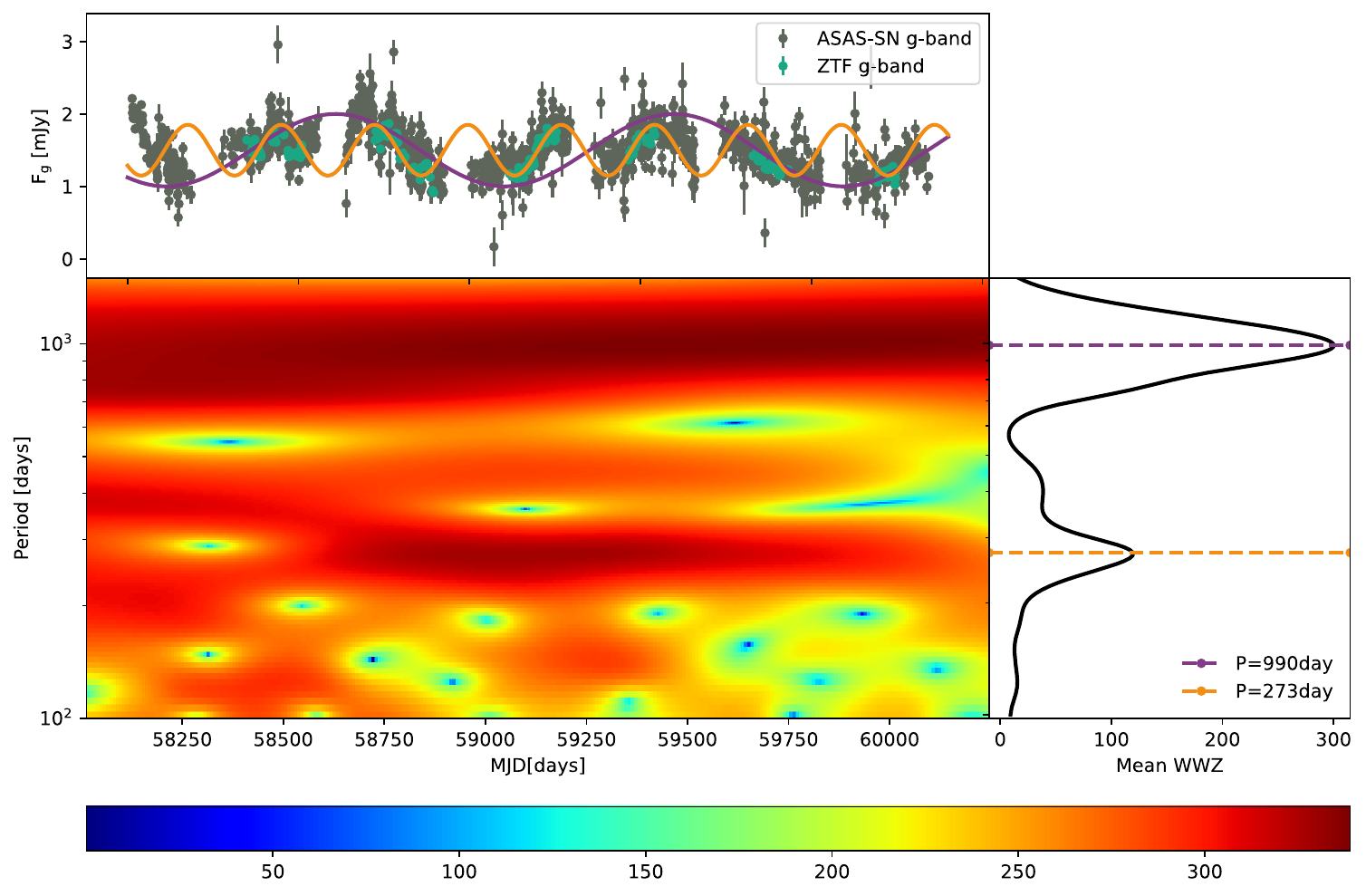}
\includegraphics[width=0.32\textwidth]{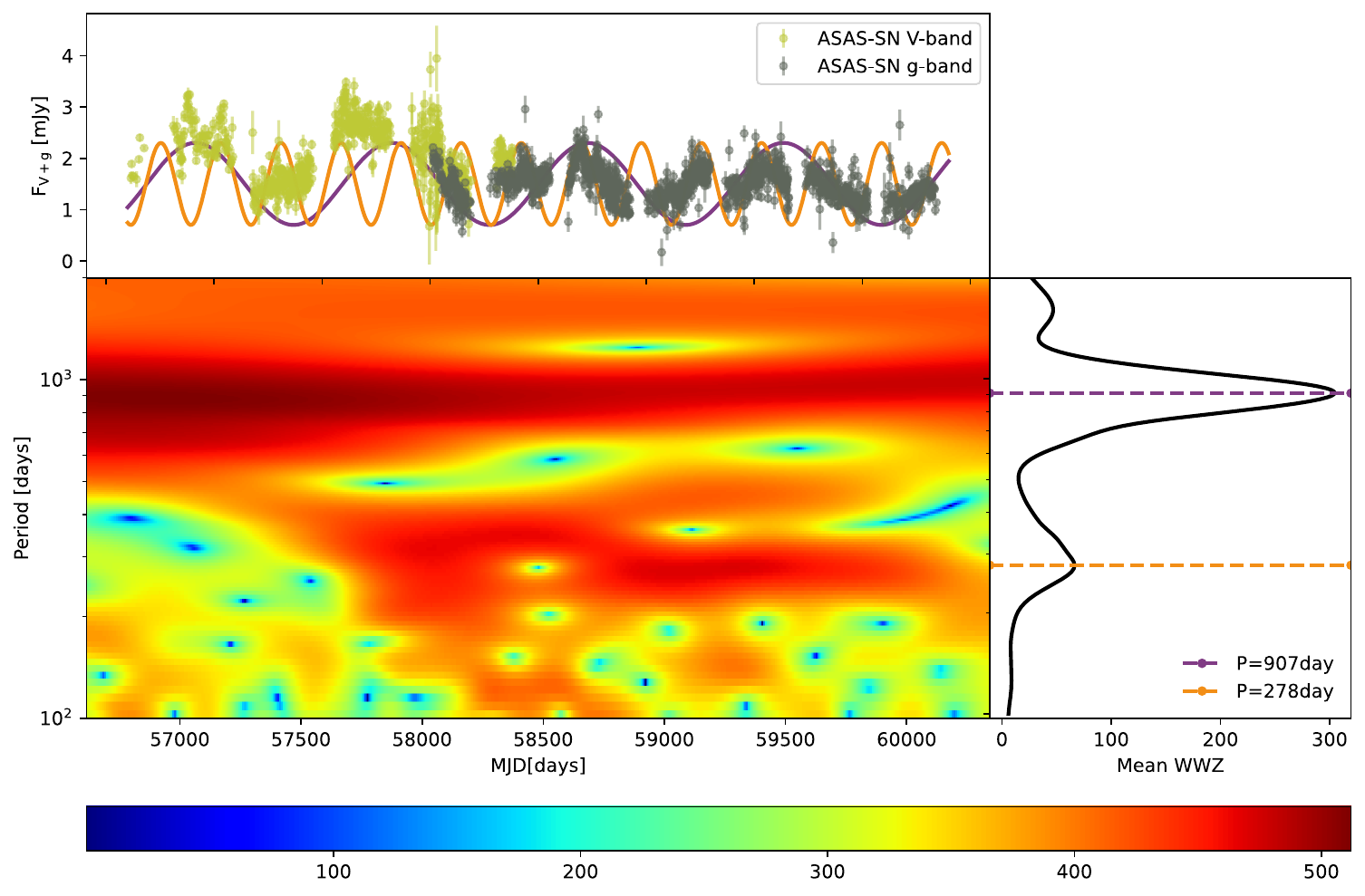}
\includegraphics[width=0.32\textwidth]{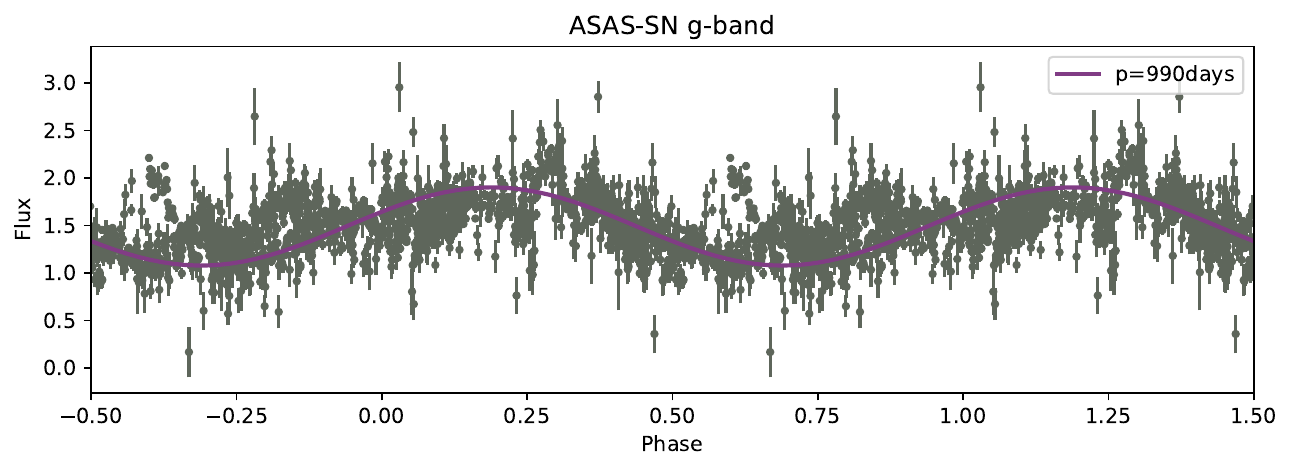}
\includegraphics[width=0.32\textwidth]{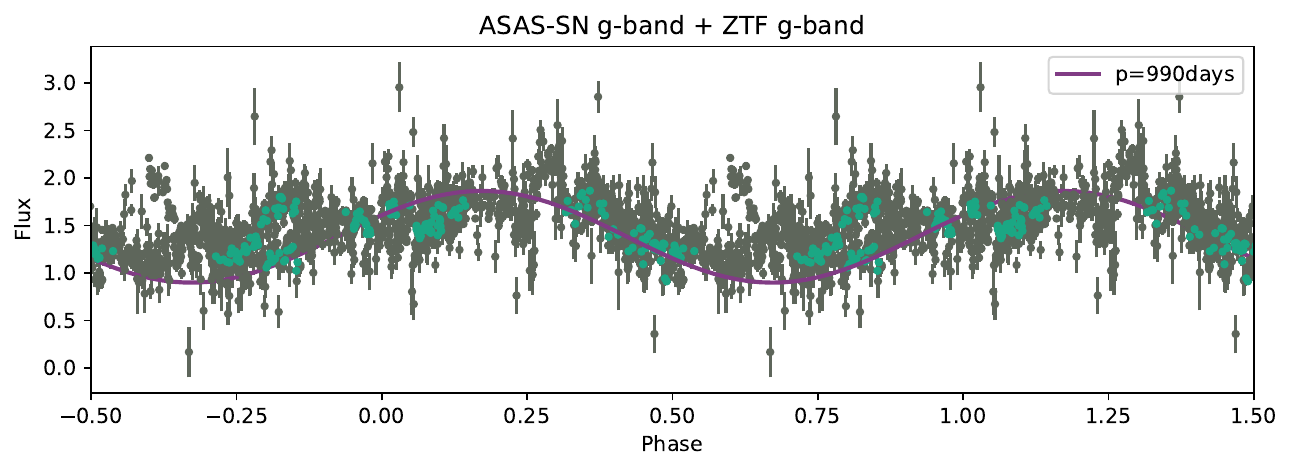}
\includegraphics[width=0.32\textwidth]{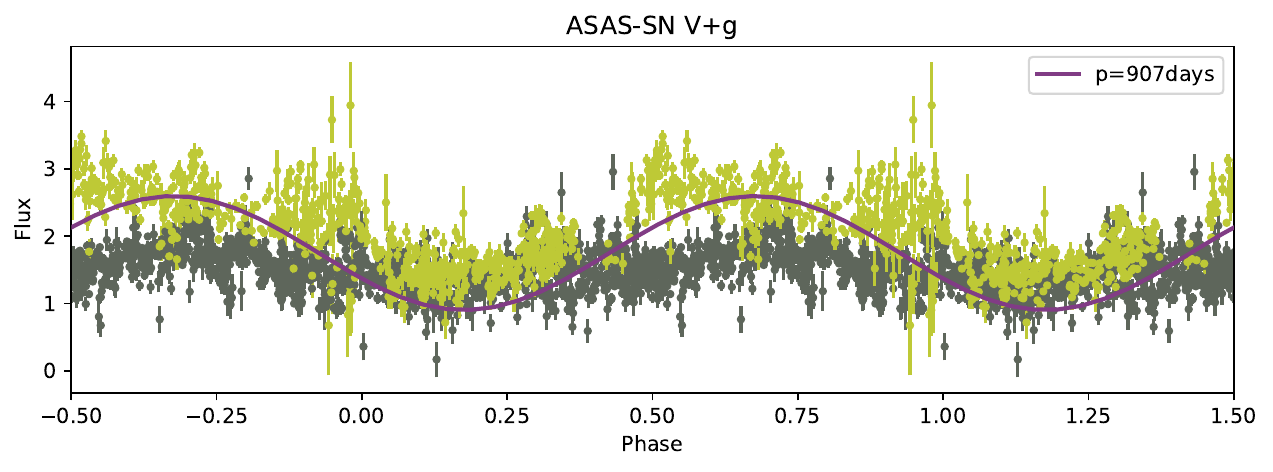}
\includegraphics[width=0.32\textwidth]{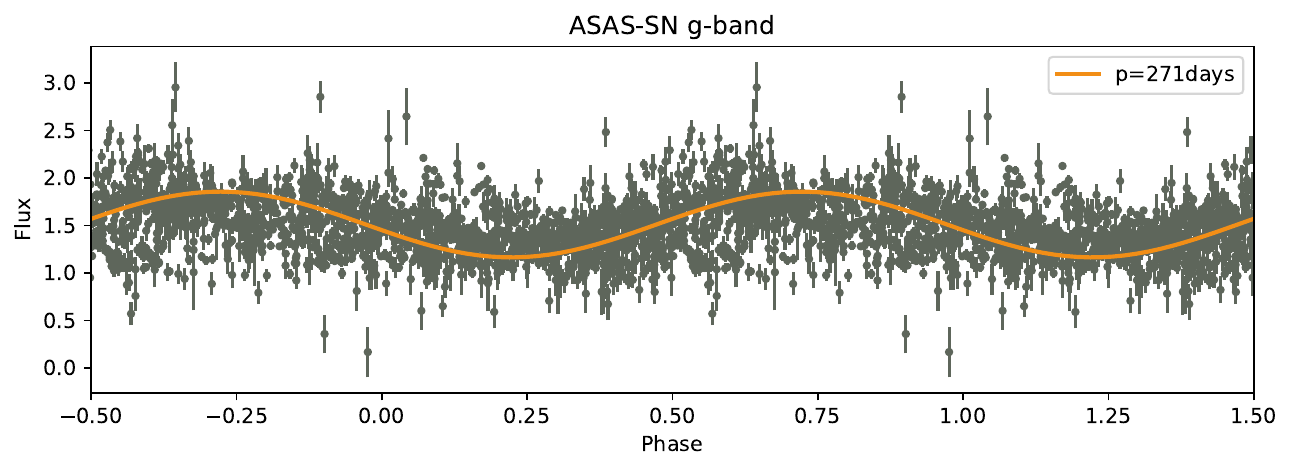}
\includegraphics[width=0.32\textwidth]{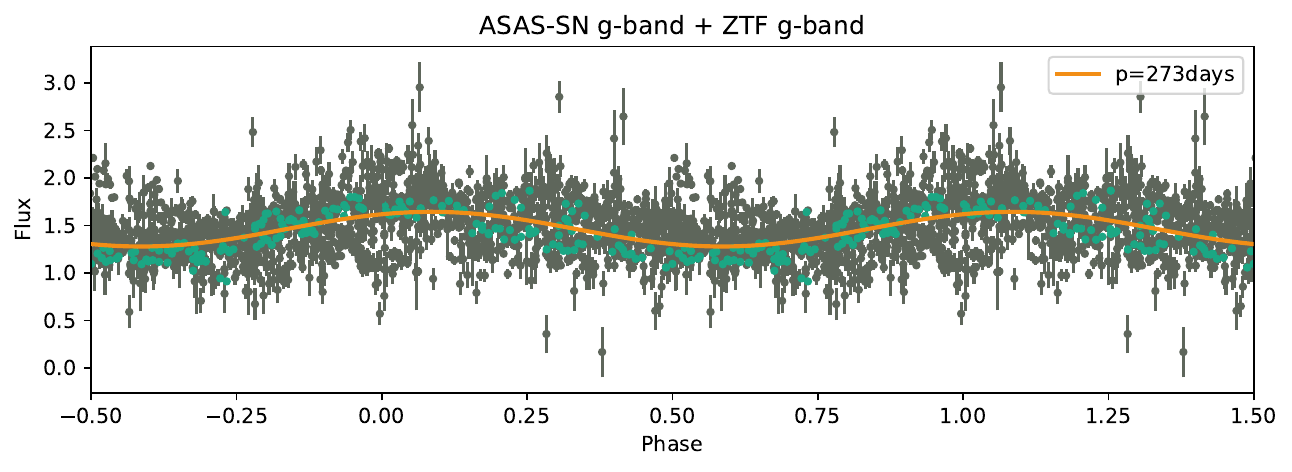}
\includegraphics[width=0.32\textwidth]{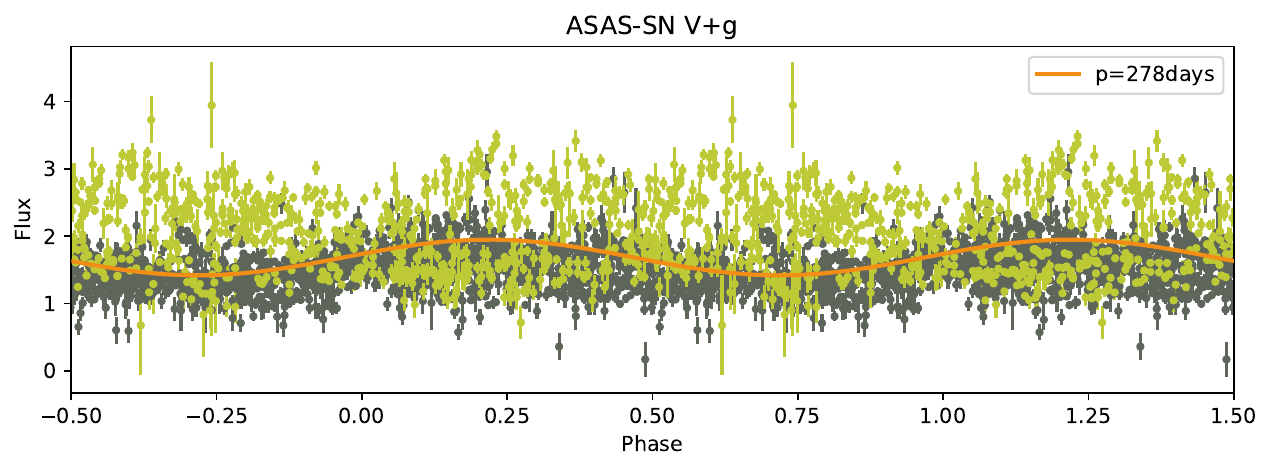}
\caption{Detection of periodic signals in ASAS-SN $g$ band, combined ASAS-SN $g$ band + ZTF $g$ band, and combined ASAS-SN $V$ band + $g$ band light curves of PKS 0301-243. The top row of each set shows the WWZ periodograms. The subpanels show (from top left, anticlockwise): the light curve and the sinusoidal curve fitting results, the WWZ time-frequency map, and the mean WWZ power spectrum. The bottom panels show the phase-folded light curves corresponding to different identified periodic signals.}
\label{fig:QPO-ASAS-SN}
\end{figure}

\begin{figure}[htbp!]
\centering

\includegraphics[width=0.48\textwidth]{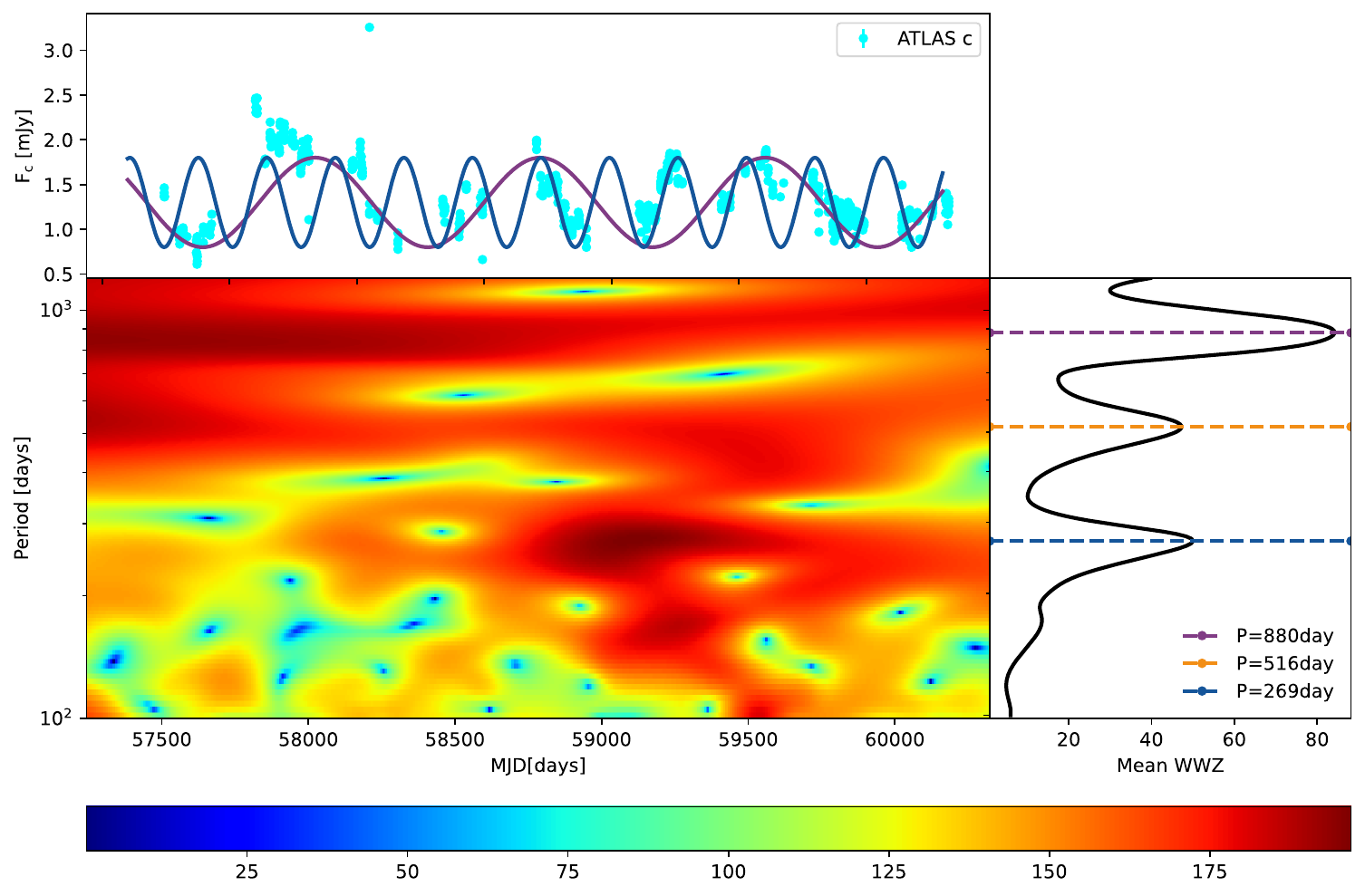}
\includegraphics[width=0.48\textwidth]{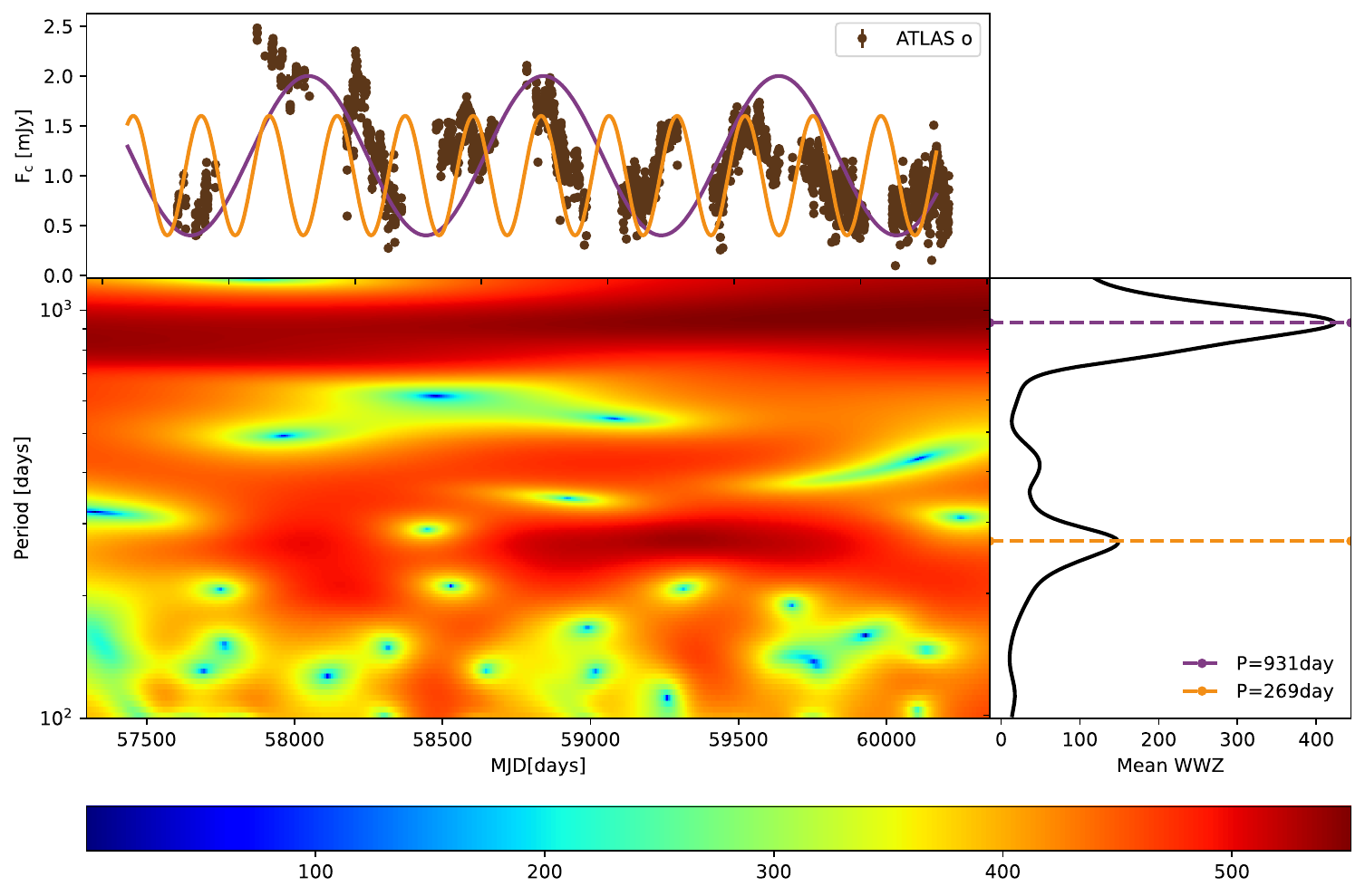}
\includegraphics[width=0.48\textwidth]{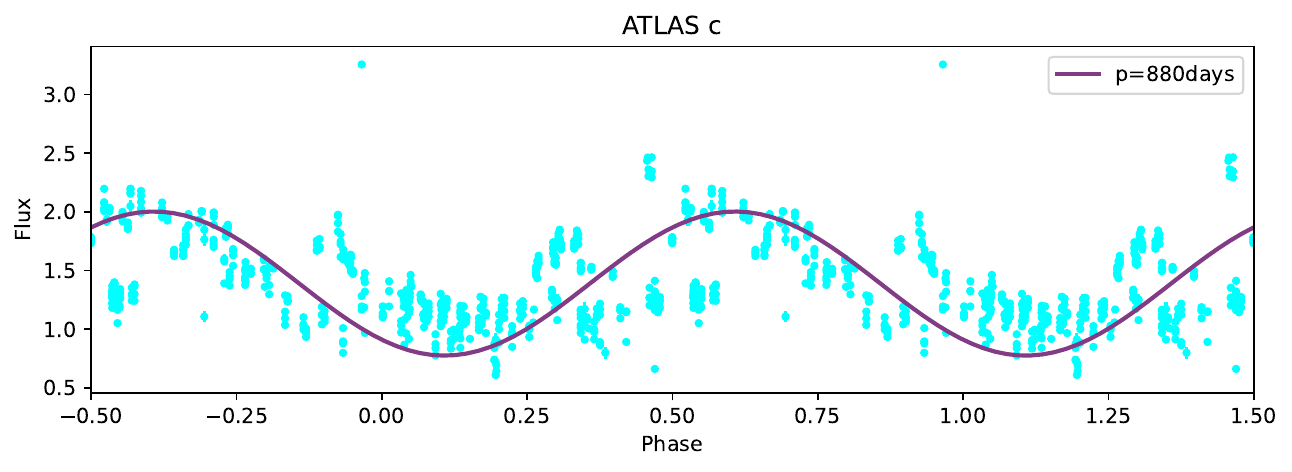}
\includegraphics[width=0.48\textwidth]{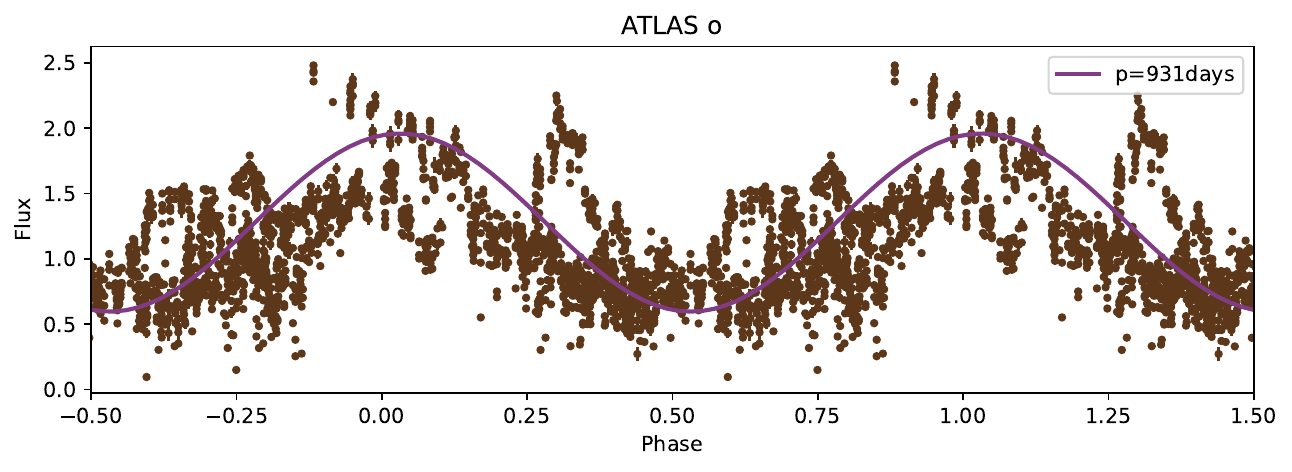}
\includegraphics[width=0.48\textwidth]{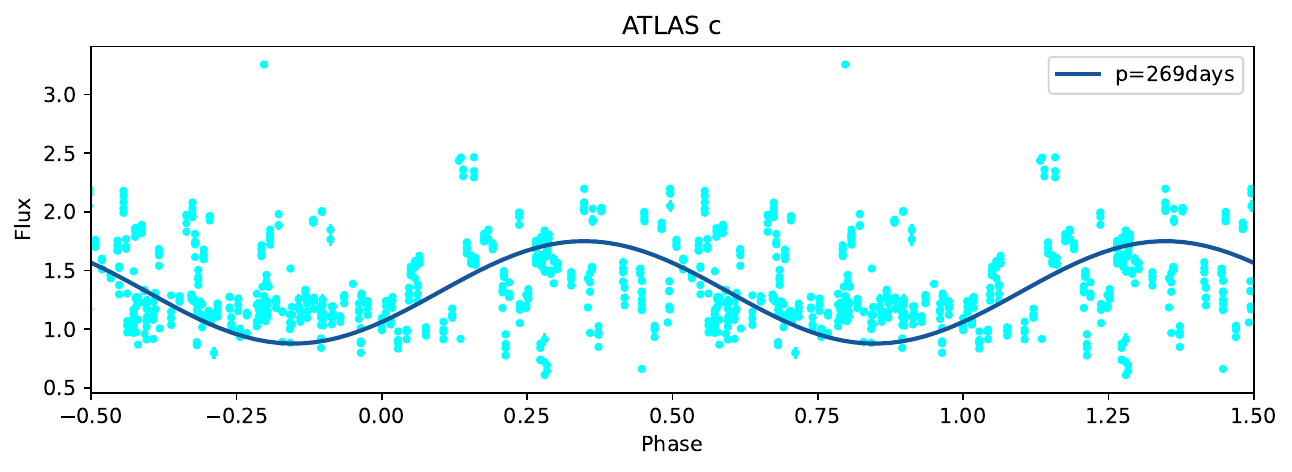}
\includegraphics[width=0.48\textwidth]{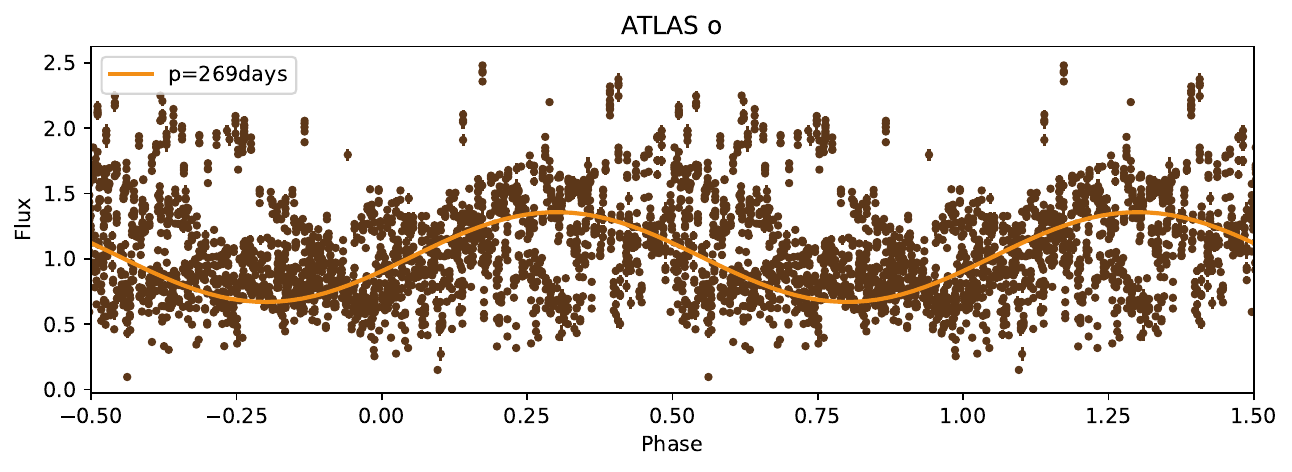}
\caption{
Same as Figure \ref{fig:QPO-ASAS-SN}, but for ATLAS c, and ATLAS o light curves. The periodic signal at 516 days in ATLAS c seems to be twice of 269 days. }
\label{fig:QPO-ATLAS}
\end{figure}

\begin{figure}[htbp!]
\centering
\includegraphics[width=0.48\textwidth]{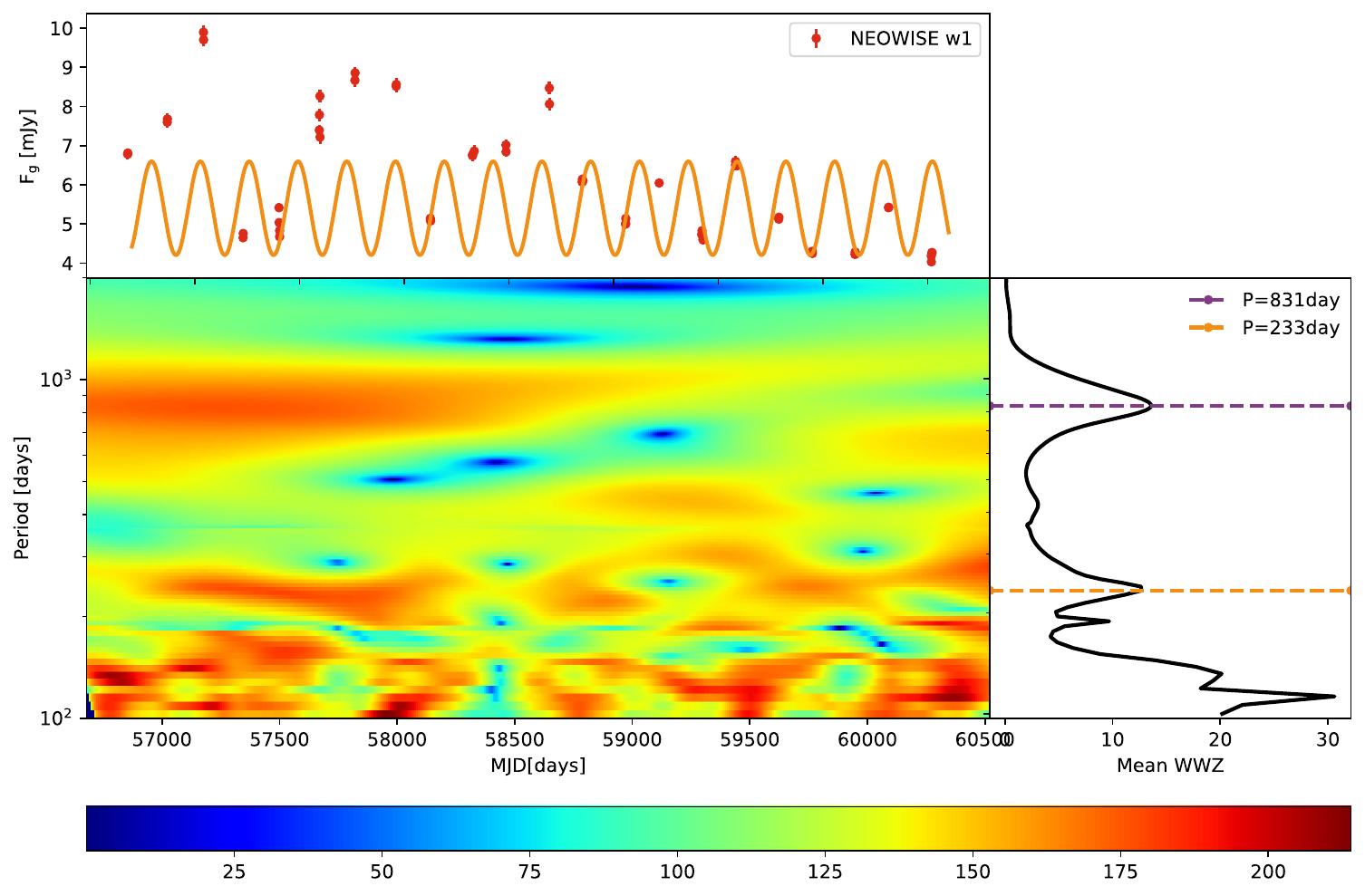}
\includegraphics[width=0.48\textwidth]{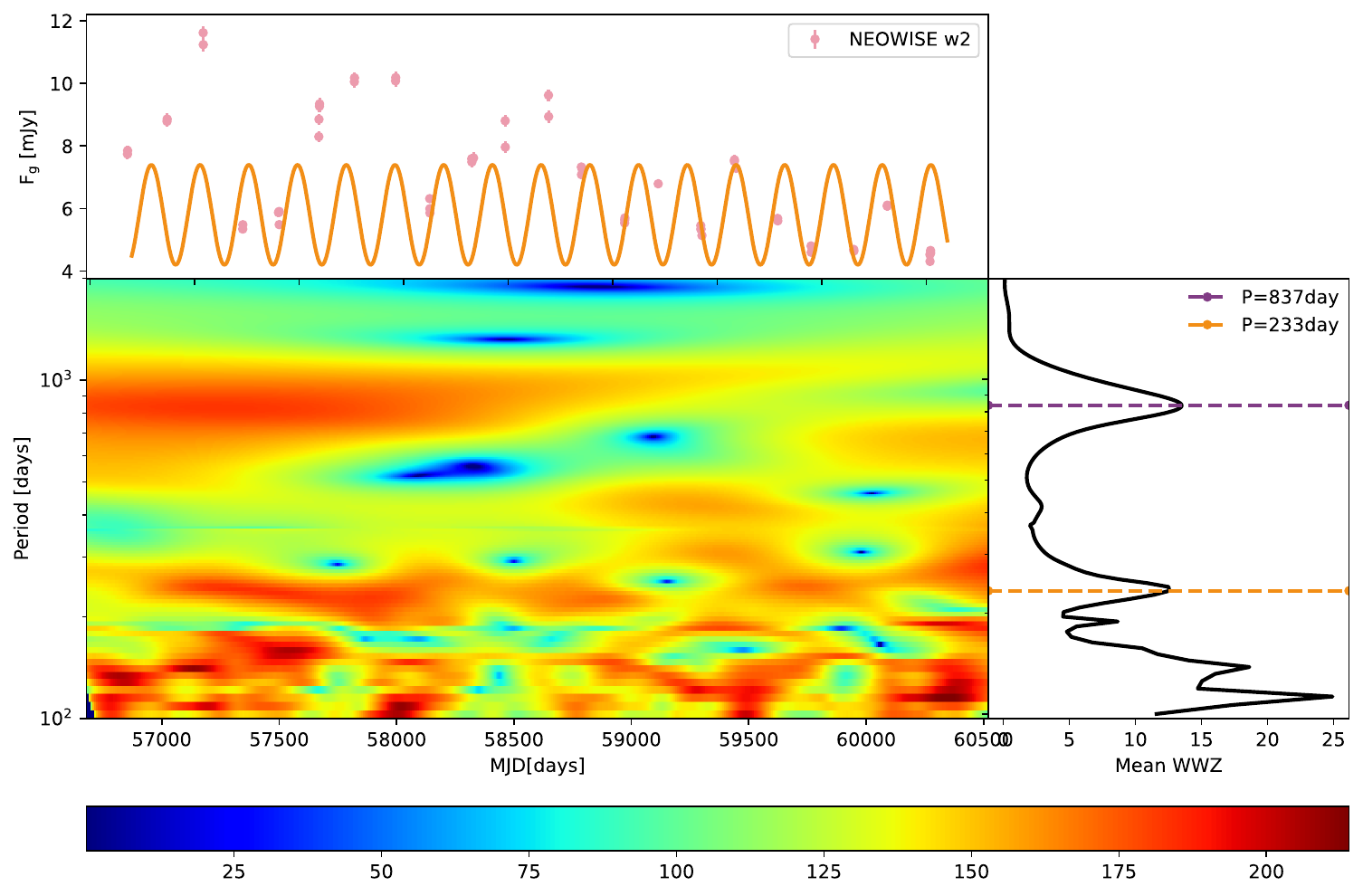}
\includegraphics[width=0.48\textwidth]{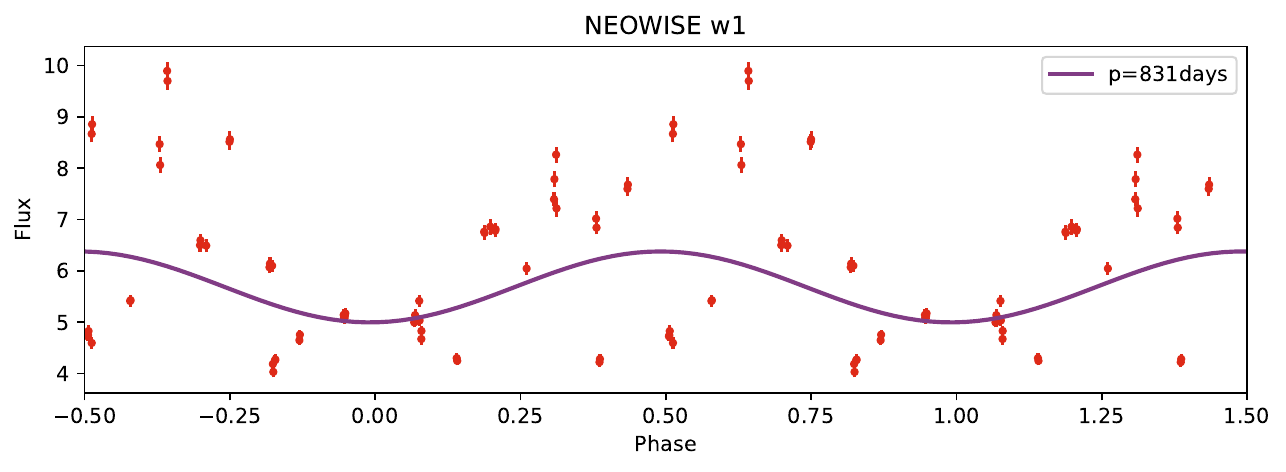}
\includegraphics[width=0.48\textwidth]{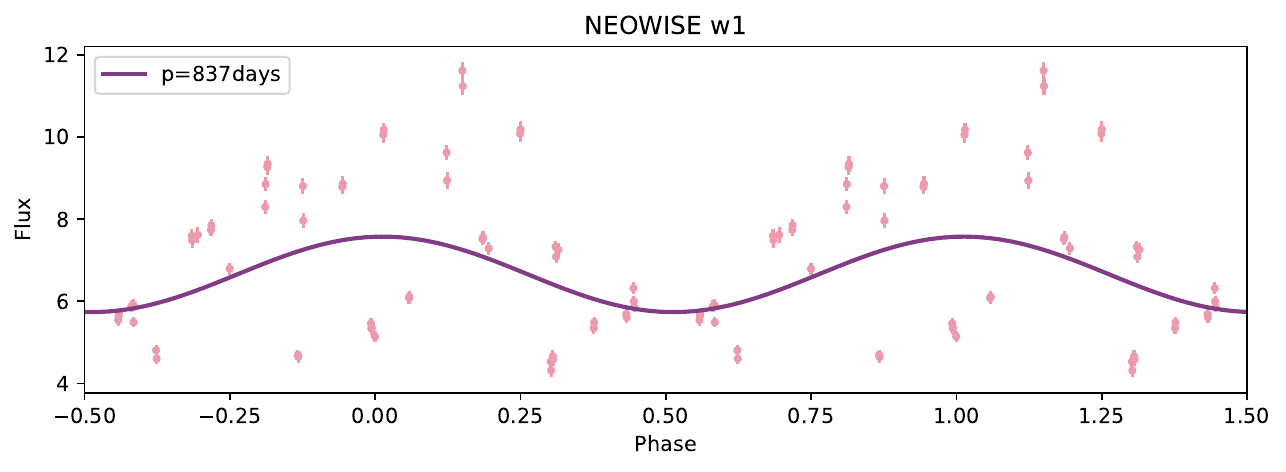}
\includegraphics[width=0.48\textwidth]{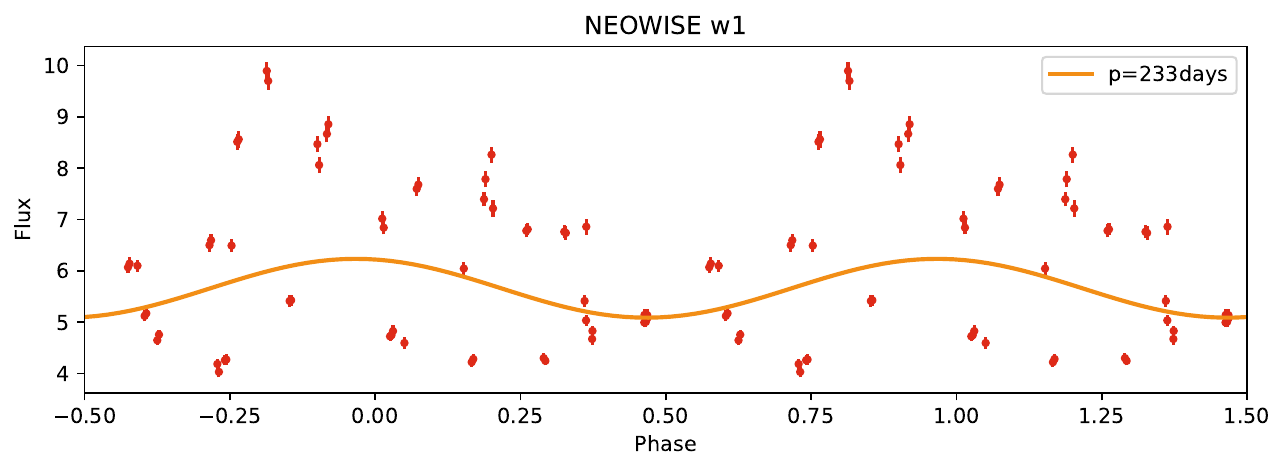}
\includegraphics[width=0.48\textwidth]{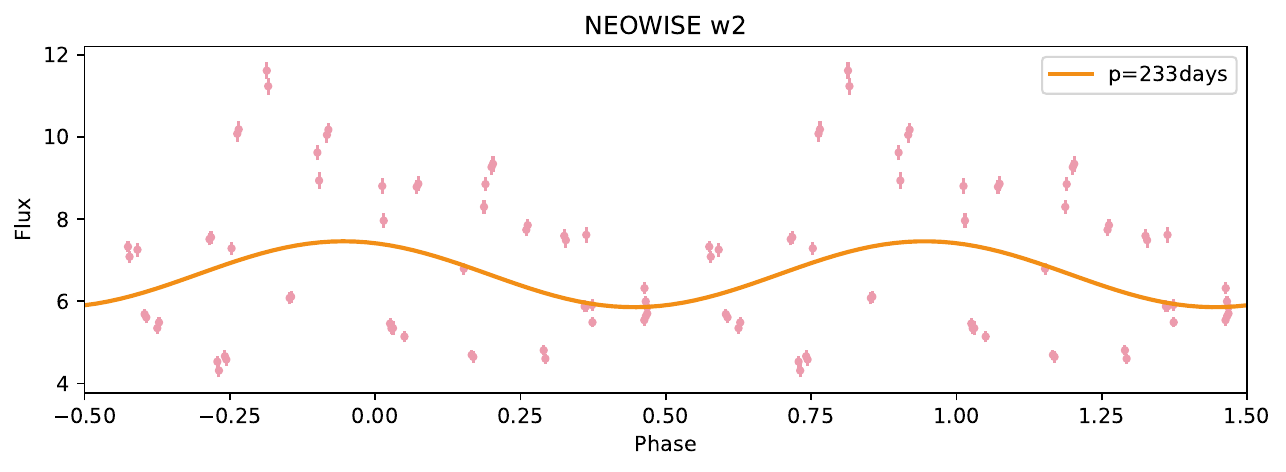}
\caption{Same as Figure \ref{fig:QPO-ASAS-SN}, but for NEOWISE $W1$ and $W2$ bands light curves.}
\label{fig:QPO-IR}
\end{figure}

\begin{figure}[htbp!]
\centering
\includegraphics[width=0.48\textwidth]{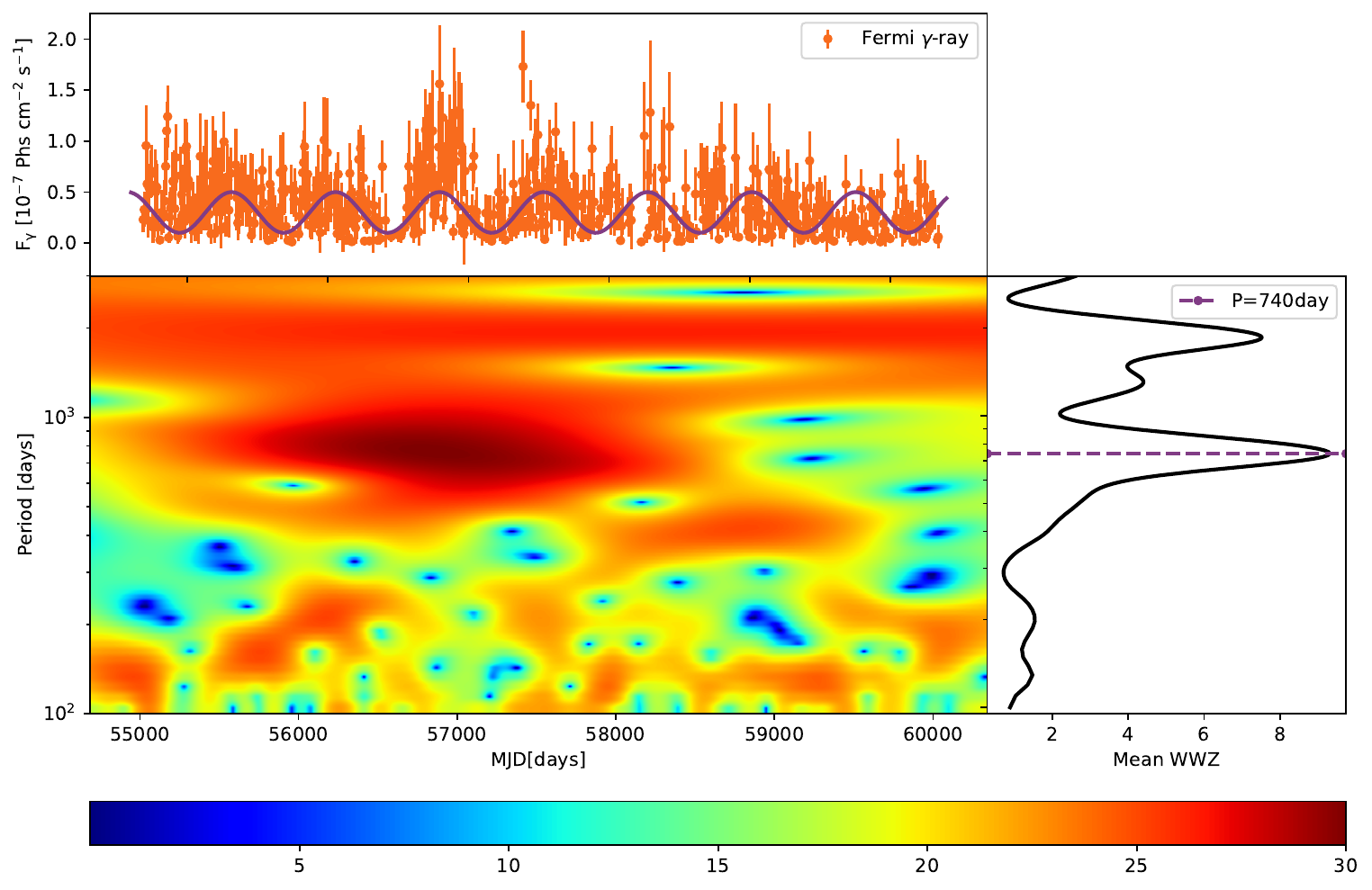}
\includegraphics[width=0.48\textwidth]{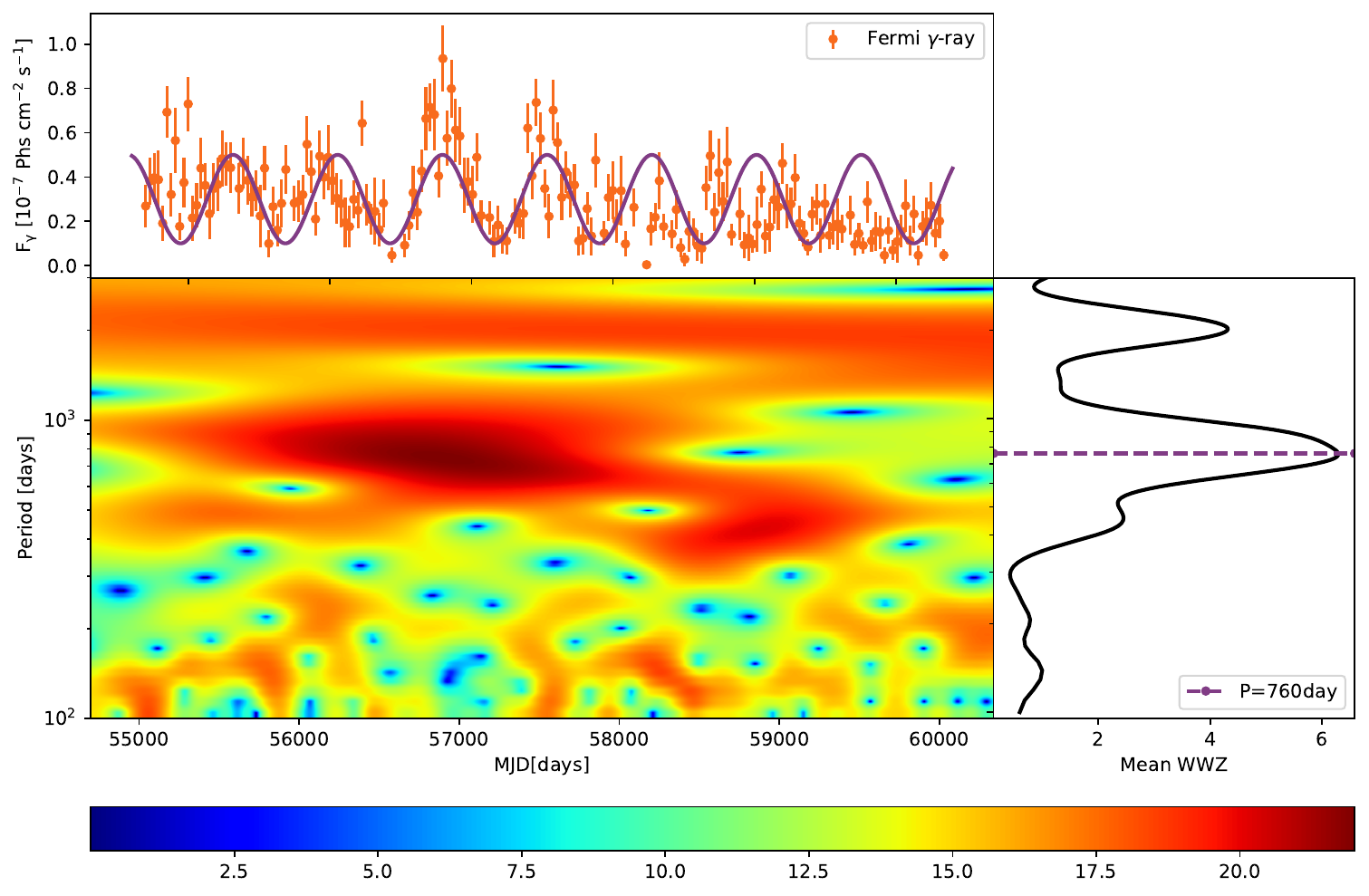}
\includegraphics[width=0.48\textwidth]{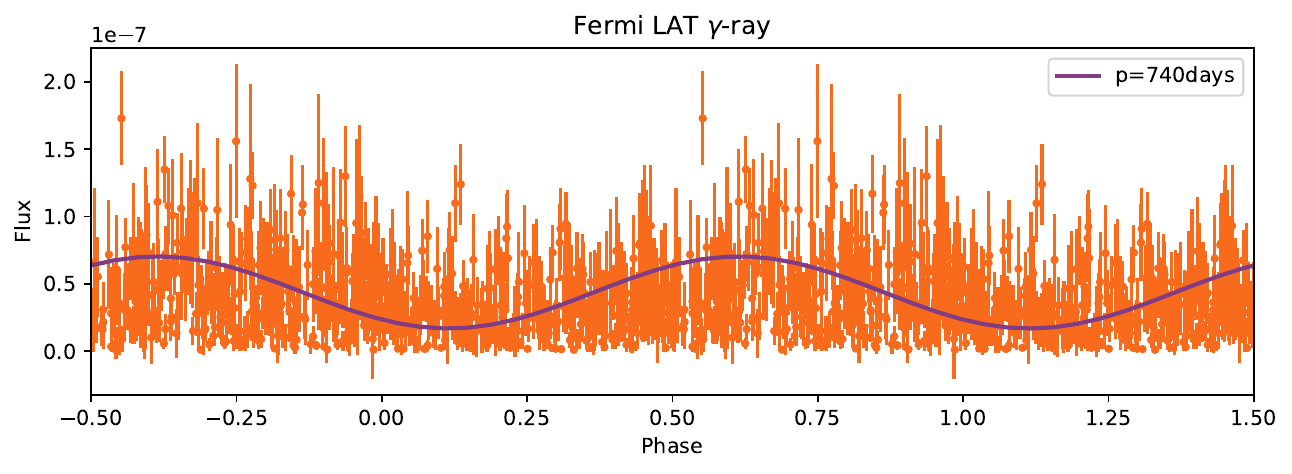}
\includegraphics[width=0.48\textwidth]{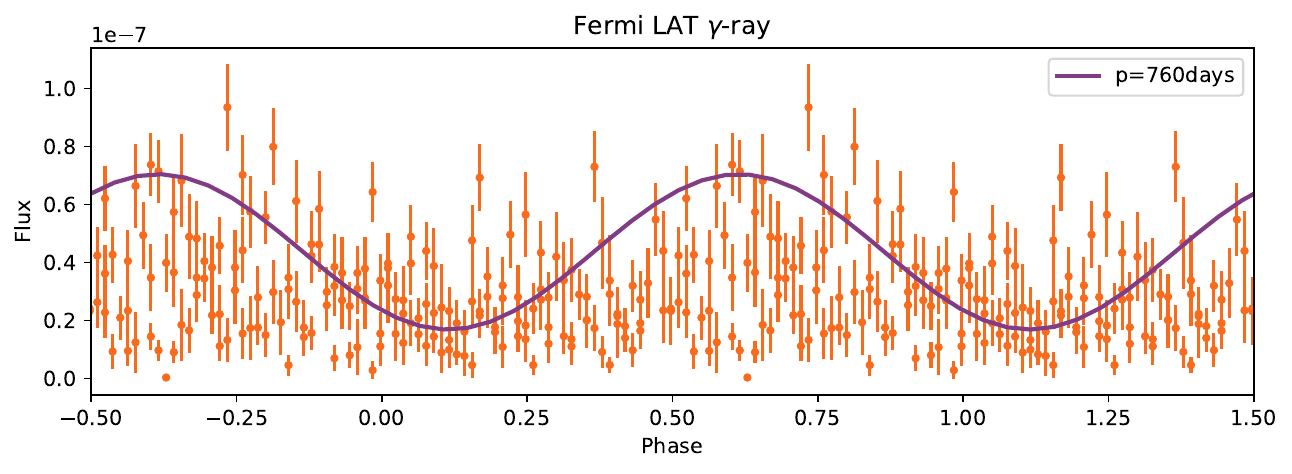}
\caption{Same as Figure \ref{fig:QPO-ASAS-SN}, but for Fermi-LAT $\gamma$-ray. Left panel shows the result of light curve with time bin of 7 days, while the right panel exhibits that with time bin of 30 days.  A prominent peak centred around 740 days is observed in the 7-day binned light curve. Similarly, a 760-day QPO is observed in the 30-day binned light curve.}
\label{fig:QPO-Fermi}
\end{figure}

Based on the Morlet wavelet\cite{Grossmann}, Foster\cite{Foster_1996} proposed the WWZ, which provides an effective method for analysing signals jointly in both the time and frequency domains. The WWZ can be applied to unevenly sampled data by projecting the signal onto a rescaled wavelet basis using weighted least squares. In addition, the WWZ method reveals the temporal evolution of any periodic signal over the duration of the observation. Both stationary and transient periodic signals at diverse timescales can be identified. 
We used the publicly available Python script developed by O'Neill et al.\cite{ONeill_2022}, performing the WWZ with 300 frequency bins for each light curves over a frequency range from 0.01 $\rm d^{-1}$ to $2/T$ $\rm d^{-1}$, where $T$ approximately equal to the total observational duration of each light curve. Figure \ref{fig:QPO-KAIT+ZTF} shows the results of the WWZ. We also employ GLSP to the supplementary data including ASAS-SN $g$ and $V$ bands, ATLAS c and o, NEOWISE $W1$ and $W2$, and Fermi-LAT $\gamma$-ray. These results are summarised in Figure \ref{fig:QPO-ASAS-SN}, \ref{fig:QPO-ATLAS}, \ref{fig:QPO-IR}, and \ref{fig:QPO-Fermi}. Similar optical periodic modulation ($\sim890$-day and $\sim270$-day) has been observed in these supplementary data. 

\subsection{Significance estimation}
\begin{figure}[htbp!]
\centering
\includegraphics[width=0.48\textwidth]{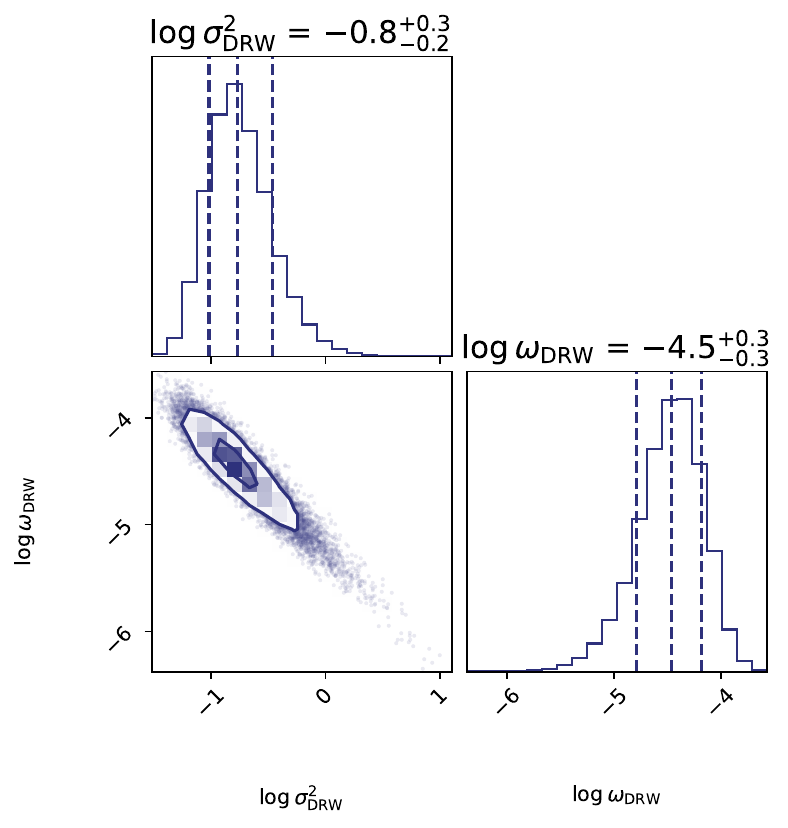} \\
\includegraphics[width=0.48\textwidth]{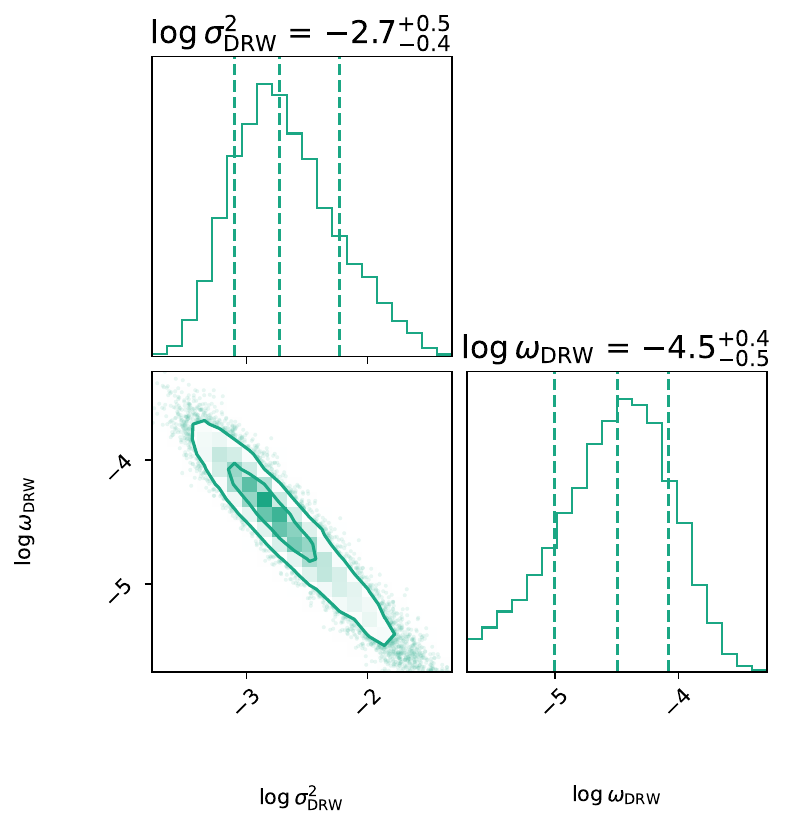} 
\includegraphics[width=0.48\textwidth]{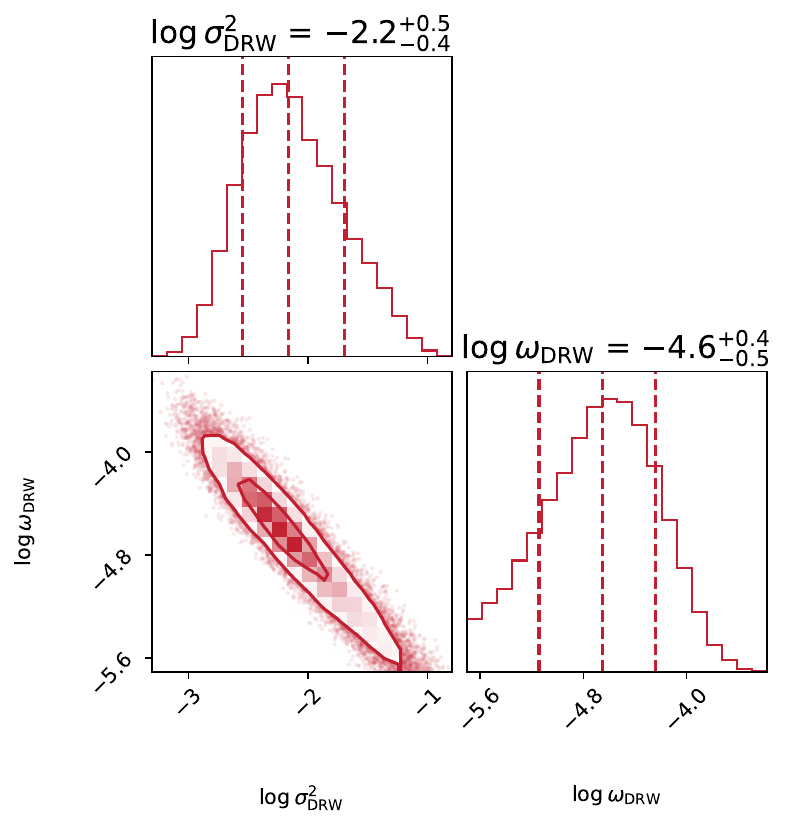}
\caption{Corner plots of the DRW fits to the KAIT, ZTF $g$ band, and ZTF $r$ band light curves. The top, lower-left, and lower-right panels show their posterior distributions. The vertical dashed lines mark the 16th, 50th, and 84th percentiles of the marginalized distributions, with the corresponding median values and uncertainties reported above each panel.}
\label{fig:DRW}
\end{figure}

Optical variability in AGN is often approximated by a damped random walk (DRW)\cite{Kelly_2009,MacLeod_2010,Xin_2024,Lin_2026}. In the present analysis, the DRW is adopted as a phenomenological model for the stochastic red-noise variability. The DRW corresponds to a stationary Gaussian process (GP; see Section \ref{sec:GP} for detailed modelling method) with an exponentially decaying covariance kernel\cite{Gurpide_2025},
\begin{equation}
    k_{\rm DRW}(\Delta t_{ij}) = \sigma_{\rm DRW}^2 e^{-\omega_{\rm DRW} \Delta t_{ij}},
\end{equation}
where $\sigma_{\rm DRW}^{2}$ is the stationary variance of the stochastic process, and $\Delta t_{ij} = |t_i - t_j|$ is the time lag between the $i$-th and the $j$-th observations. The characteristic damping timescale $\tau_{\rm DRW}$ is defined by the characteristic bending frequency
\begin{equation}
\omega_{\rm DRW}=\tau_{\rm DRW}^{-1}.
\end{equation}
Under the power spectral distribution (PSD) convention adopted here, the corresponding power spectral density is a bending power law 
\begin{equation}
    P_{\rm DRW}(\omega) = \sqrt{\dfrac{2}{\pi}} \dfrac{\sigma_{\rm DRW}^2}{\omega_{\rm DRW}} \dfrac{1}{1 + \left(\dfrac{\omega}{\omega_{\rm DRW}}\right)^2}.
\end{equation}
The spectrum is approximately flat at $\omega \ll \omega_{\rm DRW}$ and follows $P_{\rm DRW} \propto \omega^{-2}$ at $\omega \gg \omega_{\rm DRW}$. The low-frequency flattening prevents the divergent long-timescale variance associated with an unbroken power-law PSD. Figure \ref{fig:DRW} illustrates the posterior distribution plots of the DRW model for the observed light curves.

Robust significance estimation requires careful modelling of the underlying red noise which capture the feature observed in the AGN light curve. To evaluate the statistical significance of candidate periodic signals, we perform Monte Carlo (MC) simulations to calculate the local and global false alarm probabilities (FAPs) against a red noise background \cite{Wang_2020,Xin_2024,Wang_2025}. The MC recipe consists of four steps. First, we generated $10^4$ artificial light curves by drawing parameter samples from the DRW model posterior and simulating GP realisations at the observed time stamps\cite{Gurpide_2025}, with Gaussian measurement noise scaled to the median photometric uncertainty of the observed data. This procedure ensures that the simulated light curves faithfully reproduce both the red noise of the source and the irregular sampling cadence of the observation\cite{Gurpide_2025}. 
The same samplings of artificial light curves and observed light curve minimizing sampling effect in the significance analysis \cite{Chen_2024,Wang_2025qpo}. Second, the GLSP and WWZ methods are applied to each simulated light curve using identical frequency grids and algorithmic settings as for the observed data. Third, we examine the local FAP at a given frequency as the fraction of simulated light curves whose power at that frequency equals or exceeds the observed power, yielding local significance curves with thresholds corresponding to 1$\sigma$ (68.27\%), 2$\sigma$ (95.45\%), and 3$\sigma$ (99.73\%) confidence. Fourth, we compute the global FAP following the approach of O'Neill et al.\cite{ONeill_2022}, as described below. 

Distinguishing a genuine periodic signal from stochastic variability in sources dominated by red noise, such as blazar, requires careful treatment of the look-elsewhere effect. The local FAP ($p_{\rm peak}$) measures the probability that a pure noise process produces power equal to or exceeding the observed value at the specific frequency of interest, evaluated directly from the ensemble of simulated periodograms.
However, searching over a broad frequency range introduces a trial factor: the same noise realisation will, by chance, produce an apparently significant peak somewhere in the spectrum\cite{Vaughan_2005}. The global significance addresses this by assessing the rarity of the observed signal across the entire searched parameter space. In the O'Neill et. al.\cite{ONeill_2022} methodology, rather than comparing absolute maximum powers of the observed and simulated periodograms\cite{Guo_2026},  the local FAP ($p_{\rm sim}$) is first computed for the strongest peak in each simulated light curve. The global FAP is subsequently defined as the fraction of simulated light curves for which $p_{\rm sim} \le p_{\rm peak}$. By translating absolute powers into frequency-normalized probabilities before performing the global comparison, this approach effectively flattens the red noise continuum. A high-frequency signal that is strongly anomalous relative to its local noise level is evaluated on equal statistical footing with a low-frequency fluctuations, preventing the suppression of genuine high-frequency periodicities by the steep spectral slope of the noise background.

To assess whether the detected periodogram features could be produced by the irregular observing cadence, we calculated the spectral window function\cite{VanderPlas_2018}
\begin{equation}
    P_{\rm W}(f)=\left|\sum_{n=1}^{N}\exp\left(-2\pi i f t_n\right)\right|^{2},
\end{equation}
where $t_n$ denotes the time stamp of the $n$-th observation. The spectral window depends only on the temporal sampling and identifies frequencies at which seasonal gaps and other cadence structures may redistribute spectral power or generate aliases. The window spectra are plotted in the GLSP plots in Figure~\ref{fig:QPO-KAIT+ZTF}. Their amplitudes are shown only for visualization, and the physically relevant comparison is between the locations of the window-function and GLSP peaks.

The dominant feature in the window spectra of the KAIT and ZTF light curves occurs close to the annual sampling frequency, $\log_{10}(f/{\rm d}^{-1})\simeq-2.56$, corresponding to a period of approximately one year. By contrast, the principal ZTF peak at  $P\sim270$ days, or $\log_{10}(f/{\rm d}^{-1})\simeq-2.43$, is clearly offset  from the annual window peak and is independently recovered in both the ZTF $g$ and $r$ bands. Moreover, all null hypothesis simulations were evaluated at the exact observational time stamps. Consequently, spectral leakage and spurious periodogram features generated by the irregular cadence acting on the DRW process are already incorporated into the simulated GLSP and WWZ distributions and hence into the reported local and global false-alarm probabilities. We therefore conclude that the significance of the $\sim 270$-day signal, particularly in the ZTF light curves, is not driven by the sampling window.

We also preformed phase folding analysis\cite{VanderPlas_2018} to visualize significant periodicity detected in the GLSP and WWZ results. The light curves are folded using these prominent peaks identified in WWZ, resulting in phase curves that span two full cycles.

\subsection{Gaussian process and Bayesian inference} \label{sec:GP}

AGNs usually have steep PSDs, commonly referred to as red noise. Failing to account for this background noise leads the overestimation of the peak significance in the periodogram \cite{Vaughan_2005}.  GPs have emerged recently as the solution of choice to model stochastic signals in irregular sampling time-domain data \cite{Aigrain_2023,Gurpide_2025}. GP offers a flexible approach for modelling unknown time series through non-parametric models capable of capturing a large family of functions with two components, including a kernel and a mean function.  The kernel choice, which defines the GP covariance matrix, regulates the overall shapes and smoothness of functions. 
The detailed introduction of various kernels applied in time series modelling was described by G\'{u}rpide \& Middleton \cite{Gurpide_2025}, who provided a python script \texttt{ mind the gaps} for the GP modelling.
Here, we introduce the periodic kernel which is not included in G\'{u}rpide \& Middleton \cite{Gurpide_2025}. The periodic kernel describes a purely oscillatory process, and is defined as:
\begin{equation}
    k_{\rm periodic} = \sigma_p^2 \cos{\omega_0 \Delta t_{ij}},
\end{equation}
where $\sigma_p^2$ is the amplitude of the periodic component, and $\omega_0$ is periodic angular frequency.

We modelled the observed light curves using GP regression implemented in the \texttt{mind the gaps} package\cite{Gurpide_2025}, which builds upon the \texttt{celerite} framework\cite{Foreman-Mackey_2017} for computationally efficient GP inference. The GP model was fitted directly to the irregularly sampled light curves. For each light curve, the posterior distributions of the kernel parameters were sampled using the Markov Chain Monte Carlo (MCMC) sampler \texttt{emcee}\cite{Foreman-Mackey_2013}, with 100 walkers. Convergence was assessed via the integrated autocorrelation time. The fairly uninformative (uniform) priors and the strategy ensuring convergence were described in Appendix C of G\'urpide \& Middleton\cite{Gurpide_2025}. The maximum likelihood parameter estimates were obtained by initialising the MCMC walkers at the gradient-descent optimum (\texttt{fit=True}), ensuring robust identification of the global posterior maximum.

To find the best models described the periodic or aperiodic process in the observed light curves, we modelled them by adopting the DRW, simple harmonic oscillator (SHO), Matérn-3/2, Lorentzian, and periodic kernels and their combinations. The Bayesian model was utilized to evaluate the robustness of the presence of different GP models. Bayesian statistics allows precise model selection through the computation of the model evidence. The Bayesian evidence (or marginal likelihood) $Z$\footnote{The Bayesian evidence is also written as $p(D|H)$.} of a model $H$ is a normalizing constant, defined as\cite{Vaughan_2010} 
\begin{equation}
    Z = \int p(D|\theta, H) p(\theta|H) d\theta.
\end{equation}
Here, $p(D|\theta, H)$ is the likelihood and $p(\theta|H)$ is the prior for the parameters of model $H$. $D$ is the data (observed light curve), $\theta$ are the hyper-parameters for the selected model $H$. $Z$ enables the selection of the preferred model given the same dataset. For model comparison, $H_i$ and $H_j$, the Bayes factor is defined as the ratio of the evidence of the two competing models
\begin{equation}
    B_{i,j} = Z_i / Z_j.
\end{equation}

We evaluated the Bayesian evidence for a variety of GP models using nested sampling implemented in Python package \texttt{dynesty} \cite{Speagle_2020}. The GP likelihood was constructed with the \texttt{mind the gaps} package \cite{Gurpide_2025}. The kernel parameters were sampled in uniform priors. The Lorentzian quality factors $Q$ had a prior over $1.5<Q<10^{4}$. For the ZTF single period fits, the period prior covered $135-500$ days. The longer period searches in the KAIT and combined light curves used $500-1800$ days. In the two periodic component fits, the shorter and longer period priors were $135-500$ days and $500-1800$ days, respectively. Each model was evaluated with 2000 live points.

The results for the KAIT, ZTF $g$ band and $r$ band light curves are summarised in Tables \ref{tab:GP-KAIT}, \ref{tab:GP-Zg}, and \ref{tab:GP-Zr}, respectively. The Bayes factors depend strongly on the stochastic process. For KAIT light curve, the 2 $\times$ DRW model has the highest Bayesian evidence among the considered null models. Adding a Lorentzian or periodic component to this background gives Bayes factors of approximately 2.2 and 2.4, respectively. These values indicate only a modest improvement, and are consistent with the global significance. We also tested whether the KAIT light curve require a second Lorentzian component at $\sim270$ days after including the longer periodic component at $\sim890$-day. The 2 $\times$ DRW + Lorentzian model centred near 930 days has $\ln Z=-14.2$, whereas the model containing the additional second Lorentzian has $\ln Z=-15.4$. The shorter period posterior does not show the $\sim 270$-day periodic component, and the second Lorentzian component lower the evidence by $\Delta\ln Z=-1.2$. The KAIT data therefore do not support the presence of an additional $\sim 270$-day periodic component. 

For the ZTF $g$ band, the DRW model has the largest evidence among null models, but the DRW + Matérn-$3/2$ + Lorentzian model has the largest evidence considering the periodic component. The Bayes factor of relative to the corresponding null model is approximately 7.0, but it is only approximately 1.8 relative to DRW + Lorentzian. 
For the ZTF $r$ band, DRW + Lorentzian and DRW + periodic models have largest Bayesian evidences. Therefore, the DRW model provides an adequate description of the stochastic variability in the ZTF light curves, while the evidence for an additional periodic component remains modest. The period posteriors of the Lorentzian and periodic kernels in the ZTF light curves remain concentrated near 270 days when the stochastic process is changed. The Lorentzian fit also produce narrow PSD peaks with posterior $Q$ value of the Lorentzian models in the high coherence regime. The stability of the inferred period supports a common $\sim270$-day feature in the ZTF light curves.

We also investigate the the combined KAIT and ZTF $r$-band light curve\footnote{No additional flux calibration is applied because the effective passbands are similar and the flux levels agree over the overlapping time interval.}. The Bayesian evidence for the different stochastic processes are listed in Table \ref{tab:GP-Kr}. The 2 $\times$ DRW model has the largest evidence among the tested null models, consistent with the KAIT-only analysis.
Across the single-period alternatives, the inferred period is centred near 930 days, consistent with the longer empirical period of $889^{+47}_{-46}$ days. In the corresponding Lorentzian fits, the posterior medians satisfy $\ln Q>1.5$, indicating that the fitted component represents the quasi-periodic signal rather than broad noise component. For the 2 $\times$ DRW + Lorentzian model, the Bayesian factor is 3.2 relative to the null model. The two Lorentzian models also recover the longer signal near 930 days, but has a poorly constrained shorter periodic component centred at $\sim 270$ days. The Bayes factor of the 2 $\times$ DRW + 2 $\times$ Lorentzian relative 2 $\times$ DRW + Lorentzian is 0.35. The combined light curve therefore provides modest support for the single periodic component but no evidence that a second component is required. One possible explanation is that the $\sim270$-day periodic signal has a smaller amplitude and is more difficult to modelled under the irregular KAIT sampling.

\section{Least-squares Fitting of the Light Curve} \label{sec:MCMC}

We fit the observed light curves with a sinusoidal wave function
\begin{equation}
    F_{{\rm sine},i} = A \sin \left[\dfrac{2\pi}{P} (t_i-t_0)\right] + h, \label{eq:sine}
\end{equation}
where $t_i$ is observed time of the $i$-th data point. This model has 4 free parameters: amplitude $A$, period $P$, the reference time $t_0$ encoding the initial phase, and mean flux level $h$. It provides a phenomenological description of approximately coherent periodic variability expected from orbital modulation.
To account for the long-term declining trend apparent in the KAIT and ZTF light 
curves, we extend Equation (\ref{eq:sine}) by multiplying an exponential decay factor
\begin{equation}
    F_{{\rm sine+exp},i} = \left\{ A \sin \left[\dfrac{2\pi}{P} (t_i-t_0)\right] + h \right\} e^{-k(t_i-t_1)}. \label{eq:sine+exp}
\end{equation}
where $t_1$ is the epoch of the first observation, fixed throughout the fitting, and $k$ is the exponential decay rate. This extended model introduces an additional free parameter $k$, giving five parameters in total. A possible origin of the decline trend is a gradual decrease of the intrinsic jet power, potentially associated with long-term evolution of  accretion. 
We therefore apply the exponentially decayed sinusoidal model (Equation (\ref{eq:sine+exp})) to the KAIT flux. We model the ZTF light curves using the two models.

We find the best-fitting parameters by maximizing the likelihood\cite{ONeill_2022}
\begin{equation}
    \ln \mathcal{L} = - \dfrac{1}{2} \sum_{i=1}^{N} \left[ \dfrac{(F_i - F_{{\rm model},i})^2}{\sigma_i^2 + \sigma_0^2} + \ln (\sigma_i^2 + \sigma_0^2) \right],
\end{equation}
where $F_i$ is the observed flux, $F_{\rm model}$ is $F_{\rm sine}$ or $F_{\rm sine+exp}$, and $\sigma_i$ is photometric error. $\sigma_0$ represents the characteristic amplitude of an intrinsic AGN variability, which accounts for excess variability not captured by the deterministic periodic model or by the measurement errors. In AGN light curves, this term phenomenologically represents unresolved stochastic variability and other short-timescale fluctuations. We treat $\sigma_0$ as a free parameter and infer it simultaneously with the parameters of the periodic model.

Parameter uncertainties are estimated using the Affine Invariant MCMC Ensemble sampler \texttt{emcee}\cite{Foreman-Mackey_2013} with 64 walkers. The sampler is run for a maxinum of $50000$ steps. Convergence is assessed every 500 steps using the integrated autocorrelation time $\hat{\tau}$ estimated for each free parameter\cite{Gurpide_2025}. We declare convergence when two conditions are satisfied simultaneously: (i) the total chain length exceeds $50 \hat{\tau}_{\max}$, where $\hat{\tau}_{\max}$ is the longest autocorrelation time $\hat{\tau}$ across all parameters, ensuring that each walker has explored the posterior sufficiently many independent times; and (ii) $\hat{\tau}$ changed less than 1\% compared to the previous estimate. Once convergence is reached, we discard the first $2 \hat{\tau}_{\max}$ steps as burn-in and thin the remaining chains by $\hat{\tau}_{\min}/2$ to obtain approximately independent samples\cite{Foreman-Mackey_2013, Gurpide_2025}.
Parameter values and their $1\sigma$ uncertainties are reported as the 50th, 16th, and 84th percentiles of the marginalised posterior distributions, respectively.

\begin{figure}[htbp!]
\centering
\includegraphics[width=\textwidth]{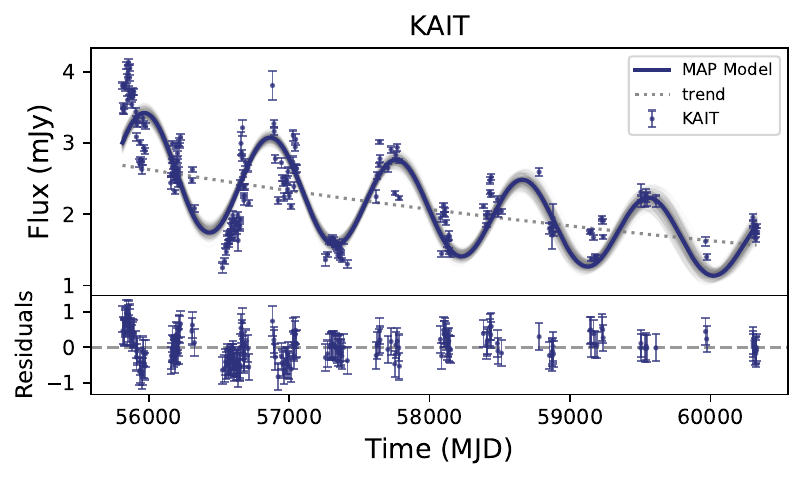} \\
\includegraphics[width=0.8\textwidth]{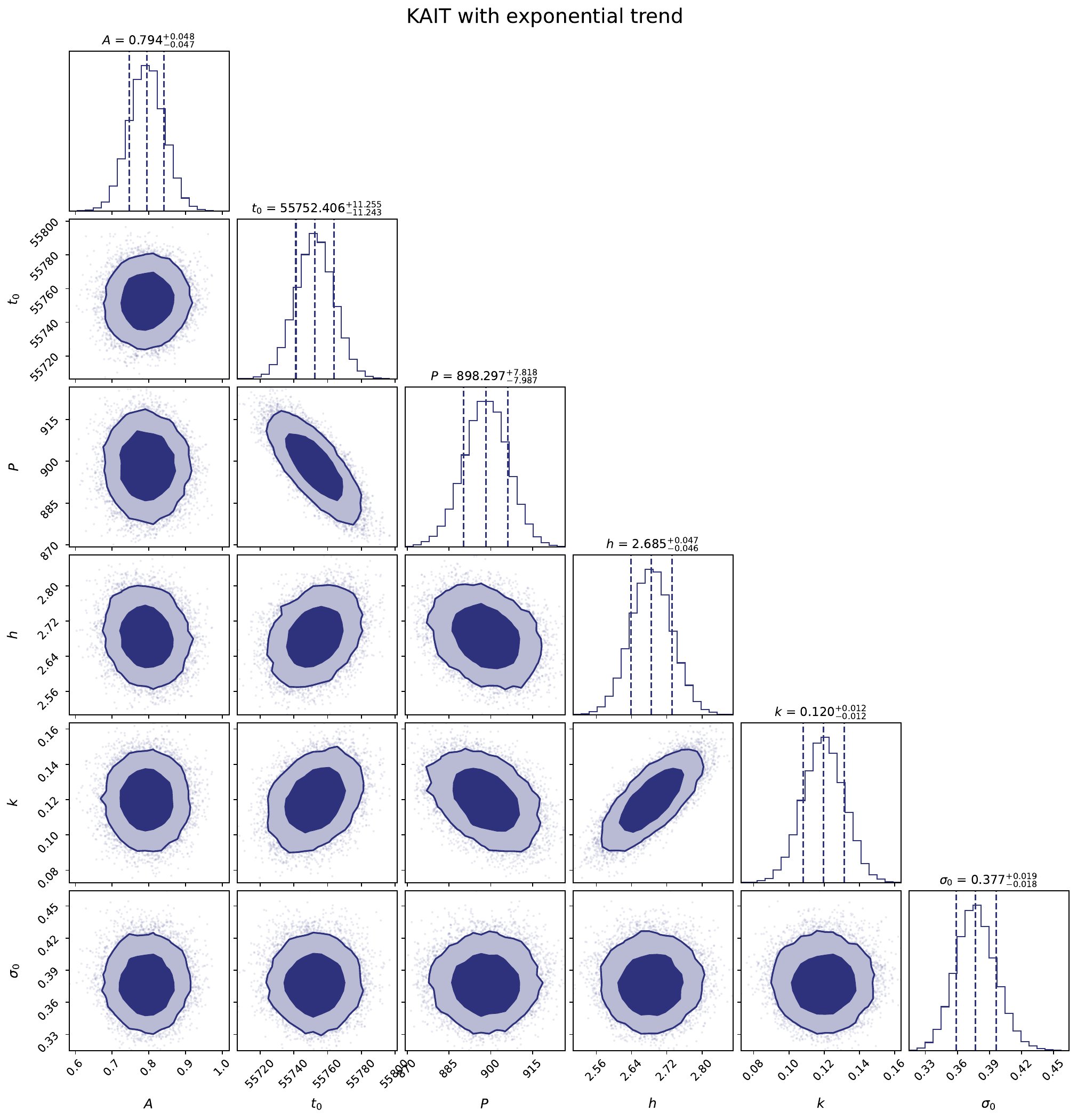}
\caption{The MCMC fitting results for KAIT light curve using the exponentially decayed sinusoidal model. Top panel: the observed light curve overlaid with the MAP best-fitting model and its residuals. Bottom panel: corner plot showing the 1D marginalised posterior distributions and 2D joint posteriors of the fitted parameters. Contours indicate the 68\% and 95\% confidence levels, and the vertical dashed lines on the marginalized histograms mark the 16th, 50th, and 84th percentiles. The sampler run for approximately 10500 steps until convergence, from which the first 310 steps are discarded as burn-in.}
\label{fig:MCMC-KAIT}
\end{figure}

\begin{figure}[htbp!]
\centering
\includegraphics[width=0.48\textwidth]{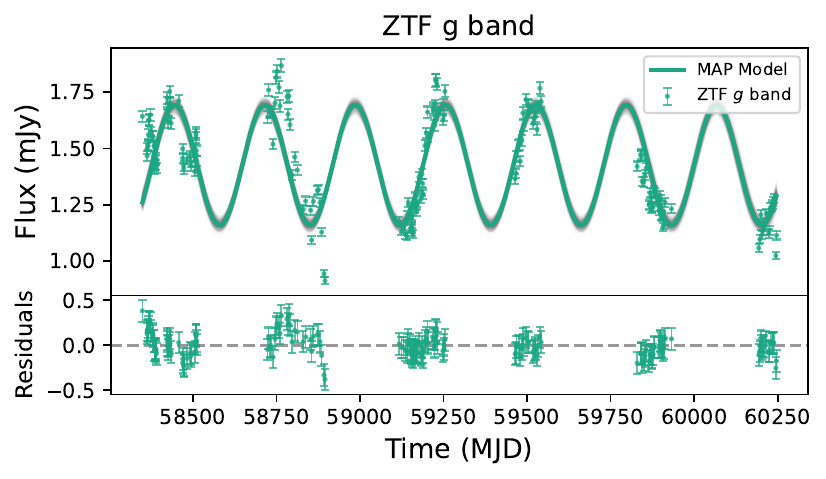} 
\includegraphics[width=0.48\textwidth]{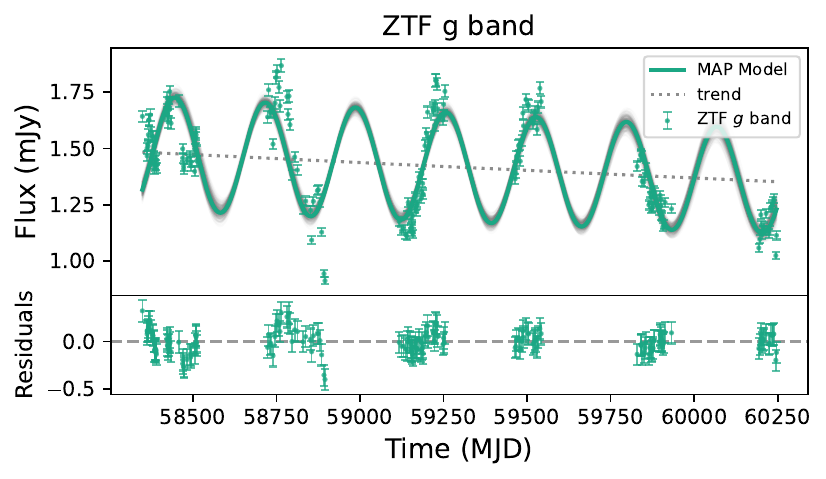} 
\includegraphics[width=0.48\textwidth]{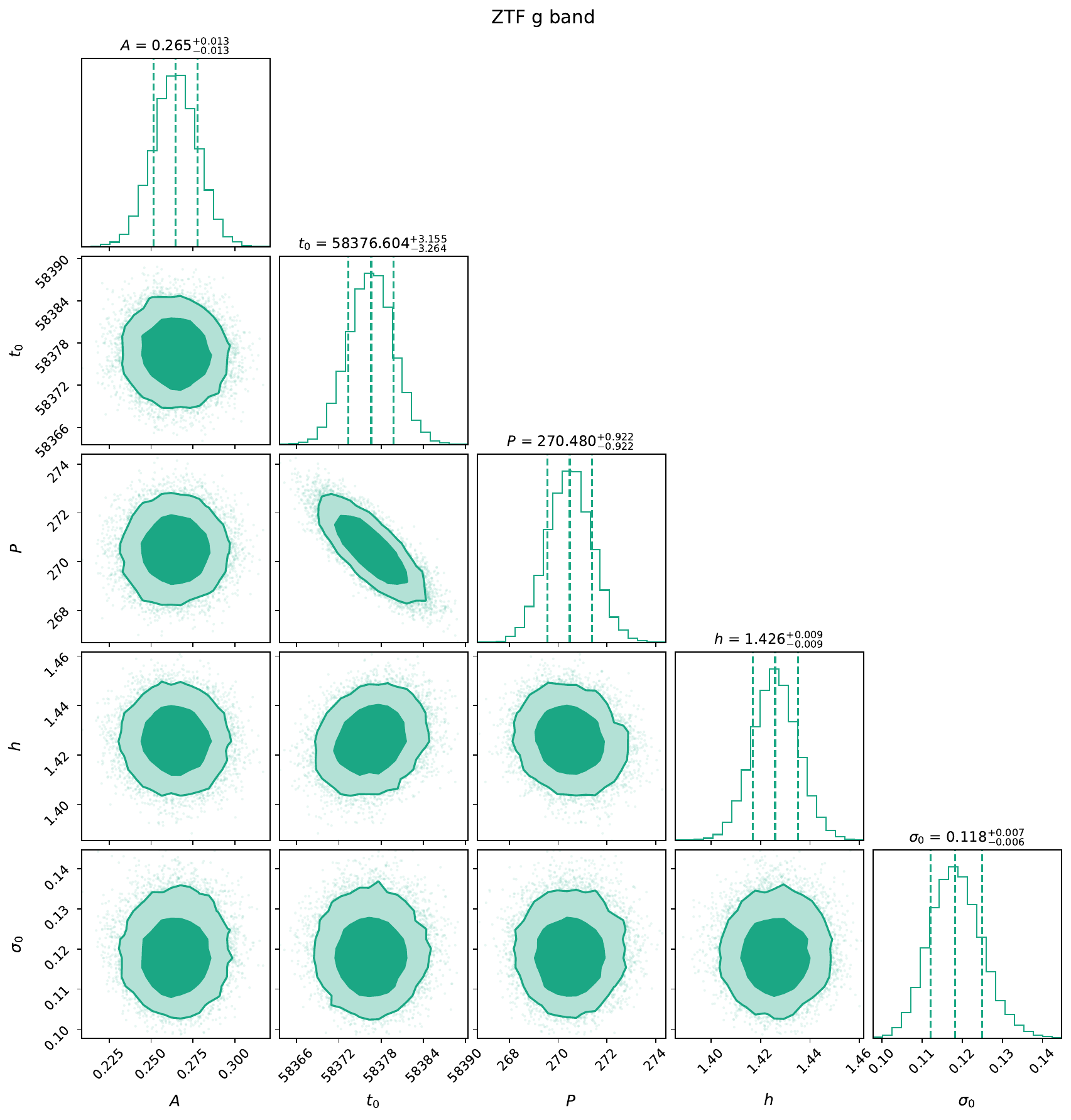} 
\includegraphics[width=0.48\textwidth]{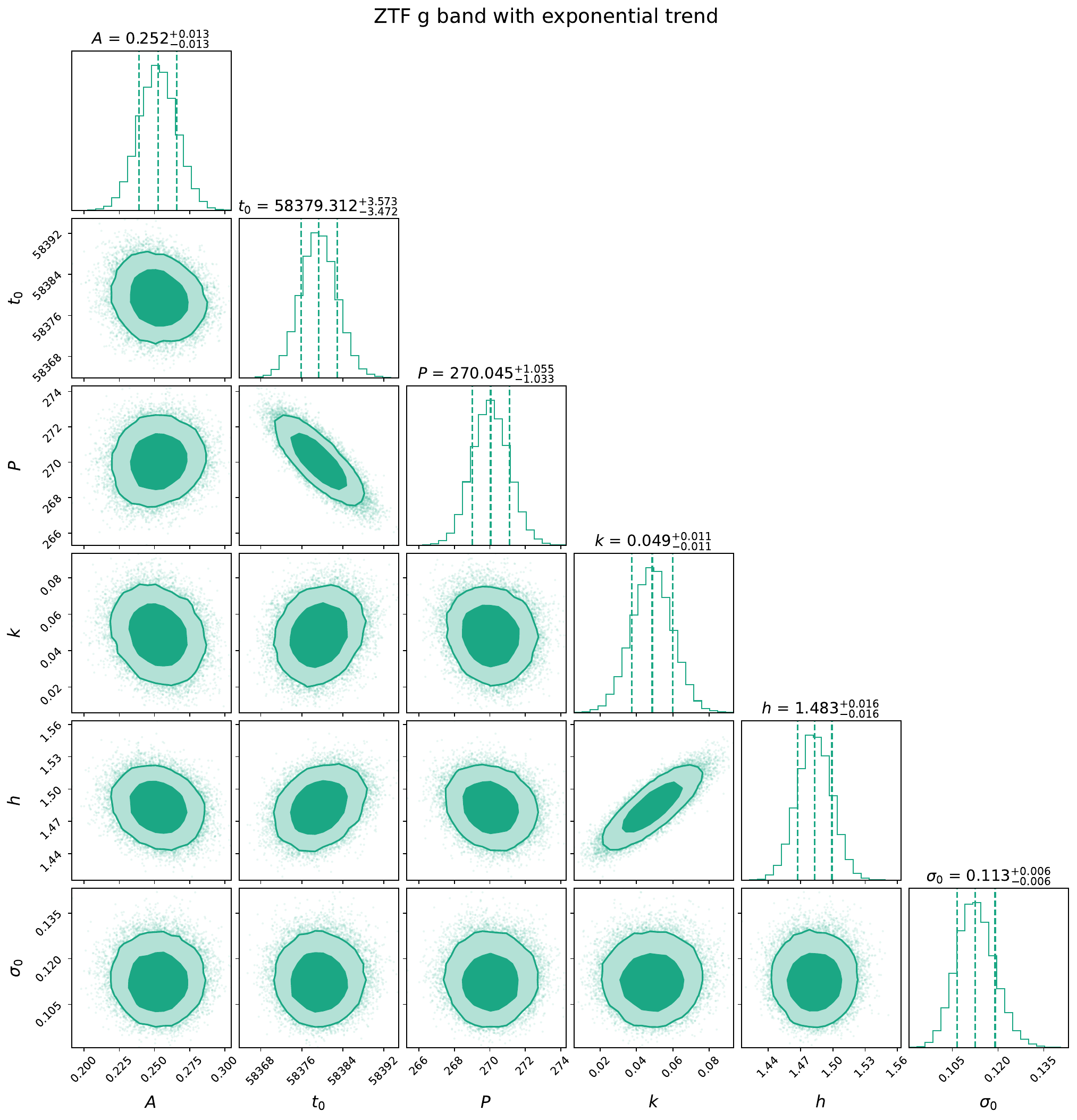}
\caption{The MCMC fitting results for ZTF $g$ band light curve using the sinusoidal model (left column) and the exponentially decayed sinusoidal model (right column). Upper panels: the observed light curve overlaid with the MAP best-fitting models and their residuals. Lower panels: corner plots showing the 1D marginalised posterior distributions and 2D joint posteriors of the fitted parameters. Contours indicate the 68\% and 95\% confidence levels, and the vertical dashed lines on the marginalized histograms mark the 16th, 50th, and 84th percentiles. The sampler run for approximately 6500 and 10500 steps until convergence for sinusoidal model and exponentially decayed sinusoidal model, respectively, from which the first 227 and 248 steps are discarded as burn-in.}
\label{fig:MCMC-Zg}
\end{figure}

\begin{figure}[htbp!]
\centering
\includegraphics[width=0.48\textwidth]{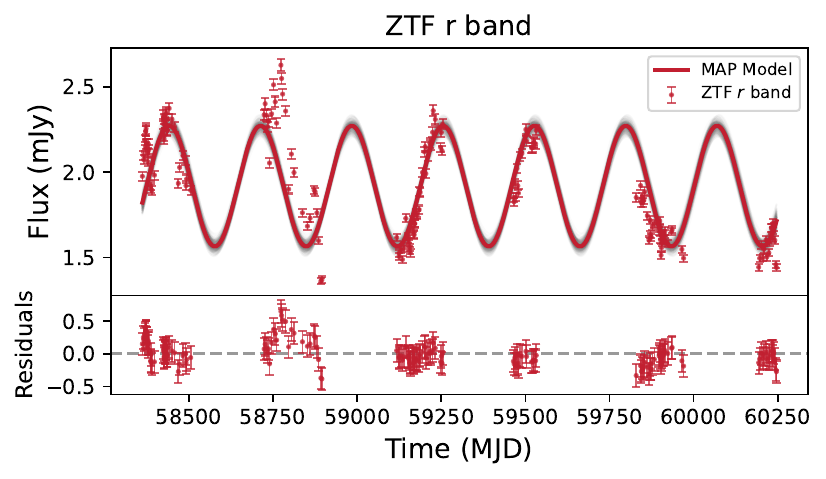} 
\includegraphics[width=0.48\textwidth]{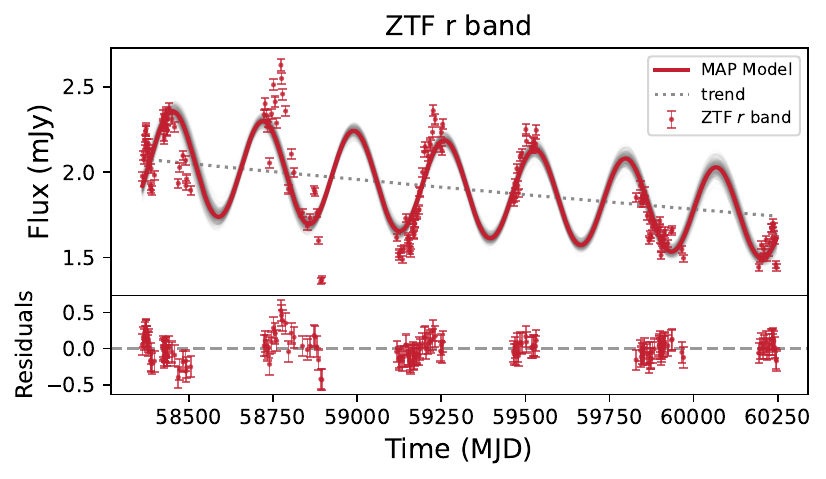} 
\includegraphics[width=0.48\textwidth]{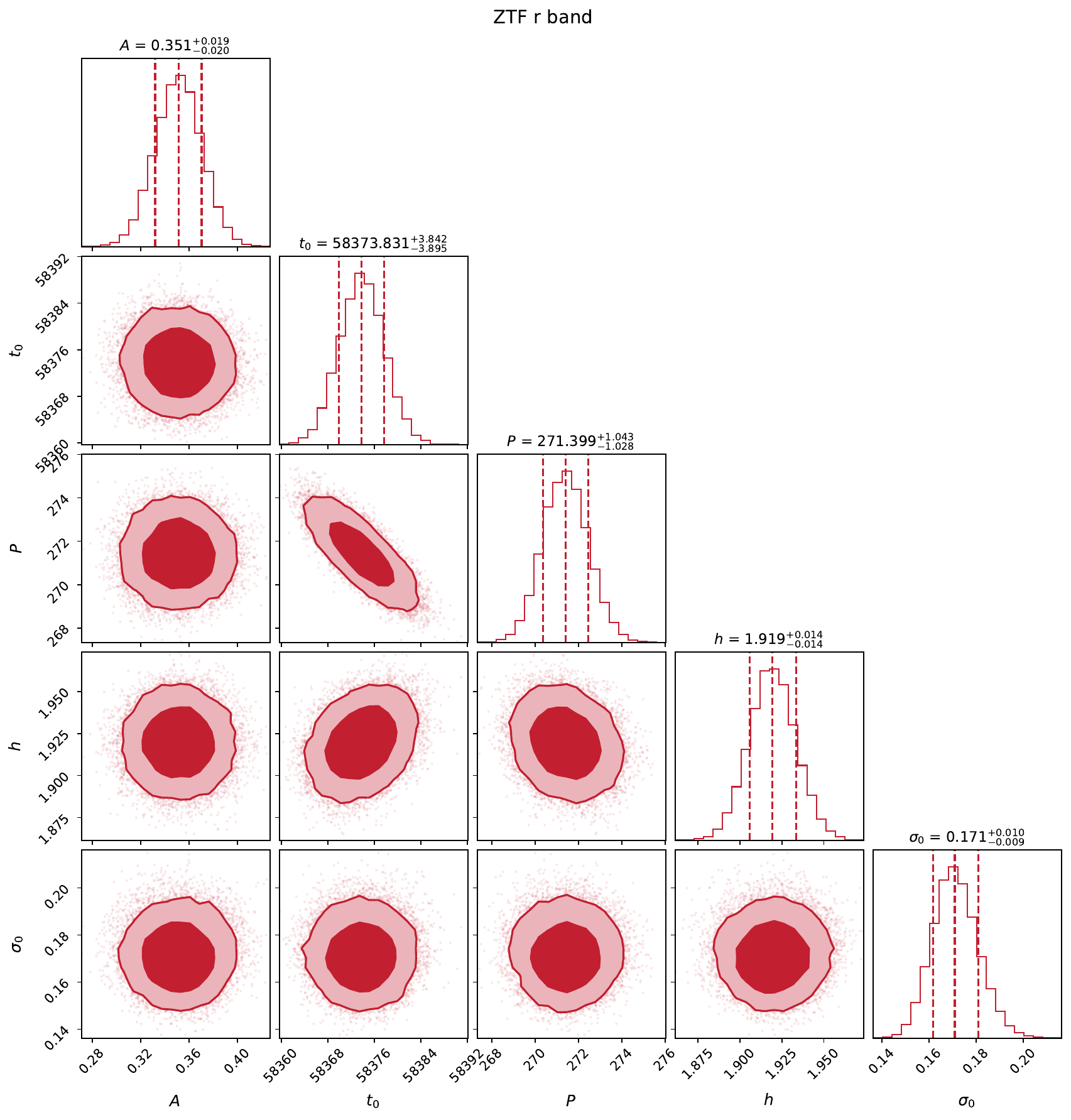} 
\includegraphics[width=0.48\textwidth]{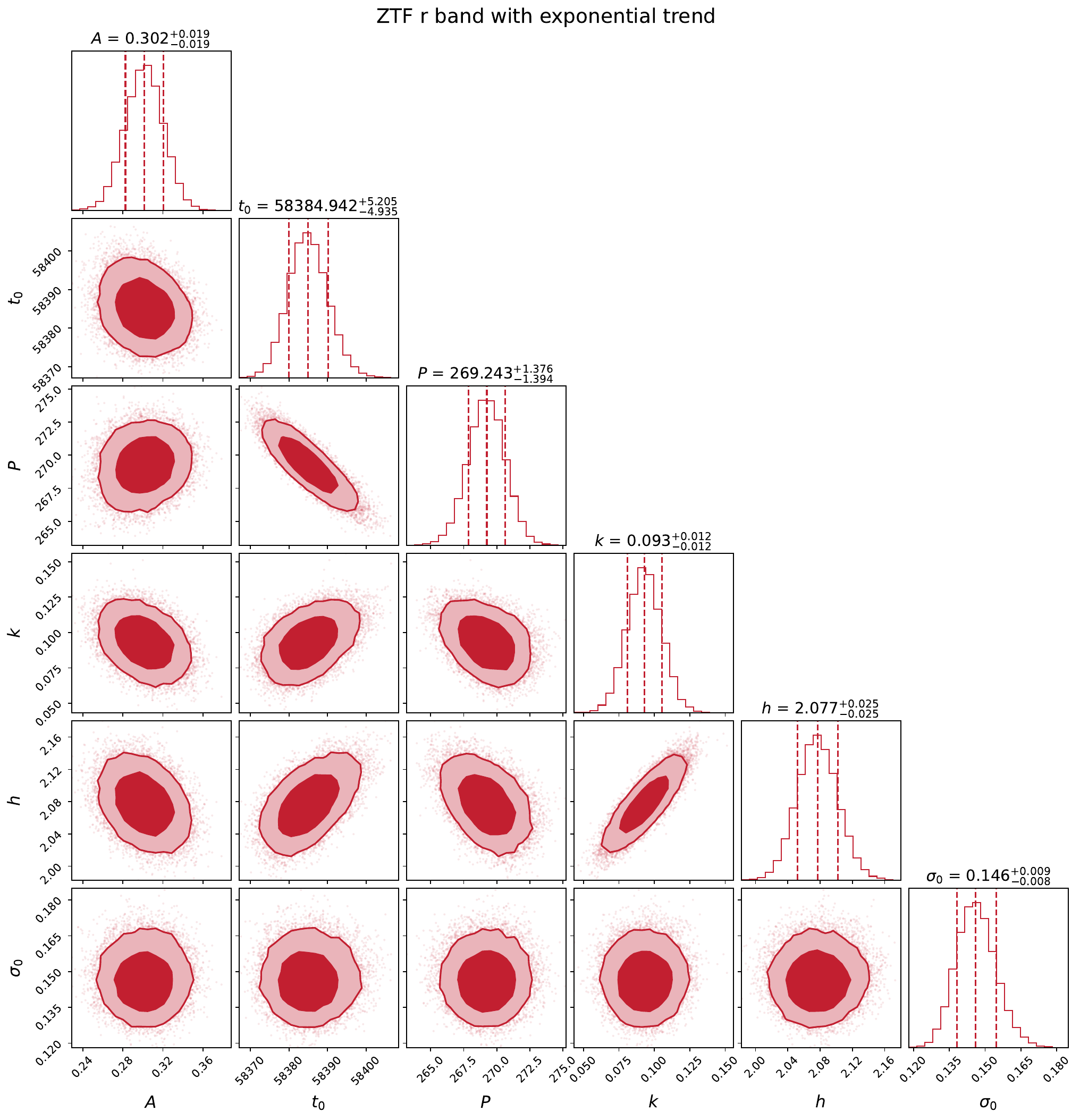}
\caption{Same as Figure \ref{fig:MCMC-Zg}, but for ZTF $r$ band light curve. The sampler run for approximately 10000 and 5500 steps until convergence for sinusoidal model and exponentially decayed sinusoidal model, respectively, from which the first 113 and 210 steps are discarded as burn-in.}
\label{fig:MCMC-Zr}
\end{figure}

All ZTF light curves are fitted by both the sinusoidal model (Equation (\ref{eq:sine})) and the exponentially decayed sinusoidal model (Equation (\ref{eq:sine+exp})), whereas the KAIT light curve, which exhibits a more pronounced declining trend, is fitted with the exponentially decayed model only.  The fitting results of parameters and their uncertainties are listed in Table \ref{tab:MCMC-LC}. The maximum a posteriori (MAP) fitting results and the residual are plotted in Figures \ref{fig:MCMC-KAIT}, \ref{fig:MCMC-Zg} and \ref{fig:MCMC-Zr}.

\begin{figure}[htbp!]
\centering
\includegraphics[width=0.8\textwidth]{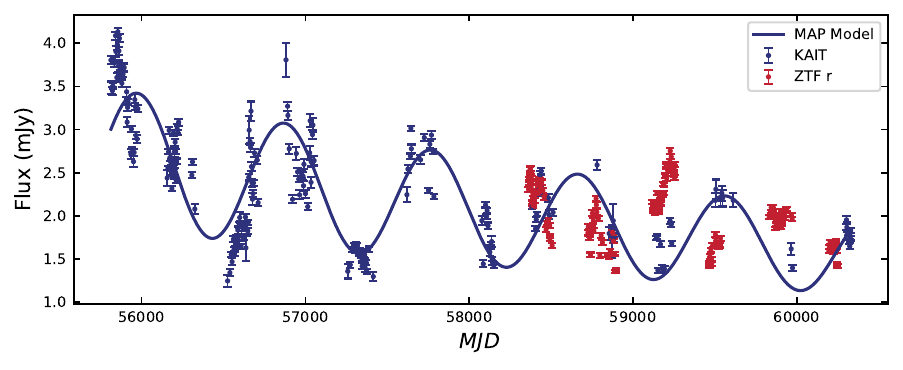}
\includegraphics[width=0.8\textwidth]{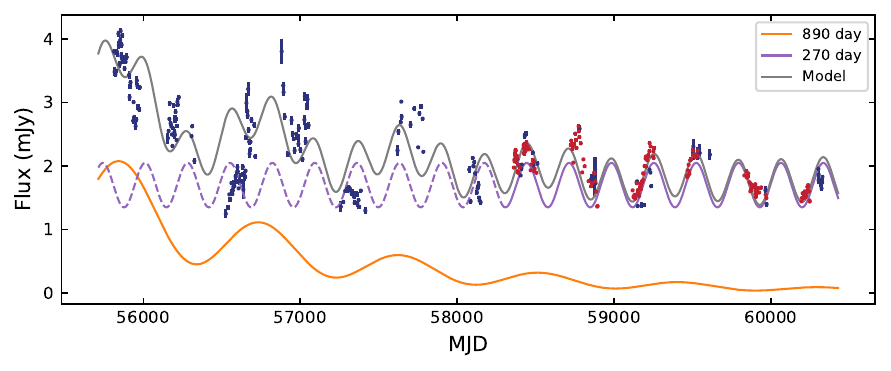}
\caption{Phenomenological model of the combined KAIT and ZTF $r$-band optical variability. The top panel shows the MAP model obtained from the MCMC fit to the KAIT light curve (Figure \ref{fig:MCMC-KAIT}) overlaid on the combined KAIT and ZTF $r$ band data. 
The bottom panel shows the same combined light curve with manually longer $\sim890$ day and shorter $\sim270$-day periodic components. The $\sim270$-day component is shown as a solid curve over the ZTF observational interval and extrapolated as a dashed curve to earlier epochs. This extrapolation is shown only for visualization and does not imply that the shorter periodicity is identified in the KAIT data.
The manually adjusted parameters are $A_{\rm jet} = 0.75$, $t_{0,{\rm jet}}=55660$, $h_{\rm jet} = 1.40$, $k_{\rm jet} = 7\times 10^{-4}$, $A_{\rm orb} = 0.35$, $t_{0,{\rm orb}}=58375$, and $h_{\rm orb} = 1.75$).}
\label{fig:decomposition}
\end{figure}

The MCMC fitting of the the KAIT light curve described above includes only the longer periodic component and therefore is not intended to reproduce all features of the combined KAIT and ZTF $r$ band light curves (see Figure \ref{fig:decomposition}).  We also construct a phenomenological two component model by combining the sinusoidal models of Equation (\ref{eq:sine}) and (\ref{eq:sine+exp}) for the shorter period and longer period components, respectively.  We fine-turn the parameters of the combined model, obtaining a better fitting result, illustrated in Figure \ref{fig:decomposition}. The resulting model reproduces the general variability trend of the combined light curve. The extension of the $\sim270$-day sinusoidal component into the KAIT observing interval is shown only to illustrate the behaviour expected if the shorter modulation persisted at earlier epochs. While the GLSP of KAIT shows a hint of power near this timescale, the signal is not independently confirmed by the WWZ analysis or GP modelling. The present data do not provide the evidence that the shorter periodicity persists throughout the complete KAIT + ZTF baseline.

\subsection{Injection-recovery test for empirical uncertainty} \label{sec:inject}

To obtain empirical estimates of the period uncertainties that account for stochastic red noise variability and the observational sampling, we perform an injection-recovery experiment. For each light curve, $10^4$ DRW light curves with the same observing times has been generated in significance estimation. A periodic component with the period and model parameters inferred from the observed light curve through the MCMC analysis (Table \ref{tab:MCMC-LC}) is then added to each DRW light curves\cite{Krishnan_2021,Penil_2025_gap,Penil_2025_flare,Lin_2026}. Each simulated light curve is then analysed using the same periodic model, likelihood function, and MCMC fitting procedure as adopted for the corresponding observed light curve.

The purpose of this experiment is to estimate the period uncertainty conditional on the injected periodic mode having already been identified. For an observing baseline $T$, and an injected period $P_{\rm inj}$, the period resolution for the first -order propagation associated with the finite baseline corresponds to a period scale\cite{Murakami_2022,Kovacevic_2026ApJS}
\begin{equation}
    \Delta P = \dfrac{P_{\rm inj}^2}{T}. \label{eq:P_resol}
\end{equation}
We consequently adopt $[P_{\rm inj} - \Delta P, P_{\rm inj} + \Delta P]$ as the period prior. This restricted interval is intended to isolate the
periodic mode associated with the injected signal and should not be interpreted as the search range of a blind period-detection experiment. 
Only fits that satisfy the adopted MCMC convergence criterion and yield a finite recovered periods are retained. For each accepted result, we record the posterior median period, denoted by $P_{{\rm rec},i}$. The distribution of the these posterior medians is used to quantify both the systematic recovery bias $\delta P= P_{\rm rec}^{\rm median} - P_{\rm inj}$, and the empirical measurement uncertainty. The median of the recovered period distribution $P_{\rm rec}^{\rm median}$ is adopted as the central estimate, while its 16th and 84th percentiles define the central $68\%$ empirical uncertainty interval. 

A noteworthy feature of the recovered period distributions is a modest accumulation near both the lower and upper limits of the adopted period interval. This boundary accumulation results from truncation by the finite period prior, whose width is determined from the period resolution scale $\Delta P$. For a small subset of simulated light curves,  the posterior support is not fully contained within the allowed interval. Restricting the  MCMC sampling to this range can therefore cause the corresponding posterior medians accumulate near its boundaries. These concentrations should not be interpreted as preferred periodicities, but rather as prior-truncated and poorly constrained recoveries. Nevertheless, the main bodies of both recovered period distributions remain centred close to the injected periods, with median biases of $\delta P \lesssim 2$ days. The boundary accumulation therefore does not appreciably shift the central recovered-period estimates. Accordingly, the derived distributions should be interpreted as conditional period recovery uncertainties,  rather than as an assessment of the ability of the fitting procedure to identify the correct period over an unrestricted parameter space.

To quantify the extent of this boundary accumulation, we define an edge region at each lower and upper limits of the adopted period prior. For a prior interval $[P_{\min},P_{\max}]$ with total width $W=P_{\max}-P_{\min}$, the lower-edge region is defined as $P_{\min}\leq P_{\rm rec}\leq P_{\min}+0.05W$, while the upper-edge region is defined as $P_{\max}-0.05W\leq P_{\rm rec}\leq P_{\max}$. Among the accepted fits, we count the numbers of recovered posterior-median periods falling within the lower and upper edge regions, $N_{\rm edge,low}$ and $N_{\rm edge,high}$, respectively. The total number of edge-affected recoveries was defined as $N_{\rm edge}=N_{\rm edge,low}+N_{\rm edge,high}$, and the corresponding edge fraction is $f_{\rm edge}=N_{\rm edge} / N_{\rm accepted}$, where $N_{\rm accepted}$ is the total number of converged fits with finite poster median periods. 
This statistic provides a direct measure of the fraction of accepted recoveries potentially affected by truncation at the adopted period prior boundaries. The $5\%$ threshold is used here as an operational definition of the boundary region and does not constitute an additional period uncertainty estimate. The recovered periods, acceptance fractions and edge fractions for the different light curves and periodic models are summarised in Table \ref{tab:MCMC-recover}. We use the ZTF $g$ band with exponentially decayed sine model to represent the results of ZTF light curves, because more fits are accepted and shows lower $f_{\rm edge}$ in this case. The representative empirical period estimates are $269^{+13}_{-14}$ days for the $\sim 270$-day modulation and $889^{+47}_{-46}$ days for the $\sim 890$-day modulation. 

\section{The schematic frame: an SMBHB system with a jet} \label{sec:SMBHB}
Periodic analyses indicate that PKS 0301-243 exhibits two distinct periodicities with different flux amplitudes in optical band. We propose an SMBHB model incorporating a jet, a circumbinary disc, and two mini-discs around two SMBHs (Figure \ref{fig:model}), which can reproduce the observed flux variation. The long-term $\sim 890$-day periodicity is attributed to the modulation of precessing jet, and the stable 269-day periodicity is plausible driven by orbital dynamics in an SMBHB system. 

\subsection{Jet precession}
In typical blazars, specially BL Lac object, the observed optical emission is dominated by the stochastic non-thermal emission from the relativistic jet. The precessing jet model has been widely applied to explain periodic variability\cite{Mohan_2015,Sobacchi_2017,Britzen_2018}. The Doppler factor varies as a function of the viewing angle  between the jet and the line of sight (see Mohan \& Mangalam\cite{Mohan_2015}, for a general relativistic treatment of the scenario). Here, we use the precessing jet model to interpret the $\sim 890$-day periodicity. Within this model, the time-dependent viewing angle $\theta(t)$ is given by\cite{Sobacchi_2017,Zhou_2018}
\begin{equation}
    \cos \theta(t) = \cos \phi \cos \psi + \sin \phi \sin \psi \cos \left( \dfrac{2\pi t}{P_{\rm jet}} \right), \label{equ:theta}
\end{equation}
where $\phi$ is the angle between the orbital angular momentum and the jet axis, $\psi$ denotes the angle between the orbital angular momentum of the primary black hole and the line of sight (Figure \ref{fig:model}), and $P_{\rm jet}$ is the observed jet periodicity. The flux is modified by Doppler boosting following $F^{\rm jet}_{\nu} \propto \delta^m F^{\rm jet^\prime}_{\nu^{\prime}}$, where $F^{\rm jet^\prime}_{\nu^{\prime}}$ is the intrinsic flux in the jet-comoving frame, $m=2$ for continuous jet, and $m=3$ for discrete jet. The Doppler factor is defined as $\delta = 1/ \Gamma (1-\beta \cos \theta)$ that depends on bulk Lorentz factor $\Gamma$, and the viewing angle $\theta$. The normalized velocity could be derived by $\beta = \sqrt{1 - 1/\Gamma^2}$. Consequently, the observed flux can be expressed as \cite{Sarkar_2021}
\begin{gather}
    F^{\rm jet}_{\nu} =  \delta^m F^{\rm jet^\prime}_{\nu^{\prime}}, \\
    \delta =\dfrac{1}{\Gamma (1+\sin\phi \sin\psi)} \left[ 1- \dfrac{\beta \cos\phi \cos\psi}{1+ \sin\phi \sin\psi} \cos \left( \dfrac{2\pi t}{P_{\rm jet}} \right) \right]^{-1} . \label{equ:F_sjet}
\end{gather}
In order to quantitatively describe the flux decaying trend of $\sim 890$-day periodicity, we introduce a decaying factor $e^{-k t}$, resulting in a time-dependent variation of the observed flux $F^{\rm jet,decay}_{\nu}$, which is given by
\begin{equation}
    F^{\rm jet,decay}_{\nu} = e^{-kt} F^{\rm jet}_{\nu}. \label{equ:cjet}
\end{equation}
The $k$ parameter can modulate the observed flux decrease trend\cite{Sarkar_2021}. The origin of the decaying factor $e^{-kt}$ could be interpreted by diverse ways. The $\gamma$-ray periodic fluctuation observed in 3C454.3\cite{Sarkar_2021} and PKS 0346-27\cite{Prince_2023_qpo} have been interpreted using the curved helical jet model with varying $\psi$, which successfully predicts the decaying multiplicative trend of flux.
Due to the fact that there is no observation in radio band constraining the value of $\phi$ and $\psi$, it is difficult to fit the KAIT light curve using this model due to the symmetry of $\phi$ and $\psi$. However, we estimated the value of $k$ in MCMC fit, given a value of $1.2\times 10^{-4}$. 

\subsection{Periodic Accretion}
The $\sim 270$-day period attributed to orbit period can be interpreted by periodic accretion onto mini-discs. Numerical simulations suggest that periodic variability can arises from the orbital motion of SMBH binaries. In a circumbinary disc, the gaseous materials are accreted onto mini-discs of SMBHs periodically, producing sawtooth shape light curves\cite{Farris_2014,Lai_2023,Lin_2026}. However, the shape of light curve also depends on orbital parameters, particularly eccentricity\cite{Miranda_2017,Charisi_2022,Lai_2023}. According to Miranda et. al.\cite{Miranda_2017} (see their Figure 5), for an equal-mass SMBHB ($q$ = 1) with viscosity parameter $\alpha$ = 0.1 and ellipticity $e$ = 0.1, the power spectrum displays two peaks at orbital period $P_{\rm orb}$ and $P_{\rm orb}/2$, and the light curve exhibits a break point during the descending phase. Similar features are observed in our GLSP and light curves in several cycles (Figure \ref{fig:QPO-KAIT+ZTF}), and the signal of $P_{\rm orb}/2$ ($151$ days) also observed in our periodograms, although the power of which is not high. If the eccentricity increases to $e$ = 0.6, the power spectrum still shows both peaks, but the light curve shows sinusoid-like shape, resembling our ZTF data profile. Moreover, for circular binaries ($e$ = 0) with $q$ = 0.2 and $\alpha$ = 0.1, the break point in the decreasing phase also appears in the light curve. These results suggest that the observed light curve of PKS 0301-243 can be reproduced with appropriate binary parameters under the periodic accretion framework. We therefore consider periodic accretion onto mini-disc to be a plausible mechanism for the $\sim 270$-day optical periodicity observed in this source.

\subsection{Orbital Doppler Boosting} \label{sec:DB}
In addition to periodic accretion in circumbinary disc, the relativistic Doppler Boosting can also produce sinusoidal light curves\cite{D'Orazio_2015,Kelley_2019,Charisi_2022,Lin_2026}.
The orbital motion of SMBHB introduces the periodic variability through Doppler boosting naturally, if the binary orbit is not strictly face-on. In the simplest Doppler boosting model, the SMBHB located in the cavity of the circumbinary disc is assumed to be on a circular orbit, and each SMBH hosts a mini-disc. For a low mass ratio system, the secondary SMBH moving at relativistic speed is closer to the inner edge of the circumbinary disc, accreting more materials. Therefore, its intrinsic luminosity is more luminous compared to the primary SMBH, dominating the observed variability\cite{D'Orazio_2015,Kelley_2019,Charisi_2022}. The relativistic Doppler boosting is expressed as\cite{D'Orazio_2015,Kelley_2019,Charisi_2022}
\begin{align}
    \Delta F &= (3-\alpha_\nu) v_2 \sin i \sin (\omega t + \Phi) /c, \\
    F^{\rm sec}_\nu &= F^{\rm sec}_{\nu,0} ( 1 + \Delta F), \label{equ:F_disc}
\end{align} 
where $F^{\rm sec}_{\nu,0}$ is the stationary flux of the secondary mini-disc, $\alpha_\nu$ is the spectral index characterizing the mini-disc, and $i$ is the inclination of the orbit with respect to the line of sight, or equivalently the angle between the line of sight and the angular momentum vector of the secondary SMBH's orbit. $\Phi$ denotes the orbital phase, and $\omega = 2\pi / P_{\rm orb}$, where $P_{\rm orb} =269^{+13}_{-14}$ days is the observed orbital period, and $c$ is the speed of light. The orbital speed of the secondary SMBH is determined by\cite{D'Orazio_2015,Charisi_2022}
\begin{equation}
    v_{2} = \dfrac{1}{1+q} \left( \dfrac{2\pi G M_{\rm tot}}{P_{\rm orb}} \right)^{\frac{1}{3}}. \label{equ:v2}
\end{equation}
Here, $q=M_2 /M_1$ is the mass ratio, $M_{\rm tot} = M_1 + M_2$ is the total binary mass, where $M_{1,2}$ are the individual masses, and $G$ is the gravitational constant. This model links the observed flux variability directly to the orbital geometry and dynamics of the SMBHB, and is particularly effective for systems with unequal mass ratios, where the Doppler modulation is most pronounced.

\begin{figure}
    \centering
    \includegraphics[width=0.7\textwidth]{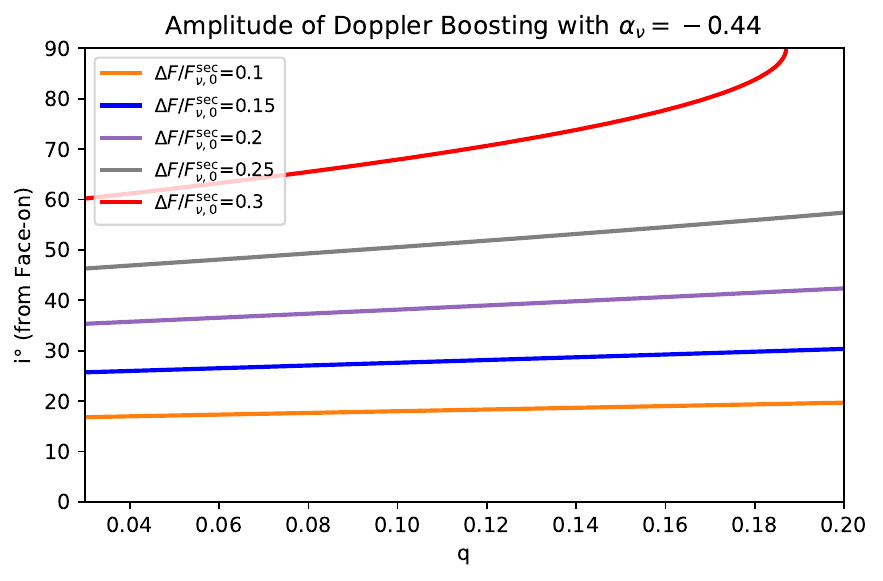}
    \includegraphics[width=0.7\textwidth]{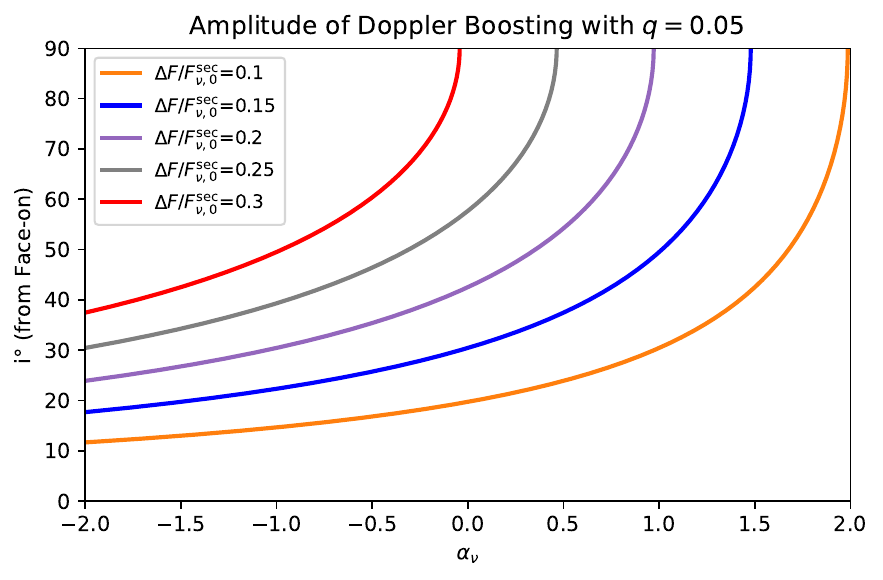}
    \includegraphics[width=0.7\textwidth]{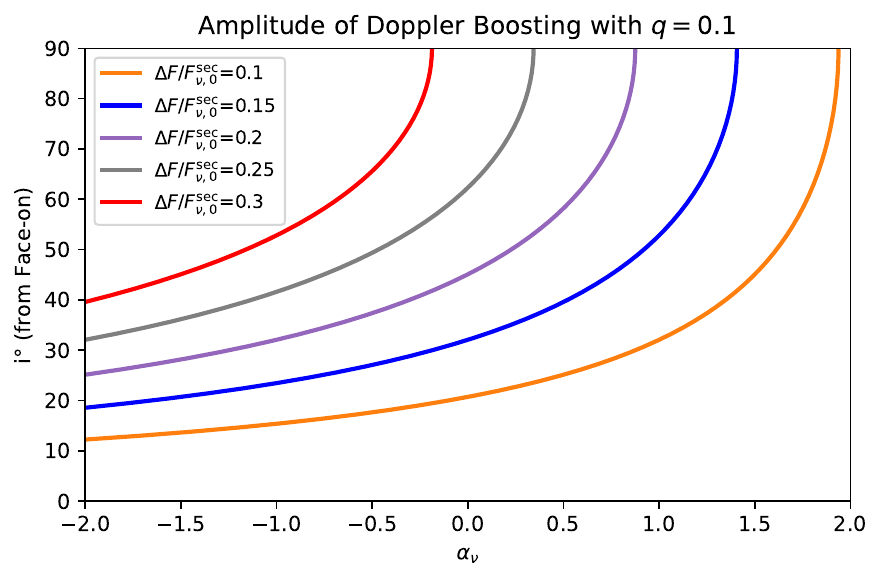}
    \caption{The parameter spaces of relativistic Doppler Boosting of orbital motion. The curves show the orbital inclination $i$, measured from the face-on direction, required to produce different fractional Doppler-boosting amplitudes, $\Delta F_\nu / F_{\nu,0}^{\rm sec}=0.10,~0.15,~0.20,~0.25$, and 0.30. The top panel presents the dependence on the binary mass ratio $q$ for a fixed spectral index $\alpha_\nu=-0.44$. The middle and bottom panels show the dependence on $\alpha_\nu$ for fixed mass ratios $q=0.05$ and $q=0.10$, respectively. Larger modulation amplitudes generally require more highly inclined orbital configurations, while the inferred inclination is also sensitive to the assumed spectral index.}
    \label{fig:DB}
\end{figure}

We should notice that, given a total binary mass of $M_{\rm tot} = 10^{8.95} M_\odot$\cite{Paliya_2021}, the varying component $\Delta F$ is determined by three parameters: mass ratio $q$, spectral index $\alpha$, and inclination $i$ (measured from face-on). And we can not determine these parameters in the light curve fit because of the degeneracy of the three parameters. In addition, there is no observation constraining the three parameters. We try to explore the parameter space that can reproduce the variation amplitude.
Within the Doppler boosting framework, assuming that the optical flux of each mini-disc scales with its accretion rate, the relative contributions of the two black holes using accretion-rate prescriptions derived from hydrodynamic simulations are estimated\cite{Charisi_2022}. They found that the Doppler-modulated emission is dominated by the secondary SMBH for $0.03 \lesssim q \lesssim 0.2$. 
Based on this, we numerically study the combination of the three parameters. First, we fix the spectral index $\alpha$ = -0.44 according to the composite quasar spectrum \cite{Vanden_2001}, and study $\Delta F / F_{\nu,0}^{\rm sec}$ in the combination of $q$ and $i$. The range of $q$ and $i$ are set as $0.03 \lesssim q \lesssim 0.2$ and $0 \degree \lesssim i \lesssim 90\degree$. Second, we fix the mass ratio $q=0.05$ or $q=0.1$, with range of other parameters $-2 \lesssim \alpha_\nu \lesssim 2$ and  $0 \degree \lesssim i \lesssim 90\degree$. The results are shown in Figure \ref{fig:DB}. The MCMC fit for ZTF $g$ and $r$ bands reveal that the values of $\Delta F / F_{\nu,0}^{\rm sec}$ are 18.8\% and 18.6\% in sinusoidal model, and the values of which are 17.4\% and 15.3\% in exponentially decayed sinusoidal model. For $\alpha_\nu = -0.44$,  an inclination $i\sim 30\degree$ is required to reproduce the observed amplitude $\Delta F / F_{\nu,0}^{\rm sec}$, and for $q=0.05$ or $q=0.1$, the range of inclination $20\degree \lesssim i\lesssim 30\degree$ is required.
However, for blazar, the orientation of jet is along our line of sight ($\theta \lesssim 10\degree$), and the jet is approximately perpendicular to the accretion disc, which makes $i\lesssim 10\degree$. If the inclination is $20\degree \lesssim i\lesssim 30\degree$, it implies that the binary orbit is not perpendicular to the blazar jet. 
This is possible since that the binary orbit is not necessary to along the circumbinary disc, and the accretion geometry can be charactered by misaligned circumbinary disc\cite{Nixon_2013,Hayasaki_2014}. Recent numerical simulation study shed light on that the accretion rate of secondary is larger than the primary in the misaligned circumbinary disc with such inclination range\cite{Yang_2026}, consistent with the assumption of Doppler Boosting model. 

Beyond the simple periodicity, relativistic Doppler boosting predicts that the amplitudes of light curves in different bands show the relation\cite{D'Orazio_2015,Xin_2020}
\begin{equation}
    \dfrac{A_{\nu_1}}{A_{\nu_2}} = \dfrac{3-\alpha_{\nu_1}}{3-\alpha_{\nu_2}},
\end{equation}
since the variability amplitude $A_\nu$ depend on the respective spectral index $\alpha_\nu$ in specific band. In our MCMC fit results, ZTF $g$ and $r$ bands exhibit different amplitude (Table \ref{tab:MCMC-LC}). However, this origin is not confirmed since no spectral indices in different bands provide more constrain on this phenomenon. The future multiwavelength observation campaigns may support further study.

\section{Evaluation of alternative periodicity models} \label{sec:Models}
In this work, we propose a scenario that incorporates a jet and a circumbinary disc to investigate the optical periodic variability of the blazar PKS 0301-243. Various physical models have been proposed in the literature to account for periodic or quasi-periodic behaviour in blazars. A detailed discussion of these models is presented in this section. 

\subsection{Instability and fluctuation of accretion disc}
The observed period may originate from instabilities or fluctuations in the accretion disc. In this scenario, processes such as oscillations in the innermost regions of the accretion disc or Kelvin-Helmholtz instabilities may lead to quasi-periodic plasma injections into the jet. These injections can produce periodic variation in the observed jet emission\cite{Bhatta_2016}. The mass of the central SMBH can be estimated via
\begin{equation}
    M = \dfrac{3.23\times10^4 P_{\rm jet}}{(r_e^{3/2} + s)(1+z)} M_\odot, \label{eq:ADins}
\end{equation}
where $P_{\rm jet}$ is the observed period (in seconds), $r_e$ is the emission region radius in units of $GM/c^2$, and $s$ denotes the dimensionless spin parameter of the SMBH. Follow Equation \ref{eq:ADins}, we estimated the mass for both Schwarzschild ($M_{\rm S}$) and Kerr ($M_{\rm K}$) cases. For $M_{\rm S}$, we adopted $r_e = 6$ and $s=0$, and for $M_{\rm K}$, we used $r_e = 1.2$ and $s=0.9982$\cite{Sharma_2025}. The resulting SMBH masses are $M_{\rm S} \sim 1.36\times10^{11} M_\odot$ and $M_{\rm K} \sim 8.64\times10^{11} M_\odot$, respectively. The estimated masses of SMBH are significantly larger than the reported values $M=10^{8.95} M_\odot$\cite{Paliya_2021}, indicating that the instability of disc is unlikely to explain the long-term optical periodicity observed in PKS 0301-243.

\subsection{Lense-Thirring precession}
\begin{figure}
    \centering
    \includegraphics[width=\textwidth]{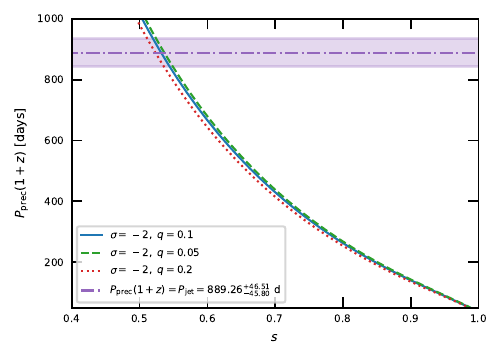}
    \caption{The precession period for PKS 0301-243 as a function of the spin parameter.  
    The purple horizontal line and shaded region mark the observed jet period of $889.26^{+46.51}_{-45.80}$ days.}
    \label{fig:LT}
\end{figure}

Long-term periodic variability observed in blazars are often attributed to an SMBHB system\cite{Ackermann_2015,ONeill_2022}, where gravitational torques from the companion induce jet precession in a misaligned accretion disc orbit\cite{Ackermann_2015}. In particular, Newtonian-driven jet precession in a close SMBHB system may be well associated with periods of $\ge$ 1 year. Alternatively, the Lense-Thirring (LT) precession of the inner edge of the disc can also cause the jet precession\cite{Britzen_2018}. Generally, under the assumption of rigid precession, the precession period of the disc/jet depends on the SMBH's angular momentum $J = GM^2|s|/c$, with the black hole spin parameter $-1 \le s \le 1$. The negative sign corresponding to retrograde rotation. The precession also depends on the precessing part of accretion disc, which spans from the inner radius $r_{\rm in} = \xi_{\rm in} R_{g}$ to the outer radius $r_{\rm out} = \xi_{\rm out} R_g$, where $R_g = GM/c^2$ is the gravitational radius. The inner radius is corresponds to the marginally stable orbit of a Kerr black hole, $r_{\rm ms} = \xi_{\rm ms}R_g$\cite{Caproni_2004}, where $\xi_{\rm ms} = 3 + A_2 \mp [(3-A_1)(3+A_1+2A_2)]^{1/2}$, with $A_1 = 1+(1-s)^{1/3}[(1+s)^{1/3}+(1-s)^{1/3}]$ and $A_2 = (3s^2+A_1^2)^{1/2}$. The disc precession period also depends on the surface density profile of the disc. Assuming the disc surface density depends only on the radius $\Sigma_{\rm d} = \Sigma_0 \xi^\sigma$, we take $\sigma = -2$ to represent a decreasing surface density with the distance from the black hole. Thus, the precession period may be obtained following\cite{Caproni_2004}
\begin{align}
    P_{\mathrm{prec}}&=\frac{2 \pi G M_1}{c^3} \frac{\int_{\xi_{\mathrm{ms}}}^{\xi_{\rm{out}}} \Sigma_{\mathrm{d}}(\xi)[\Upsilon(\xi)]^{-1} \xi^3 \mathrm{d} \xi}{\int_{\xi_{\mathrm{ms}}}^{\xi_{\rm{out}}} \Sigma_{\mathrm{d}}(\xi) \Psi(\xi)[\Upsilon(\xi)]^{-2} \xi^3 \mathrm{d} \xi},
    \\
    \Upsilon(\xi) &= \xi^{3/2} + s, \\
    \Psi(\xi) &= 1 - (1-4s\xi^{-3/2} + 3s^2\xi^{-2})^{1/2}.
\end{align}
In the circumbinary disc scenario, following Roedig et al.\cite{Roedig_2014}, the outer radius of the primary mini-disc is given by $r_{\rm out}^1 = 0.27 q^{-1/3} a$. Adopting the total mass $M_{\rm tot}=10^{8.95} M_\odot$ and the mass ratio $q = 0.1$, we obtain $r_{\rm out} = 48.9 r_{\rm ms}$. For comparison, $r_{\rm out} = 46.7 r_{\rm ms}$ and $r_{\rm out} = 53.4 r_{\rm ms}$ for $q=0.05$, and $q=0.2$, respectively. The LT precession period $P_{\rm prec}$ corresponding observed period $P_{\rm jet}$ is shown in Figure \ref{fig:LT} as a function of the dimensionless spin parameter $s$. The horizontal line and shaded region indicate the observed jet precession period of PKS 0301-243, $P_{\rm jet} = 889^{+47}_{-46}$ days, and its uncertainty, respectively. As shown in Figure \ref{fig:LT}, assuming a steep density profile with $\sigma=-2$, the observed $\sim 890$-day periodicity can be reproduced by LT precession for $0.5 \lesssim s \lesssim 0.55$,  within the mass ratio $0.05\leq q \leq 0.2$. Although the relevant physical parameters cannot be precisely constrained\cite{Britzen_2018}, this consistenncy suggests that LT precession provides a physically plausible explanation for the observed $\sim 890$-day jet precession timescale in PKS 0301-243. Nevertheless, the agreement in precession timescale alone does not uniquely establish an LT origin, and additional constraints on the disc structure and black hole spin would be required to distinguish this scenario from other possible jet precession mechanisms.

\section{GW emission estimation} \label{sec:GW}

SMBHB systems emit GW throughout their inspiral, with significant emission in the nHz range once the separation shrinks  below 0.1 pc\cite{Burke-Spolaor_2019}. Such signals have been detected by PTA collaborations\cite{Xu_2023,EPTA_2023A&A,Zic_2023,Agazie_2023}. The GW frequency in rest frame is twice the intrinsic orbital frequency, $f_{\rm GW} = 2f_{\rm int} = 2/P_{\rm int}$, giving $f_{\rm GW} = 1.09^{+0.06}_{-0.05} \times10^{-7}$ Hz. The GW strain amplitude from a circular orbit, averaged over binary inclination, is estimated as\cite{Maggiore_2007}:
\begin{equation}
    h_{\rm GW} = \dfrac{4(G {\mathcal{M}})^{5/3}(\pi f_{\rm int})^{2/3}}{c^4 D_{\rm L}},
\end{equation}
where $\mathcal{M}$ is the chirp mass, and $D_L$ is the luminosity distance. We estimate the luminosity distance of this target as $D_{\rm L} = 1352.47$ Mpc.
The chirp mass $\mathcal{M}$ is defined as:
\begin{equation}
    \mathcal{M} = \dfrac{(M_1 M_2)^{3/5}}{(M_1 + M_2)^{1/5}}.
\end{equation}
Thus, the GW strain amplitude is computed as $h_{\rm{GW}} = 8.61^{+0.26}_{-0.30} \times 10^{-17}$. The coalescence time for a circular SMBHB system is\cite{Peters_1964}
\begin{equation}
    t_{\rm c} = \dfrac{5 c^5 a^4}{256 G^3 M_1 M_2 (M_1+M_2)},
\end{equation}
here, $a$ is the semimajor axis. Under the assumption of circular orbit, we take $a \sim r$ , yielding $t_{\rm c} = 1122.09^{+143.96}_{-144.14}$ years.

The advent of various GW detectors has opened a new window for multimessenger astronomy through combined gravitational and electromagnetic observations\cite{Mingarelli_2025}. We collect the sensitivity curves of current and planned GW observatories. Except for ASTROD\cite{Ni_2013} and Taiji\cite{Liu_2023}, the sensitivity curves of other GW observatory, along with strain frequency and amplitude ranges of distinct GW sources, are taken from Gravitational Wave Detectors and Sources website\cite{Moore2015}. These sensitivity profiles of characteristic strain amplitude $h_c$ are shown in Figure \ref{fig:GW_detect}. We also simulate the GW evolution utilizing the Python package {\textit{riroriro}}\cite{Buskirk_2019,van_Zeist_2021}. The waveform implementation in \texttt{riroriro} follows the analytic framework\cite{Buskirk_2019}. In this framework, the inspiral phase is described using post-Newtonian expansions, while the late merger is represented through the implicit rotating source approximation, providing an analytic description of the binary waveform across its evolution. For the compact SMBHB considered here, we use the each value of mass ratio ($q$ = 0.05, 0.1, and 0.2) and the luminosity distance $D_{\rm L} = 1352.47$ Mpc as inputs to estimate the corresponding GW properties and assess their potential observability. The evolution track of PKS 0301-243 is overlaid on the sensitivity profiles of $h_c$ (see Figure \ref{fig:GW_detect}), where the time zero corresponds to the moment of SMBHB coalescence.

An additional consideration is the probability of observing a system at such a late stage of binary evolution. For a GW-driven orbital evolution, the frequency evolution in the source rest frame is given by \cite{Haiman_2009,Xin_2021}
\begin{equation}
    \dot{f}_{\rm GW} = \dfrac{96 \pi^{8/3}G^{5/3}}{5c^5} \mathcal{M}^{5/3} f_{\rm GW}^{11/3},
\end{equation}
where $\mathcal{M}$ denotes the chirp mass. The binary residence time in the rest frame can be estimated as\cite{Haiman_2009,Xin_2021}
\begin{equation}
    t_{\rm r}^{\rm int} = \dfrac{3}{8} \frac{f_{\rm GW}}{\dot{f}_{\rm GW}},
\end{equation}
and the observed residence time is $t_{\rm r} = (1+z) t_{\rm r}^{\rm int}$. For the fiducial SMBHB parameters assumed for PKS 0301-243, $q=0.1$, we obtain the residence time of $t_{\rm r}^{\rm int}\approx1110^{+144}_{-146}$ years in the rest frame. The estimated observed residence time is of $t_{\rm r}\approx1406^{+182}_{-185}$ years. For a GW-driven binary, this residence time is essentially identical to the merger time estimated above. This timescale is several orders of magnitude shorter than the typical active lifetime of a quasar, which is generally estimated to be of order a few $\times10^7$ years\cite{Yu_2008,Xin_2021}. The corresponding duty cycle is therefore of order $t_{\rm r}^{\rm int}/(2\times10^7) \sim 5 \times 10^{-5}$, implying that only a very small fraction of active SMBHBs would be expected to be observed in this evolutionary stage at any given time.
Although intrinsically rare, such systems are not necessarily unexpected in modern time-domain surveys. Large-scale periodicity searches have examined samples of order $10^5-10^6$ quasars\cite{Graham_2015_CRTS,Chen_2024_ZTF}. Applying the above duty-cycle estimate suggests that the expected number of such systems is at most of order unity to several, depending on the intrinsic SMBHB occurrence rate and observational selection effects.
Consequently, while the SMBHB interpretation of PKS 0301-243 implies an uncommon evolutionary state, the detection of a small number of candidates in current surveys is not inconsistent with the short residence time inferred for this source\cite{Cavaliere_2019}.

\clearpage

\clearpage

\begin{table}
\caption{Periodic analysis and significance of optical light curves. $\sigma_{\rm local}$ and $\sigma_{\rm global}$ denote local significance and global significance, respectively.}
\centering
    \begin{tabular}{lcccccc}
		\hline
        {Time series} &  & {GLSP} &  & & {WWZ}\\
		   &{period [d]} & $\sigma_{\rm local}$ & $\sigma_{\rm global}$  & {period [d]} & $\sigma_{\rm local}$ & $\sigma_{\rm global}$\\
		\hline
		KAIT &  $921$, $266$ & 99.97\%, 99.63\% & 97.53\%, 84.91\% & $907$ & 98.83\% & 96.68\%\\
        ZTF $g$ band & $271$ & 99.98\% & 98.77\% & $269$ & 99.99\% & 99.70\%\\
		ZTF $r$ band & $273$ & 99.96\% & 98.05\% & $269$ & 99.94\% & 98.53\%\\
		\hline
	\end{tabular}
\label{table:QPO}
\end{table}

\begin{table}
\caption{Significant periodic signals for various light curves.}
\centering
    \begin{tabular}{lc}
		\hline
		{Time series} & {period [WWZ(days)]}\\
		\hline
		Fermi-LAT 0.1-100 GeV (7-day) $\gamma$-ray & $740$\\
            Fermi-LAT 0.1-100 GeV (30-day) $\gamma$-ray & $760$\\
        ASAS-SN $g$ band & $990$,$271$\\
        ASAS-SN $g$ band + ZTF $g$ band & $990$,$273$\\
        ASAS-SN $V$+$g$ band & $907$,$278$\\
        ATLAS c & $880$,$269$\\
        ATLAS o & $931$,$269$\\
        NEOWISE $W1$& $233$\\
        NEOWISE $W2$& $233$\\
		\hline
	\end{tabular}
\label{table:additional-QPO}
\end{table}

\begin{table} [h]
        \centering
        \caption{Bayesian statistical results of KAIT for different stochastic variability models $H$. $H_0$ is pure stochastic process (null), $H_1$ is null + Lorentzian model, and $H_2$ is null + periodic model.}
        \begin{tabular}{cccccc}
        \hline
            Stochastic noise & $\ln Z_0$ & $\ln Z_1$ & $\ln Z_2$ & $B_{1,0}$ & $B_{2,0}$\\
        \hline
            DRW & -19.47 & -17.31 & -18.73 & 8.64 & 2.08 \\
            Matérn-3/2 & -70.29 & -19.37 & -52.82 & $1.3\times 10^{22}$ & $1.0 \times 10^8$\\
            SHO & -87.32 & -21.83 & -67.45 & $2.8\times 10^{28} $ & $4.3\times 10^8$ \\
            DRW + Matérn-3/2 & -15.63 & -15.49 & -15.57 & 1.15 & 1.06 \\
            DRW + SHO & -15.70 & -15.53 & -15.76 & 1.19 & 0.95 \\
            DRW + DRW & -15.00 & -14.21 & -14.10 & 2.20 & 2.44 \\ 
         \hline
        \end{tabular}
        \label{tab:GP-KAIT}
    \end{table}

    \begin{table} [h]
        \centering
        \caption{Same as Table \ref{tab:GP-KAIT}, but for the ZTF $g$-band light curve.}
        \begin{tabular}{cccccc}
        \hline
            Stochastic process & $\ln Z_0$ & $\ln Z_1$ & $\ln Z_2$ & $B_{1,0}$ & $B_{2,0}$\\
        \hline
            DRW & 211.25& 212.39& 211.87& 3.13& 1.86\\
            Matérn-3/2 & 191.96& 213.99& 205.28& $e^{22}$& $e^{13}$\\
            SHO & 183.56& 213.15& 199.56& $e^{30}$& $e^{16}$\\
            DRW + Matérn-3/2 & 211.03& 212.98& 212.16& 7.03& 3.10\\
            DRW + SHO & 210.92& 212.49& 211.86& 4.81& 2.56\\
            DRW + DRW & 211.00& 211.94& 211.78& 2.56& 2.18\\ 
         \hline
        \end{tabular}
        \label{tab:GP-Zg}
    \end{table}

    \begin{table} [h]
        \centering
        \caption{Same as Table \ref{tab:GP-KAIT}, but for the ZTF $r$-band light curve.}
        \begin{tabular}{cccccc}
        \hline
            Stochastic process& $\ln Z_0$ & $\ln Z_1$ & $\ln Z_2$ & $B_{1,0}$ & $B_{2,0}$\\
        \hline
            DRW & 161.29& 162.01& 162.22& 2.05& 2.63\\
            Matérn-3/2 & 129.39& 155.59& 143.72& $e^{26}$& $e^{14}$\\
            SHO & 120.30& 155.12& 136.16& $e^{35}$& $e^{16}$\\
            DRW + Matérn-3/2 & 160.38& 161.01& 161.23& 1.88& 2.34\\
            DRW + SHO & 160.33& 161.04& 161.08& 2.03& 2.12\\
            DRW + DRW & 160.84& 161.72& 162.07& 2.41& 3.42\\ 
         \hline
        \end{tabular}
        \label{tab:GP-Zr}
    \end{table}

    \begin{table} [h]
        \centering
        \caption{Bayesian statistical results for the combined KAIT + ZTF $r$-band light curve with different stochastic variability models $H$. $H_0$ is pure stochastic model (null), $H_1$ is null + Lorentzian model, $H_2$ is null + periodic model, and $H_3$ is null + Lorentzian + Lorentzian model.}
        \begin{tabular}{cccccccc}
        \hline
            Stochastic process & $\ln Z_0$ & $\ln Z_1$ & $\ln Z_2$ & $\ln Z_3$ & $B_{1,0}$ & $B_{2,0}$ & $B_{3,0}$\\
        \hline
            DRW & 71.60& 81.64& 72.32& 80.28& $e^{10}$& 2.05 & $e^{8.7}$\\
            Matérn-3/2 & -67,78& 63.63& -40.32& 61.59& $e^{131}$& $e^{27}$ & $e^{129}$\\
            SHO & -99.32& 57.60& -72.47& 55.66& $e^{158}$& $e^{27}$ & $e^{155}$\\
            DRW + Matérn-3/2 & 82.21& 82.50& 82.19& 81.57& 1.34& 0.98& 0.53\\
            DRW + SHO & 81.80& 82.09& 81.50& 81.30& 1.33& 0.74 & 0.61\\
            DRW + DRW & 82.72& 83.89& 83.52& 82.83& 3.22& 2.23 & 1.12\\ 
         \hline
        \end{tabular}
        \label{tab:GP-Kr}
    \end{table}

\begin{table}
    \centering
    \renewcommand{\arraystretch}{1.5}
    \caption{The MCMC fitting results of light curves. For light curves, 'Data-jet' is KAIT + ZTF $r$ band subtract jet. For models, 'sine' denotes the sinusoidal model (Equation (\ref{eq:sine})), and 'sine\&exp' is the exponentially decayed sinusoidal model (Equation (\ref{eq:sine+exp}). For clarity, the values of $k$ multiply 1000.}
    \begin{tabular}{ccccccc}
    \hline
       Light curve + model& $A$ & $P$ [d] & $t_0$ & $h$ & $k$ & $\sigma_0$ \\
    \hline
        KAIT + sine\&exp  & $0.16^{+0.01}_{-0.01}$ &  $889.93^{+6.99}_{-7.02}$ & $55770.00^{+9.85}_{-9.84}$ & $0.41^{+0.01}_{-0.01}$ & $0.14^{+0.01}_{-0.01}$ & 
        $0.065^{+0.003}_{-0.003}$\\
        ZTF $g$ band + sine & $0.26^{+0.01}_{-0.01}$ & $270.47^{+0.93}_{-0.92}$ & $58376.61^{+3.28}_{-3.26}$ & $1.43^{+0.01}_{-0.01}$ & NAN &
        $0.12^{+0.01}_{-0.01}$\\
        ZTF $g$ band + sine\&exp & $0.25^{+0.01}_{-0.01}$ & $270.05^{+1.04}_{-1.02}$ & $58379.34^{+3.55}_{-3.50}$ & $1.48^{+0.02}_{-0.02}$ & $0.05^{+0.01}_{-0.01}$ &
        $0.11^{+0.01}_{-0.01}$\\
        ZTF $r$ band + sine & $0.35^{+0.02}_{-0.02}$ & $271.39^{+1.03}_{-1.00}$ & $58373.86^{+3.77}_{-3.82}$ & $1.92^{+0.01}_{-0.01}$ & NAN
        & $0.17^{+0.01}_{-0.01}$\\
        ZTF $r$ band + sine\&exp & $0.30^{+0.02}_{-0.02}$ & $269.24^{+1.34}_{-1.28}$ & $58384.84^{+5.28}_{-4.80}$ & $2.08^{+0.02}_{-0.02}$ & $0.09^{+0.01}_{-0.01}$&
        $0.25^{+0.01}_{-0.01}$\\
        Data-jet + sine & $0.17^{+0.02}_{-0.02}$ & $281.03^{+0.92}_{-0.93}$ & $55817.04^{+9.34}_{-9.63}$ & $1.63^{+0.01}_{-0.01}$ & 
        NAN&$0.25^{+0.01}_{-0.01}$ \\
    \hline
    \end{tabular}
    \label{tab:MCMC-LC}
\end{table}

\begin{table}
    \centering
    \renewcommand{\arraystretch}{1.5}
    \caption{The injection-recovery test results using MCMC fitting. For models, 'sine' denotes the sinusoidal model (Eq. (\ref{eq:sine})), and 'sine\&exp' is the exponentially decayed sinusoidal model (Eq. (\ref{eq:sine+exp})). The period priors are determined by period resolution (Eq. (\ref{eq:P_resol}))}
    \begin{tabular}{cccccccc}
    \hline
       Light curve + model & Period Prior & $P_{\rm inj}$ [d] & $P_{\rm rec}$ [d] & $N_{\rm accepted}$ & $N_{\rm edge}$ & $f_{\rm edge}$ & $\delta P$ [d]  \\
    \hline
        KAIT + sine\&exp  & [714,1065] & 889.93 & $889.26^{+46.51}_{-45.80}$ & 9081 & 578 & 6.36\% & -0.67   \\
        ZTF $g$ band + sine & [231,309] & 270.47 & $270.53^{+10.33}_{-10.91}$ & 9594 & 580 & 6.05\% & 0.06 \\
        ZTF $g$ band + sine\&exp & [231,309] & 270.05 & $269.05^{+12.78}_{-13.61}$ & 9475 & 494 & 5.21\% & -1.00 \\
        ZTF $r$ band + sine & [231,309] & 271.39 & $271.89^{+10.18}_{-8.96}$ & 9677 & 521 & 5.38\% & 0.50 \\
        ZTF $r$ band + sine\&exp & [231,309] & 269.24 & $268.04^{+16.07}_{-15.15}$ & 9418 & 643 & 6.83\% & -1.20 \\
    \hline
    \end{tabular}
    \label{tab:MCMC-recover}
\end{table}

\bibliographystyle{nsr}
\bibliography{Ansr_sample}

\end{document}